\documentclass[aps,prl,notitlepage,reprint,twocolumn]{revtex4-2}
\usepackage[utf8]{inputenc}
\usepackage[T1]{fontenc}
\usepackage{epsfig}
\usepackage{epstopdf}
\usepackage{epsfig}
\usepackage{dcolumn}
\usepackage{color}
\usepackage{amsmath}
\usepackage{amssymb}
\usepackage{graphicx}
\usepackage{hyperref}
\usepackage{nameref}
\usepackage{indentfirst}
\usepackage{color}
\usepackage{changes}

\usepackage{bm}
\usepackage{marvosym}
\usepackage{setspace}
\usepackage{pifont}

\usepackage{lineno}

\begin{document}
\title{Anomalous Wave-Packet Dynamics in One-Dimensional Non-Hermitian Lattices}


\author{Yanyan He$^1$}\email{he.yanyan.c6@tohoku.ac.jp}

\author{Tomoki Ozawa$^{1}$}
\email{tomoki.ozawa.d8@tohoku.ac.jp}

\affiliation{$^1$Advanced Institute for Materials Research (WPI-AIMR), Tohoku University, Sendai 980-8577, Japan}

\begin{abstract}
Non-Hermitian (NH) systems have attracted great attention due to their exotic phenomena beyond Hermitian domains. Here we {systematically} study the wave-packet dynamics in general one-dimensional NH lattices and uncover several unexpected phenomena. The group velocity of a wave packet during the time evolution in such NH lattices is not only governed by the real part of the band structure but also by its imaginary part. 
The momentum also evolves due to the imaginary part of the band structure, which can lead to a self-induced Bloch oscillation in the absence of external fields. Furthermore, we discover the wave-packet dynamics can exhibit disorder-free NH jumps even when the energy spectra are entirely real, {which we call the \textit{NH wave-packet jump}}. Finally, we show that the {NH wave-packet jumps} can lead to both positive and negative temporal Goos–Hänchen shifts at the edge.
\end{abstract}

\maketitle
\textit{Introduction}---Wave dynamics lies at the heart of a variety of fundamental phenomena in lattice systems, such as Bloch oscillations \cite{bloch1929quantenmechanik,zener1934theory}, dynamic localization \cite{holthaus1992collapse}, and Anderson localization \cite{anderson1958absence,anderson1985question,schwartz2007transport,segev2013anderson}. Although these phenomena were originally discovered for electronic wavefunctions, similar phenomena turn out to exist also in classical light. Beyond lattice physics, wave dynamics also exhibits various unique phenomena for light. A prominent example is the propagation of a light beam at an interface between media with different refractive indices, where the reflected beam undergoes a spatial displacement, known as the Goos--Hänchen shift \cite{goos1947neuer,rechtsman2011negative,grosche2016spatial,miller1972shifts,huang2008goos,de2010observation,wu2011valley,soboleva2012giant,jiang2015topological}, leading to an important application in sensing \cite{yin2006goos,zhu2024label,berguiga2025complete}.
Recently, the counterpart of this effect in the time domain, i.e., temporal Goos–Hänchen shift (TGHS), was proposed in a dispersive linear
medium with a time boundary \cite{ponomarenko2022goos} and subsequently demonstrated in Hermitian synthetic discrete-time
heterolattices \cite{qin2024temporal}.

On the other hand, considerable attention has been devoted to wave dynamics in non-Hermitian (NH) systems with dissipation, which are described by NH Hamiltonians. Recent studies have shown that when disorder originates from dissipation rather than on-site potentials, conventional Anderson localization is replaced by a quantized jump behavior, termed the NH jump \cite{weidemann2021coexistence,tzortzakakis2021transport,leventis2022non,sahoo2022anomalous,longhi2023anderson,kokkinakis2024anderson,chen2024dynamic,turker2024funneling,ghatak2024diffraction,kokkinakis2025dephasing,li2025universal,shang2025spreading,xue2025non}. This phenomenon stems from complex energy spectra and localized eigenstates, which exhibits a unique feature of NH systems. Another intriguing example is the NH skin effect \cite{lee2016anomalous,yao2018edge,kunst2018biorthogonal,alvarez2018non,lee2019anatomy,yokomizo2019non,brandenbourger2019non,longhi2019probing,xiao2020non,weidemann2020topological,helbig2020generalized,ghatak2020observation,liu2022complex,longhi2022self,shen2025observation,zhao2025two,longhi2025erratic}, where eigenstates become localized at the boundary under open boundary conditions (OBCs) in NH lattices with asymmetric couplings.  This effect has been shown to originate from a nontrivial point-gap topology defined in the complex energy plane \cite{gong2018topological,zhang2020correspondence,okuma2020topological}. It has been noted that the skin effect can give rise to unconventional wave reflections at boundaries where the reflected velocity of a wave packet deviates from the conventional case, called the dynamic skin effect \cite{li2022dynamic,guo2022theoretical,li2024observation}. In addition, self-acceleration of wave packets accompanied by momentum changes has also been found in NH lattices with or without skin effects \cite{silberstein2020berry,solnyshkov2021quantum,longhi2022non,hu2023wave,xue2024self,dong2025non}. Despite these advances, a general theory of wave dynamics in NH lattices is still not fully developed, and a variety of unique physical phenomena remain largely unexplored.  

In this work, we {systematically} study the wave-packet dynamics in general one-dimensional NH lattices. The group velocity of a Gaussian wave packet in such systems is governed not only by the slope of the real part of band structure but also by its imaginary part, when the wave packet evolves with time. Based on this anomalous group velocity, we find a self-induced Bloch oscillation without electric fields. We further discover disorder-free NH jumps in NH lattices under OBCs with entirely real spectra, {which we call the \textit{NH wave-packet jump}, and it} can be explained by the complex band structure under periodic boundary conditions (PBCs).
We also find that when a wave packet is close to the edge, {the NH wave-packet jumps} can also lead to positive and negative TGHSs, accompanied by corresponding frequency blue and red shifts, respectively. {We also find the phase transitions among dynamic skin effects, TGHSs, and NH wave-packet jumps, thereby establishing a fundamental connection between the NH skin effect and observable wave dynamics.}

\begin{figure}[ht!]
\centering
\includegraphics[width=\linewidth]{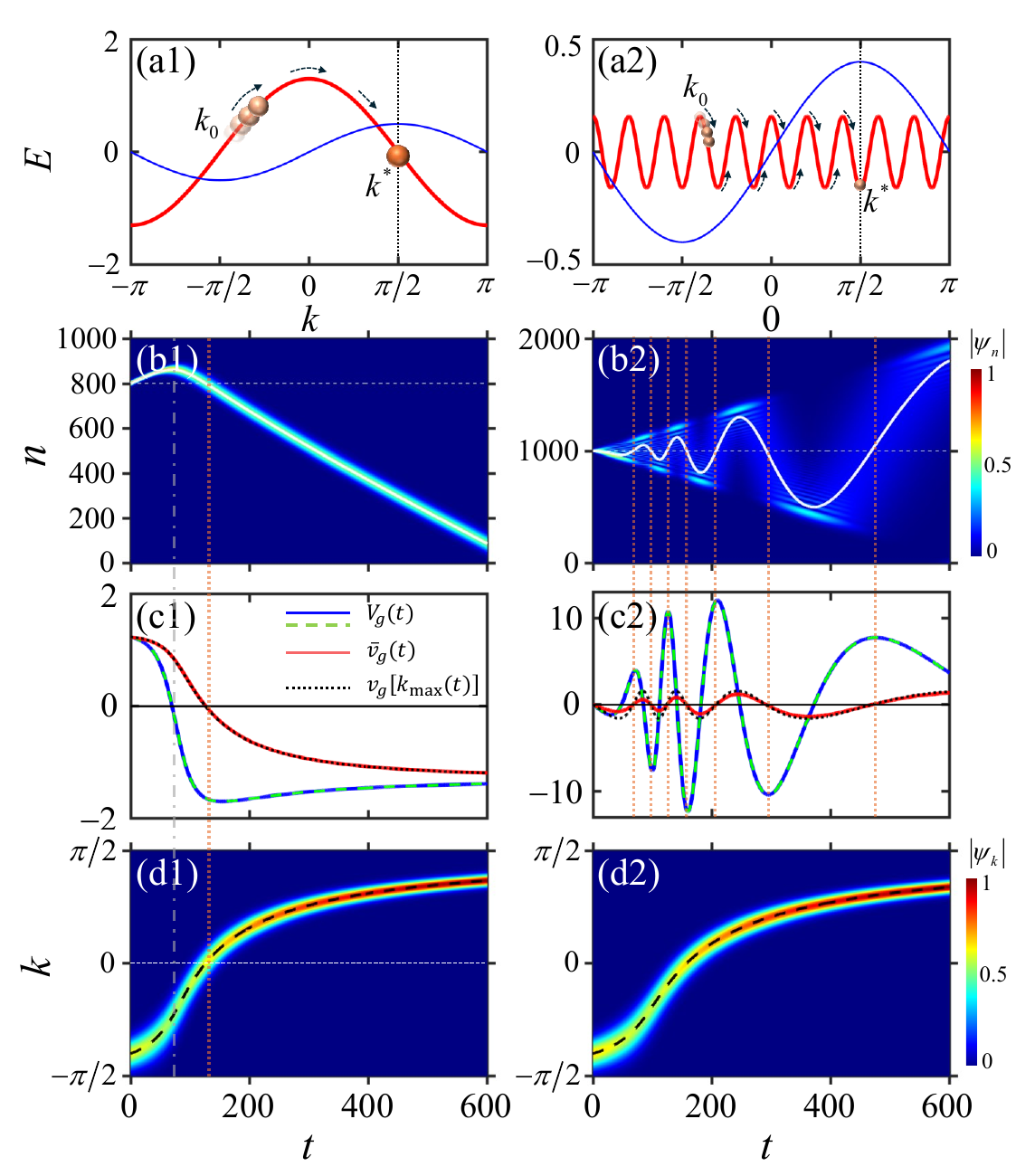}
\caption{(a1,~a2) Band structures, $E_R(k)$ (red lines) and $E_I(k)$ (blue lines). (b1,~b2) The evolution of a wave packet in NH lattices, where the white solid lines denote the predicted center of mass $\bar{n}$ according to Eq.~(\ref{n}). The horizontal white dashed lines are at $n_0$. (c1,~c2) The evolution of group velocities $\bar{v}_g(t)$ (red lines), $v_g[k_\mathrm{max}(t)]$ (black dotted lines), the numerical $V_g(t)$ (blue solid lines), and the predicted $V_g(t)$ (green dashed lines) according to Eq.~(\ref{v_g}). The vertical gray dash-dotted line in (b,~c) denotes $V_g=0$, while the red dotted lines highlight the agreement of zero points of $\bar{n}(t)=n_0$ with $\bar{v}_g(t)=v_g[k_\mathrm{max}(t)]=0$. (d1,~d2) The evolution of the wave packet in momentum space with the theoretical $k_{\mathrm{max}}(t)$ denoted by the black dashed lines. The parameters are $J_1^L=0.9,~J_1^R=0.4$, and $J_{m>1}^{L,R}=0$ in (a1-d1), while $J_1^L=0.2,~J_1^R=-0.2$, and $J_{10}^{L}=J_{10}^{R}=0.08$ in (a2-d2). $k_0=-0.4\pi$ and $\sigma=5$. }
\label{fig1}
\end{figure}

\textit{Anomalous group velocity in NH lattices}---We consider general one-dimensional single-band NH lattice models. (As we will point out, most of the results hold also for multi-band models.) The Hamiltonian we consider in real space is 
\begin{align}
    H=\sum_{n}Va^{\dagger}_{n}a_{n}+\sum_{m,n}\left(J_m^Ra^{\dagger}_{n+m}a_n+J_m^La^{\dagger}_{n}a_{n+m}\right), \label{eq:ham}
\end{align}
 where $a_{n+m}^{\dagger}~ (a_{n+m})$ is the creation (annihilation) operator for the $(n+m)$-th lattice site with the coupling order $m=1,2,3...$ ranging all the sites. $J_m^{R(L)}$ denotes the coupling strength from left to right (right to left) and can take complex values, while $V=\Delta+i\gamma$ is the on-site energy, with $\Delta$ ($\gamma$) being the real (imaginary) potential. The corresponding band structure calculated from the Bloch Hamiltonian under the PBC reads $E(k)=V+\sum_m(J_m^Re^{-imk}+J_m^Le^{imk})=E_R(k)+iE_I(k)$, with $E_R(k)~[E_I(k)]$ being the real (imaginary) part of the band structure and $k$ being the Bloch momentum. Since $V$ just contributes to the overall shift of the energy in complex plane, and does not affect the wave-packet dynamics, we set $V=0$ without loss of generality in the rest of the paper.

We discuss the wave-packet dynamics under this lattice obeying the Schr\"odinger-type equation \cite{longhi2022non,li2022dynamic,xue2024self}
\begin{align}
    i\frac{d}{dt}\vec{\psi}(t) = H\vec{\psi}(t), \label{eq:schrodinger}
\end{align}
where $\vec{\psi}(t)$ is the wavefunction describing the amplitude of the wave at each site at time $t$.
We assume a Gaussian wave packet at $t = 0$, whose amplitude at site $n$ is $\psi_n(0)=\frac{1}{(2\pi\sigma^2)^{1/4}}e^{-\frac{(n-n_0)^2}{4\sigma^2}}e^{ik_0(n-n_0)}$. Here $k_0$, $n_0$, and $\sigma$ are the initial momentum, central position, and width of the wave packet.

We define the momentum-space wave packet by the Fourier transformation of $\psi_n(t)$ as $\psi_k (t) \propto \sum_{n}\psi_n(t)e^{-ikn}$.
Assuming infinite-size lattice, we can analytically solve Eq.~(\ref{eq:schrodinger}) and find $\psi_k(t)
=Ce^{-ikn_0}e^{-iE_R(k)t}e^{-\sigma^2(k-k_0)^2+E_I(k)t}$, where $C$ is a constant.
We call the momentum at which $\psi_k(t)$ is maximum $k_\mathrm{max}(t)$. This $k_\mathrm{max}(t)$ will evolve in time $t$ as \cite{muschietti1993real,gong2018topological,silberstein2020berry,tao2025imaginary} [see also Supplemental Material (SM) \cite{supp}] 
\begin{align}
k_\mathrm{max}(t)=k_0+\frac{t}{2\sigma^2} \left.\frac{dE_I(k)}{dk}\right|_{k = k_\mathrm{max}}.\label{k}
\end{align}
Thus $k_\mathrm{max}$ will change with time if the imaginary part of the band is non-zero $[E_I(k)\ne0]$, and it will increase (decrease) for ${d E_I(k)}/{dk}>(<)0$, and eventually approaches $k^*$ with the maximum $E_I(k)$.

Since the change of $k_\mathrm{max}$, the center of mass of the wave packet $\bar{n}$ will evolve with time as \cite{supp,silberstein2020berry,longhi2022non,hu2023wave,xue2024self}
\begin{align}
\bar{n}(t)&=\frac{\sum_nn|\psi_n(t)|^2}{\sum_n|\psi_n(t)|^2}=n_0+\bar{v}_g(t) t, \label{n}
\end{align}
where $\bar{v}_g(t) \equiv \langle v_g \rangle$, with $v_g \equiv dE_R(k)/dk$ and the momentum-space average of a function $f(k)$ is defined by $\langle f(k) \rangle \equiv \int_{-\pi}^\pi dk |\psi_k(t)|^2 f(k)/\int_{-\pi}^\pi dk |\psi_k(t)|^2$. Note that $\bar{v}_g(t)$ is nothing but the ordinary group velocity for Hermitian systems, which is given by the average of $dE(k)/dk$.
We, however, find that the corresponding group velocity of this non-Hermitian system is
\begin{align}
V_g(t) \equiv \frac{d\bar{n}(t)}{dt} = \bar{v}_g(t)+\frac{d \bar{v}_g (t)}{dt}t,\label{v_g}
\end{align}
which is different from the Hermitian system by the second term. This second term can be rewritten as $\frac{d \bar{v}_g (t)}{dt}t = 2t[\langle E_I v_g\rangle-\langle E_I\rangle\bar{v}_g]$ \cite{supp}, which we call the \textit{anomalous group velocity}.
Note that the existence of anomalous group velocity and the evolution of momentum $k_\mathrm{max}$ [Eq.~(\ref{k})] were first found by L. Muschietti and C. T. Dum in 1993 \cite{muschietti1993real}.


We now numerically demonstrate the anomalous wave packet dynamics considering the Hatano-Nelson model~\cite{hatano1996localization}, which is Eq.~(\ref{eq:ham}) with $J_m^{R(L)} \neq 0$ only for $m = 1$.
For such a model, $E_I(k)$ has the maximum value at $k^*=\pi/2$ [Fig.~\ref{fig1}(a1)]. We simulate the real-space dynamics of a wave packet and plot the normalized wavefunction $|\psi_n(t)|/\sqrt{\sum_n|\psi_n(t)|^2}$ \cite{weidemann2021coexistence,tzortzakakis2021transport,leventis2022non,sahoo2022anomalous,li2022dynamic,longhi2022non} in Fig.~\ref{fig1}(b1), with the center of mass $\bar{n}(t)$ highlighted by a white solid line.
The maximum momentum $k_\mathrm{max}(t)$ of the wave packet evolves towards $k^*$, consistent with the theoretical prediction of Eq.~(\ref{k}) (black dashed line) [Fig.~\ref{fig1}(d1)]. 

In Fig.~\ref{fig1}(c1), we plot the group velocities of the corresponding time evolution. One sees that $\bar{v}_g(t)$ (red solid line) agrees well with $v_g[{k_\mathrm{max}(t)]}$ (black dotted line), which means that the momentum-space average of $v_g = dE_R(k)/dk$ is well approximated by its value at $k_\mathrm{max}(t)$. However, as we noted earlier, the true group velocity of NH system should be different from $\bar{v}_g(t)$. Indeed, in a solid blue line, we plot the true group velocity calculated from the numerically obtained values of $\bar{n}(t)$, and we see a clear deviation from $\bar{v}_g(t)$. The numerically obtained group velocity agrees perfectly with the theoretically predicted value of $V_g(t)$ as given in Eq.~(\ref{v_g}), plotted in a green dashed line.
We note an intrinsic relation between $\bar{v}_g$ and $V_g$; they are related by $\bar{v}_g(t)=(1/t)\int_0^{t}V_g(t^\prime)dt^\prime$. This relation implies that the average velocity of the wave packet evolving to time $t$ is exactly $\bar{v}_g (t)$ [see also Eq.~(\ref{n})]. Since $\bar{v}_g(t) \approx v_g[{k_\mathrm{max}(t)}]$ we see that the average velocity over time $t$ is given by the instantaneous value of $v_g = dE_R(k)/dk$ at $k = k_\mathrm{max}(t)$. We note that Eqs.~(\ref{k}-\ref{v_g}) also hold in multi-band cases (see Appendix A). {We note that the momentum is not conserved here even though the system has a translational symmetry \cite{supp}.}

\textit{Self-induced Bloch oscillations}---The combined effect of momentum evolution and anomalous group velocity allows for a self-induced Bloch oscillation in the absence of an external electric field. This phenomenon happens when $E_R(k)$ oscillates faster than $E_I(k)$. To achieve such a situation, we consider Hermitian long-range couplings $J_{10}^R = J_{10}^L$ and non-Hermitian nearest-neighbor coupling $J_1^L = -J_1^R$. The band structure is plotted in Fig.~\ref{fig1}(a2). According to Eq.~(\ref{k}), the wave packet moves in momentum space toward $k^*=\pi/2$, at which $E_I(k)$ becomes maximum.


Figure~\ref{fig1}(b2) shows the wave-packet evolution in real space along with the center of mass $\bar{n}(t)$, indicating a self-induced Bloch oscillation.
The envelope amplitude of the Bloch oscillation increases linearly with time, unlike the Hermitian case where it remains constant \cite{bloch1929quantenmechanik,zener1934theory}, and its slope is determined by the maximum value of $v_g = dE_R(k)/dk$ \cite{supp}.
The period of oscillation is not constant, and this is because the momentum of the wave packet does not change linearly but becomes slower in time, as one can see from Fig.~\ref{fig1}(d2). 

\begin{figure}[t!]
\centering
\includegraphics[width=0.9\linewidth]{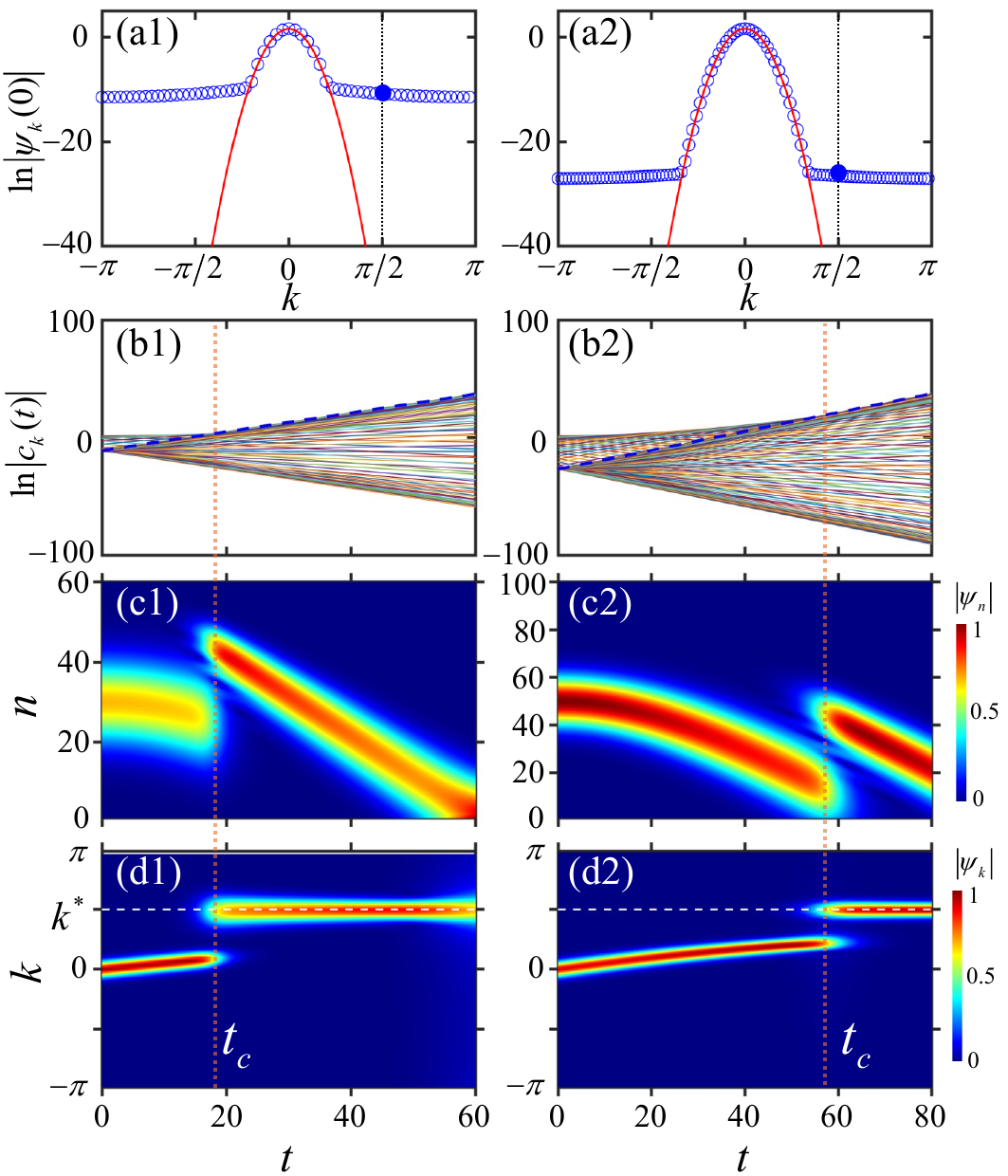}
\caption{(a1,~a2) Initial wave packet in momentum space with a finite (blue circles) and infinite size (red lines). The blue dots indicate the values for $k=k^*=\pi/2$. (b1,~b2) The evolution of $c_k(t)$, where $c_{k=k^*}(t)$ is highlighted by the blue dashed lines. The evolution of the wave packet in (c1,~c2) real and (d1,~d2) momentum spaces under OBCs, where the horizontal white dashed lines denote $k=k^*=\pi/2$. The vertical red dotted lines in (b,~c,~d) highlight the time $t_c$ when the {NH wave-packet jump} occurs. The sizes in (a1-d1) and (a2-d2) are $N=60$ and $N=100$, respectively. Other parameters are $J_1^L=0.9,~J_1^R=0.1,~J_{m>1}^{L,R}=0,~k_0=0$,~and $\sigma=5$.}
\label{fig2}
\end{figure}

\begin{figure}[t!]
\centering
\includegraphics[width=0.9\linewidth]{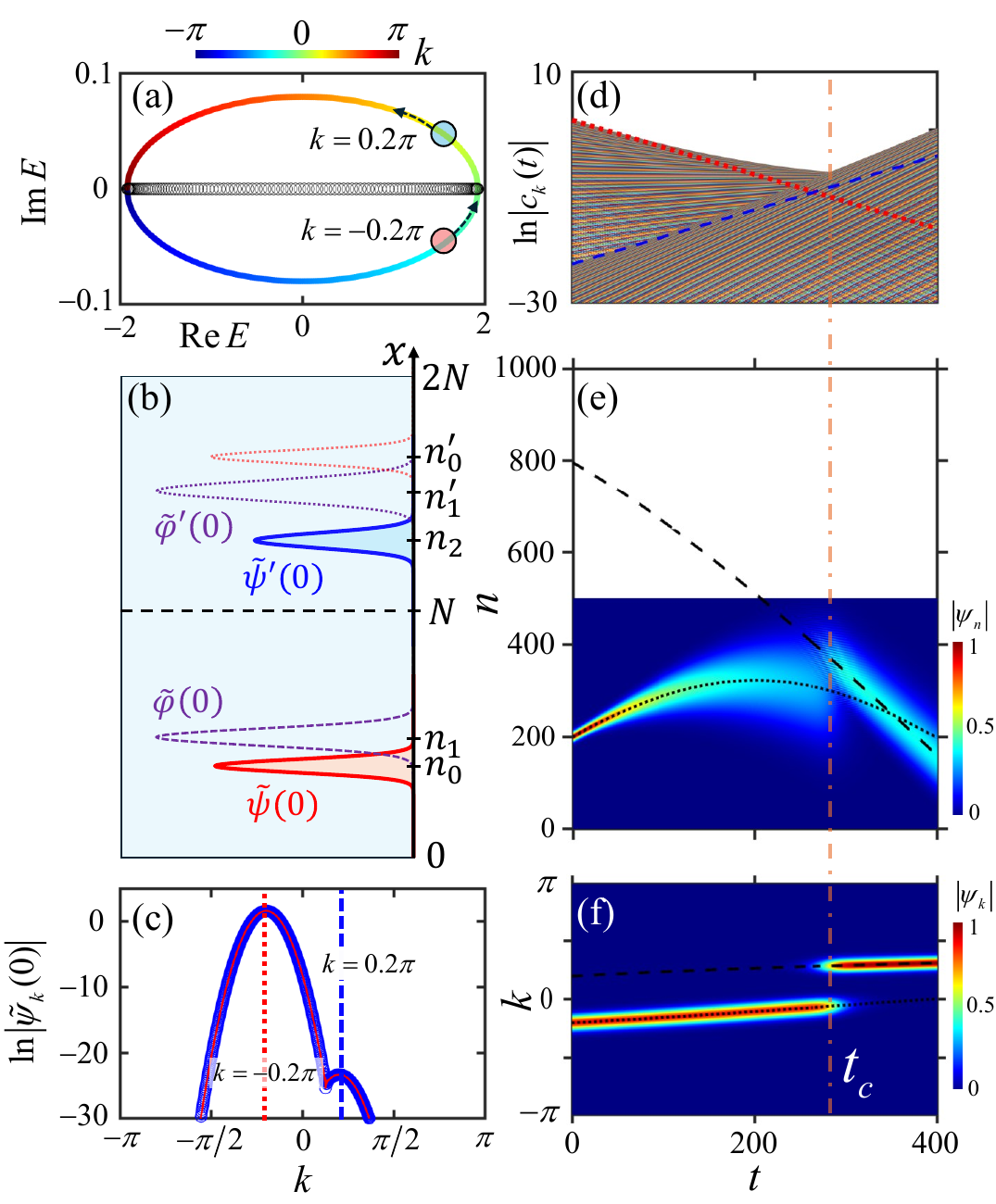}
\caption{(a) Energy spectra under the PBC (colored line) and OBC (black dots) for a finite lattice of size $N=500$. The red (blue) circle denote the initial state of the initial (auxiliary) wave packet, which moves along the spectra denoted by the black dashed arrow. (b) Relative position of different wavefunctions. (c) The initial wave packet $\tilde{\psi}(0)+\tilde{\psi}'(0)$ in momentum space, where the blue dots (red line) denote the result under the finite lattice of size $2N$ (infinite case). (d) The evolution of $c_k(t)$, where the red dotted (blue dashed) line denotes the case for $k_0=-0.2\pi$ ($-k_0=0.2\pi$). (e) The evolution of the wave packet in real space. (f) The evolution of the wave packet in momentum space. The black dotted (dashed) line in (e) and (f) are theoretical center of mass $\bar{n}$ and $k_{\mathrm{max}}(t)$ for the initial (auxiliary) wave packet using Eq.~(\ref{n}) and Eq.~(\ref{k}), respectively. The parameters are $J_L=1,~J_R=0.92,~k_0=-0.2\pi$, and $\sigma=5$.}
\label{fig3}
\end{figure}

\textit{{NH wave-packet jumps} with real spectra}---So far, we have considered lattices which are relatively large compared to the width $\sigma$ of the wave packet. When the lattice is smaller, {NH wave-packet jumps} occur. 
In Figs.~\ref{fig2}(c1) and~\ref{fig2}(c2), we plot the time evolution of a wave packet in a lattice of size 60 and 100, respectively. We observe that jumps of the wave function occur during the evolution, which are the {NH wave-packet jump} we are going to analyze.
We note that the {NH jumps} found here are { associated with wave-packet dynamics in systems with entirely real spectra under OBCs [see black dots in Fig.~\ref{fig3}(a)], which are in contrast to previous studies of NH jumps with disorders and complex spectra~\cite{weidemann2021coexistence,tzortzakakis2021transport,leventis2022non,sahoo2022anomalous,longhi2023anderson,kokkinakis2024anderson,chen2024dynamic,turker2024funneling,ghatak2024diffraction,kokkinakis2025dephasing,li2025universal,shang2025spreading,xue2025non}.}


We consider the same wave packet $\psi_n(0)$ at $t = 0$ in real space as before. In momentum space, when the lattice is finite, the wave function is $\psi_k'(0)=\sum_{n}\psi_n(0)e^{-ik_nn}$,
where $k_n=0,2\pi/N,2\times 2\pi/N,...,(N-1)\times 2\pi/N$.
In Fig.~\ref{fig2}(a1), we show the finite-size momentum-space wave function $\psi_k'(0)$ for $N = 60$ and the infinite-size counterpart $\psi_k(0)$. We see that $\psi_k'(0)$ overlaps with $\psi_k(0)$ near the initial momentum $k_0=0$, and deviates from it for other momenta.  The evolution of the wave packet in momentum space is then $\psi_k'(t)=c_k(t)e^{-iE_Rt}$, with $c_k(t)=\psi_k'(0)e^{E_I(k)t}$ being the amplitude of each momentum at time $t$.
The momentum distribution function $c_k(t)$ with $c_{k=k^*}(t)$ highlighted by the blue dashed line is shown in Fig.~\ref{fig2}(b1). One sees that $c_{k=k^*}(t)$ gradually increases and exceeds other $c_{k\ne k^*}(t)$ near $t_c\approx17$, and then dominate the evolution. Consequently, the wave packet jumps to the state with $k = k^*=\pi/2$, as shown in the real and momentum-space evolution in Figs.~\ref{fig2}(c1) and~\ref{fig2}(d1), respectively. 

Since the initial momentum distribution $\psi_k'(0)$ is related to the size of the lattice, the jump time $t_c$ depends on the size $N$. This is demonstrated in Figs.~\ref{fig2}(a2)-~\ref{fig2}(d2), where we show another case with $N=100$ and keep other parameters the same.
In addition to the jump time, jump position can also be identified \cite{supp}. It should be noted that the wave dynamics remains the same under both OBCs and PBCs, as the wave packet propagates in the bulk. {The NH wave-packet jumps disappear when the lattice size is larger than $2\pi\sigma^2$. We also find that such NH wave-packet jumps occur in a wide range of NH lattices, not restricted to the ones with asymmetric couplings~\cite{supp}.} For multi-band lattices, {NH wave-packet jumps} can occur between different bands (see Appendix A).

\textit{{NH wave-packet jumps} close to the edge}---We now discuss that, { when a wave packet approaches the edge of a system exhibiting the skin effect and has a purely real spectrum under the OBC, NH wave-packet jumps can also occur, but with a different mechanism.}

We consider, as an example, the Hatano-Nelson model (i.e., $J^L_m = J^R_m = 0$ for $m \neq 1$) with $J_R=J^R_1 = 0.92$ and $J_L=J^L_1 = 1$. The energy spectra under the OBC (black dots) and PBC (colored line) are plotted in Fig.~\ref{fig3}(a), where we see that the OBC spectra are real and different from the PBC spectra with a counterclockwise loop, indicating the nontrivial point-gap topology and existence of the skin effect at the left boundary under the OBC \cite{gong2018topological,zhang2020correspondence}. In Figs.~\ref{fig3}(e) and ~\ref{fig3}(f), we plot the time evolution of a wave packet with the initial central momentum $k_0=-0.2\pi$ in real and momentum spaces under the OBC, respectively, for a wave packet closer to and directed toward the edge. We see that a {NH wave-packet jump} occurs at around $t_c \approx 280$. Expanding the wave function right after the {NH wave-packet jump} in eigenstates under PBC, which is allowed even under OBC because PBC eigenstates form a complete bi-orthogonal basis, we see that the wave function is jumped to a state not with the maximum $E_I (k)$, but to a momentum opposite to the initial momentum [Fig.~\ref{fig3}(f)]. Since the {NH wave-packet jump} in Fig.~\ref{fig2} is to a state with the largest $E_I(k)$, a different mechanism is behind the {NH wave-packet jump} in Fig.~\ref{fig3}.

As we now discuss, the {NH wave-packet jump} in Fig.~\ref{fig3} can be explained by the \textit{reflection} at the edge. To understand the {NH wave-packet jump} in terms of reflection, we first perform a similarity transformation to convert the NH Hamiltonian into a Hermitian form. For the Hermitian Hamiltonian, we can use knowledge of ordinary reflection of a wave packet at the edges. Let us assume that the edge is placed at $n = 500$. A reflection at the edge can be understood by considering a hypothetically larger lattice as in the upper half $n > 500$ part of Fig.~\ref{fig3}(e). Reflected wave packet can be identified with an \textit{auxiliary} wave packet created at $n > 500$ moving toward $n < 500$ region.
In Fig.~\ref{fig3}(b), we plot as $\tilde\psi(0)$ the initial wave packet moving up the lattice, and as $\tilde\varphi (0)$ the initial wave packet after the similarity transformation to a Hermitian Hamiltonian. Note that the center of the wave packet changes after the similarity transformation. The wave packet $\tilde\varphi^\prime(0)$ is an auxiliary one obtained by inverting $\tilde\varphi (0)$ at the original edge at $n = 500$. The wave packet $\tilde\psi^\prime(0)$ is the auxiliary wave packet after inverse-similarity transformation, with the opposite central momentum $-k_0$ and shifted position, which we denote by $n_2$, which is different from the mirror-symmetric point of $\tilde{\psi}(t)$, which we denote by $n_0'$. The reflection, in this language, is nothing but the wave packet $\tilde\psi (t)$ moving to $n > 500$ region while the auxiliary wave packet $\tilde\psi^\prime (t)$ moving into $n < 500$ region. In Fig.~\ref{fig3}(c), we plot the momentum-space distribution of $\tilde\psi(0) + \tilde\psi^\prime(0)$.
Since the initial (auxiliary) wave packet with the central momentum $k_0$ ($-k_0$) experiences loss (gain) [see the red and blue circles in Fig.~\ref{fig3}(a)], the amplitude $|\tilde\psi (t)|~ \left(|\tilde\psi^\prime (t)|\right)$ will decrease (increase). When the amplitude of the auxiliary wave packet $\tilde\psi^\prime (t)$ exceed that of the initial wave packet $\tilde\psi (t)$, as described by the momentum distribution function $c_k(t)$ in Fig.~\ref{fig3}(d), {the NH wave-packet jump} occurs and the auxiliary wave packet starts to dominate the evolution. More quantitative argument is given in Appendix B. The jump time $t_c$ dependents on the ratio $J_R/J_L$, the width of the wave packet $\sigma$, and the size $N$ (see SM \cite{supp}).

\begin{figure}[t!]
\centering
\includegraphics[width=\linewidth]{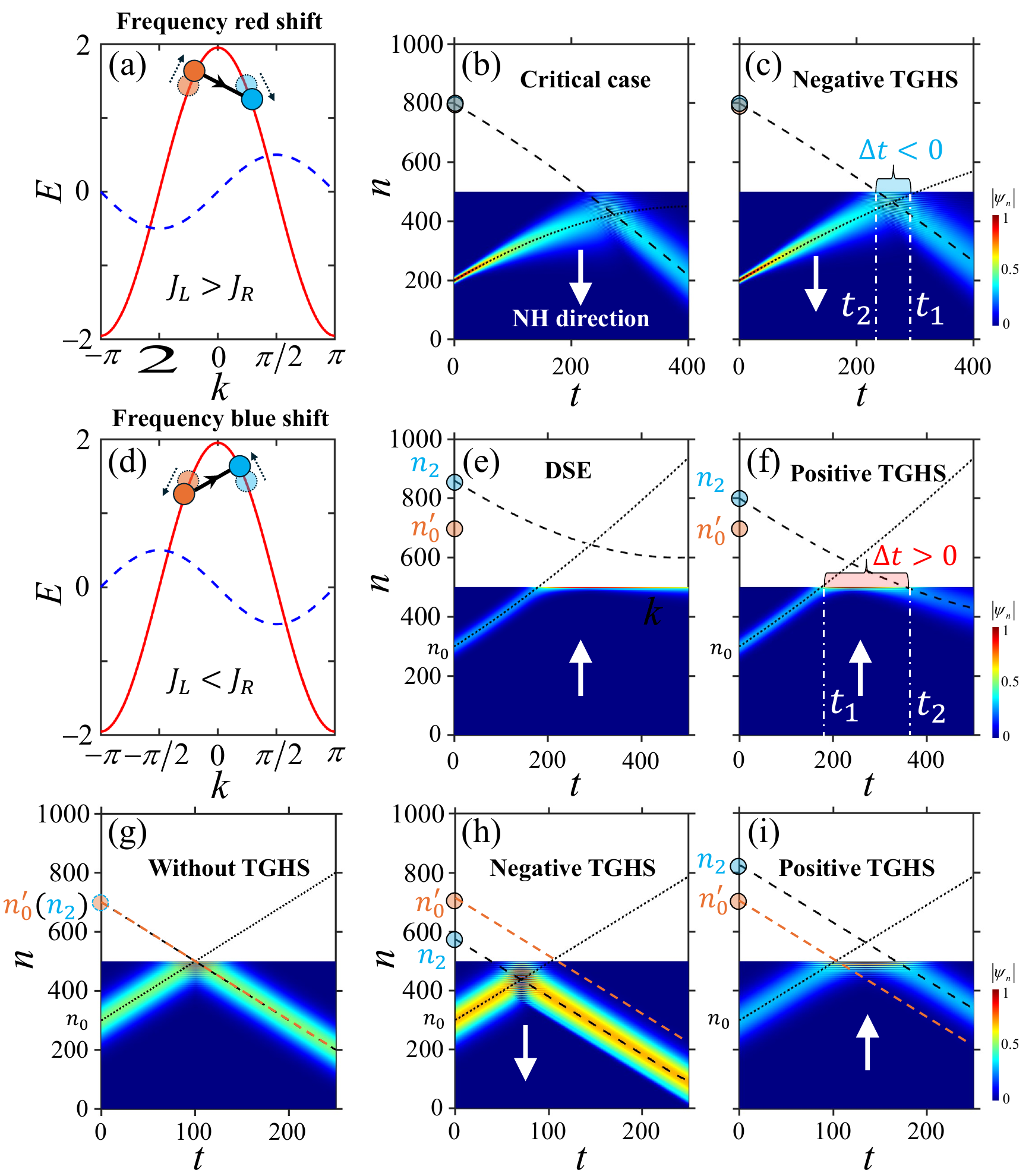}
\caption{(a,~d) Typical band structures, $E_R(k)$ (red solid lines) and $E_I(k)$ (blue dashed lines) with (a) $J_L>J_R$ and (d) $J_L<J_R$. (b,~c,~e,~f,~g-i) The evolution of the wave packet in real space under the OBC. The parameters are $J_L=1,~J_R=0.96$ (b), $J_L=1,~J_R=0.98$ (c), $J_L=0.7,~J_R=1$ (e), $J_L=0.8,~J_R=1$ (f), 
$J_L=J_R=1$ (g),~
$J_L=1,~J_R=0.95$ (h), and $J_L=0.95,~J_R=1$ (i). $k_0=-0.2\pi,~\sigma=5$ for (b,~c), $k_0=-0.2\pi,~\sigma=15$ for (e,~f), and $k_0=-0.5\pi,~\sigma=35$ for (g,~h,~i). The red (blue) circles denote $n_0'$ ($n_2$). The red dashed lines in (g-i) denote the Hermitian cases.
}
\label{fig4}
\end{figure}

\textit{TGHSs induced by skin effects}---
The model we have just considered is the Hatano-Nelson model with $J_R/J_L = 0.92$. As we make the ratio $J_R/J_L = 0.92$ closer to 1, namely making the system closer to the Hermitian limit, {the NH wave-packet jump} will turn into TGHS. In Figs.~\ref{fig4}(b) and~\ref{fig4}(c), we plot the reflection of a wave packet for $J_R/J_L = 0.96$ and $J_R/J_L = 0.98$. When $J_R/J_L= 0.92$, {the NH wave-packet jump} occurs away from the boundary, but as the ratio $J_R/J_L$ becomes 0.96, the reflection point touches the boundary [Fig.~\ref{fig4}(b)]. As the ratio gets even larger (such as 0.98), we observe the negative TGHS [Fig.~\ref{fig4}(c)], namely $\Delta t=t_2-t_1<0$ with $t_2$ and $t_1$ being the time that the center of mass of $\tilde{\psi}'(t)$ and $\tilde{\psi}(t)$, respectively, touches the boundary. Note that when the ratio reaches 1, negative TGHS disappears and a simple reflection at the boundary occurs.

\begin{figure}[t!]
\centering
\includegraphics[width=\linewidth]{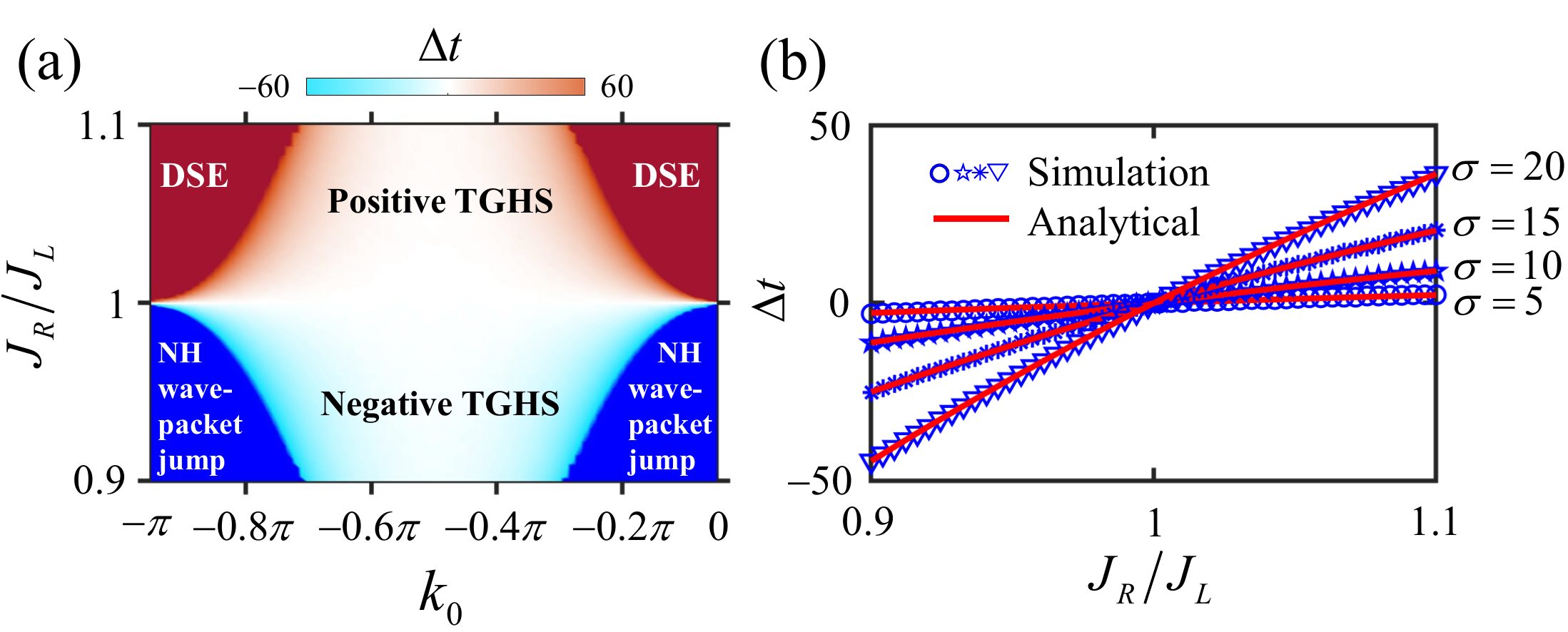}
\caption{{(a) Phase diagram under different choices of $k_0$ and $J_R/J_L$. Other parameters are $\sigma=5$ and $N=n_0+200$. (b) Relation of TGHS $\Delta t$ with $J_R/J_L$ for different $\sigma$ at $k_0=-0.5\pi$.}}
\label{fig5}
\end{figure}

Now let us consider what happens when we make the ratio $J_R/J_L$ greater than 1. In Figs.~\ref{fig4}(e) and ~\ref{fig4}(f), we plot the cases for $J_R/J_L = 1/0.7$ and $J_R/J_L = 1/0.8$, respectively. When $J_R/J_L = 1/0.8$, we observe the positive TGHS. This is because, as described in Fig.~\ref{fig4}(f), the initial and auxiliary wave packets cross at the extended region beyond the edge. As the non-Hermiticity is increased to $J_R/J_L = 1/0.7$, reflection ceases to occur [Fig.~\ref{fig4}(e)]; this phenomenon was previously found and termed dynamic skin effect (DSE) ~\cite{li2022dynamic,li2024observation}. {Fig.~\ref{fig5}(a) is the numerically obtained phase diagram in the $(k_0, J_R/J_L)$ plane, which shows the phase transitions among the DSE, NH wave-packet jump, and TGHS.}

For the case of $k_0=-0.5\pi$, where $E_I(k_0)$ takes an extreme value, the momentum remains unchanged, allowing an analytical TGHS 
\begin{align}
\Delta t=\frac{n_2-n_0'}{v_g(k_0)}=\frac{2\sigma^2 \mathrm{ln}({J_R/J_L})}{J_L+J_R},    
\end{align}
which shows $\Delta t>0~(<0)$ for $J_R>(<)~J_L$, and $\Delta t=0$ for the Hermitian case with $J_R=L_L$ [see Figs.~\ref{fig4}(g,~h,~i)]. {The agreement between the analytical and numeral results for different $\sigma$ is demonstrated in Fig.~\ref{fig5}(b).}

We have further found that, the peak momentum $k_\mathrm{max}$ exhibits unequal amplitudes during the reflection process, accompanied by a frequency (energy) shift. As we described in Figs.~\ref{fig4}(a) and ~\ref{fig4}(d), the momenta of the incident and auxiliary states move from the red and blue dotted circles along the $+k~(-k)$ direction to the solid circles, where the reflection occurs, indicating the frequency red (blue) shift for $J_L>J_R~(J_L<J_R)$ (see  Appendix C). {Note the above NH wave-packet jumps, DSEs, and TGHSs with phase transitions can occur in other models beyond the Hatano-Nelson model \cite{supp}}.

\textit{{Conclusion}}---We have found that, {beyond the static properties of point-gap topologies of the complex energy bands and eigenstate localizations of the skin effects}, even in a simple setting of wave packet propagation on a one-dimensional lattice without any external force, non-Hermitian Hamiltonian leads to phenomena quite distinct from Hermitian Hamiltonians, such as wave-packet evolution with an anomalous group velocity, self-induced Bloch oscillations, {NH wave-packet jumps} without disorder and with completely real spectra, and TGHSs. {Our work hence bridges the gap of static nontrivial point-gap topology with skin effects and observable wave propagation phenomena. 
Adding external forces, temporal and spatial modulations, effects of other bands, and nonlinearity can lead to further exotic phenomena, which are the topics of future works.} Extending the results to two and higher dimensions, including effects of (both real and imaginary) magnetic fields, is also an interesting perspective. Our results can be verified in a variety of experimental platforms.
{The most promising one might be the synthetic temporal photonic lattices, where it is possible to engineer the asymmetric couplings \cite{weidemann2020topological}, open boundary conditions \cite{wang2025nonlinear}, and Gaussian wave packets \cite{qin2024temporal}. Other platforms where our proposal can be tested include} photonic ring resonators \cite{longhi2015non,zhu2020photonic,xin2023manipulating}, synthetic momentum lattices \cite{dong2024quantum,liang2022dynamic}, acoustic lattices \cite{zhang2021acoustic,gu2022transient}, and photonic quantum walks \cite{xiao2020non,xue2024self}. 

{\textit{Note added}---After submission of our paper, the papers~\cite{jana2025solution,beck2025wave,kokkinakis2025non} appeared. The papers \cite{jana2025solution,beck2025wave} find the similar NH wave-packet jumps for wave packets propagating near the edge on a Hatano-Nelson lattice, where this phenomenon is termed "non-Hermitian reflection" \cite{beck2025wave}. The paper \cite{kokkinakis2025non} also discusses NH jumps under purely real spectrum, but in the presence of disorders. }

\begin{acknowledgments}
\textit{Acknowledgments}—--This work is supported by JSPS KAKENHI Grant No. JP24K00548 and JST PRESTO Grant No. JPMJPR2353.\\
\end{acknowledgments}

\twocolumngrid
\bibliography{reference}
\bigskip

\bigskip
\appendix


\textit{Appendix A: Anomalous group velocity and {NH wave-packet jumps} in multi-band NH lattices}---In this section, we discuss that the anomalous group velocity and {NH wave-packet jumps} can also exist in multi-band lattices. We consider the general NH Su-Schrieffer-Heeger (SSH) model \cite{lee2016anomalous,lieu2018topological,yao2018edge,kunst2018biorthogonal} as an example, whose momentum-space Hamiltonian is given by
\begin{align}
    H_k=\begin{bmatrix}
\Delta_a+i\gamma_a  & v+\gamma/2+w e^{-ik}\\  v-\gamma/2+w e^{ik} & \Delta_b+i\gamma_b
\end{bmatrix},
\end{align}
where $\Delta_a ~(\Delta_b)$ and $i\gamma_a~(i\gamma_b)$ represent the on-site real and imaginary potentials for the sublattice $a~(b)$ and $\gamma$ denote the strength of asymmetric couplings. The general evolution of the wave function in momentum space is 
\begin{align}
\psi_k(t)=c_1(0)e^{-iE_1(k)t}|u_1(k)\rangle+c_2(0)e^{-iE_2(k)t}|u_2(k)\rangle,
\end{align}
where $|u_1 (k)\rangle~(|u_2(k)\rangle)$ is the right eigenstate for the eigenenergy $E_1(k)~[E_2(k)]$. The initial coefficient is $c_{1(2)}(0)=\langle v_{1(2)}(k)|\psi_k(0)\rangle$, with $|v_1 (k)\rangle$ and $|v_2(k)\rangle$ being the left eigenstates. They satisfy the biorthogonal condition $
    \langle v_m|u_n\rangle=\delta_{mn}$.

We first discuss the wave-packet dynamics in this lattice where the initial wave packet is prepared in the upper band $E_1(k)$ in the form
\begin{align}
    \begin{bmatrix}
        \psi_{k,a}(0)\\\psi_{k,b}(0)
    \end{bmatrix}=W_k|u_1\rangle,
\end{align}
where $W_k=Ce^{-\sigma^2(k-k_0)^2}e^{-ikn_0}$. The real-space wave packet can be obtained by taking the Fourier transformation of this function.
\begin{figure}[t!]
\centering
\includegraphics[width=\linewidth]{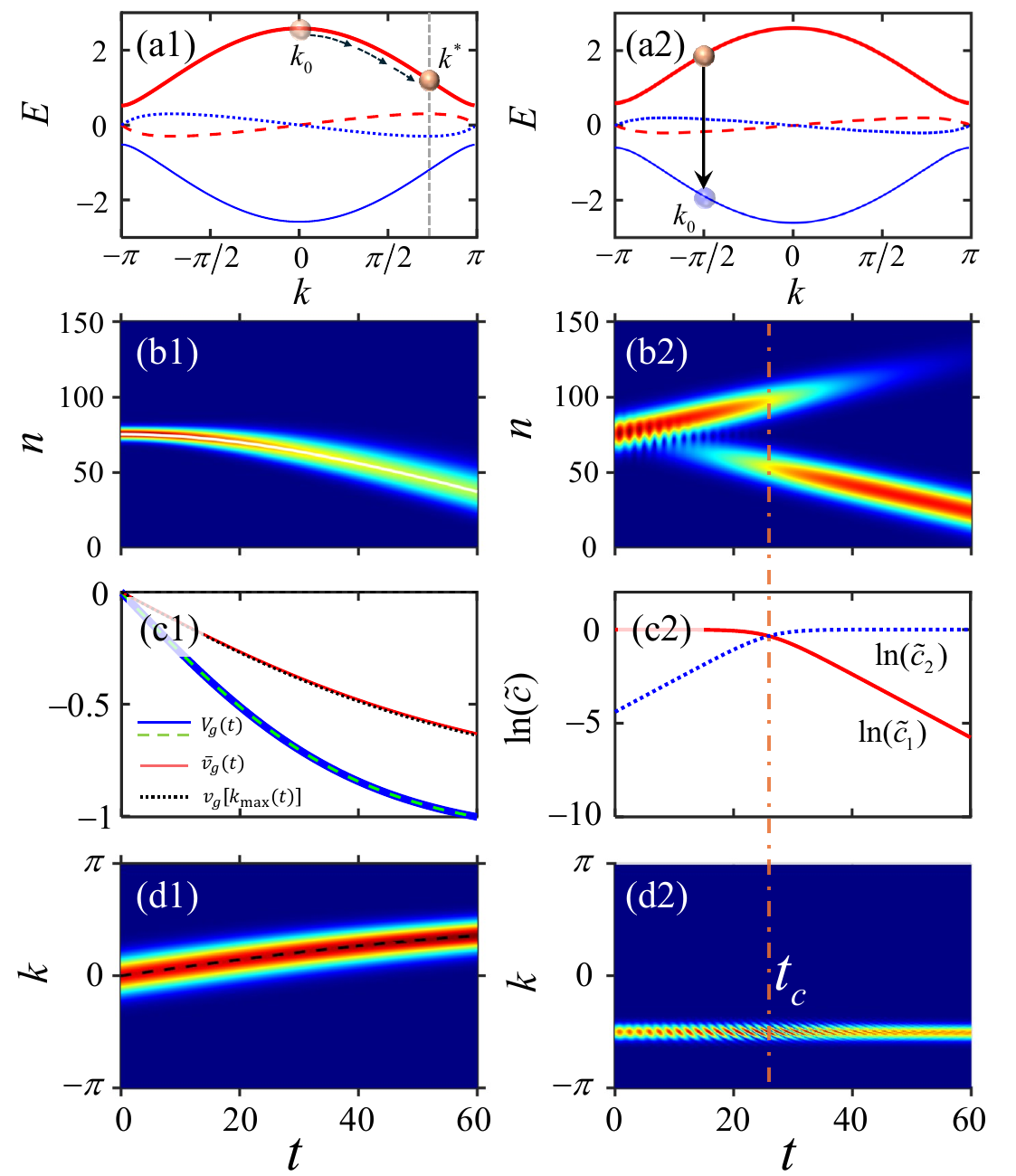}
\caption{(a1,~a2) Band structures, where the solid (dashed) lines denote the $\mathrm{Re}(E_{1,2})$ [$\mathrm{Im}(E_{1,2})$] and red (blue) lines represent the upper (lower) bands. $\mathrm{Im}(E_{1,2})$ is enlarged by $4$ times in (a2). The gray dashed line in (a1) denotes the $k=k^*$. The black solid arrow in (a2) denotes the jump from the state in the upper band to the state on the lower band. (b1,~b2) The evolution of the normalized $\psi_{n,a}(t)$ in real space. (c1) The evolution of group velocities. (c2) The evolution of normalized participation ratio of the upper (lower) band $\tilde{c}_1$ ($\tilde{c}_2$). (d1,~d2) The evolution of the wave packet in momentum space $\psi_{k,a}(t)$. The parameters in (a1-d1) are $v=1,~w=1.6,~\gamma=0.6,~k_0=0$, and $\sigma=2$, while the parameters in (a2-d2) are $v=1,~w=1.6,~\gamma=0.1,~k_0=-0.5\pi$, and~$\sigma=5$.}
\label{fig_EM1}
\end{figure}
In Figs.~\ref{fig_EM1}(a1)-~\ref{fig_EM1}(d1), we show the evolution of the wave packet. One sees that the state with an initial momentum at $k_0=0$ evolves along the upper band $\mathrm{Re}(E_1)$ to the state at $k^*$, which has the maximum $\mathrm{Im}(E_1)$. The corresponding evolutions in real and momentum spaces for the sublattice $a$ are shown in Figs.~\ref{fig_EM1}(b1) and~\ref{fig_EM1}(d1), where we see that the momentum gradually increases from $k_0$ to $k^*$. The group velocity of the exact dynamics $V_g(t)$ agrees well with the predicted result in Eq.~(\ref{v_g}), which is different from the $v_g[k_\mathrm{max}(t)]$ and $\bar{v}_g(t)$ [see Fig.~\ref{fig_EM1}(c1)]. The dynamics for the sublattice $b$ is the same to that for the sublattice $a$. 

We then discuss the case of the wave dynamics where the initial wave packet is prepared in both bands in the form 
\begin{align}
    \begin{bmatrix}
        \psi_{k,a}(0)\\\psi_{k,b}(0)
    \end{bmatrix}=c_1(0)W_k|u_1\rangle+c_2(0)W_k|u_2\rangle,
\end{align}
with $c_1(0)~[c_2(0)]$ being the component on the upper (lower) band. We consider that the initial wave packet is mainly prepared in the upper band with $c_1(0)=0.9,~ c_2(0)=0.1$. The evolutions of the wave packet in both real and momentum spaces are shown in Figs.~\ref{fig_EM1}(b2) and~\ref{fig_EM1}(d2), where a {NH wave-packet jump} is observed in real space, while the momentum remains the same before and after the jump. Such an {NH wave-packet jump} can be understood from the competition of the two bands during the time evolution. We calculate the normalized participation ratio of each band $\tilde{c}_{1(2)}(t)=|c_{1(2)}(t)|/\sqrt{|c_1(t)|^2+|c_2(t)|^2}$ with time in Fig.~\ref{fig_EM1}(c2). We see that $\tilde{c}_2(0)\approx1$, indicating the initial wave packet is mainly located in the upper band. With increasing of time, the component of the upper band $\tilde{c}_1(t)$ gradually decreases, as the state in the upper band is lossy with $\mathrm{Im}(E_1)<0$, while the component of the lower band $\tilde{c}_2(t)$ gradually increases, as the state in this band has gain with $\mathrm{Im}(E_2)>0$ [see Fig.~\ref{fig_EM1}(a2)]. Therefore, at time $t_c$ where $\tilde{c}_2(t_c)=\tilde{c}_1(t_c)$, the wave packet is dominated by the lower band, indicating the {NH wave-packet jump} in real-space dynamics [see Fig.~\ref{fig_EM1}(b2)]. During this process, the momentum nearly remains the same, indicating the jumps between bands, as shown in Fig.~\ref{fig_EM1}(d2) [see also Fig.~\ref{fig_EM1}(a2)].

\textit{Appendix B: Similarity transformation for {NH wave-packet jumps} close to the edge---} The evolution of the wave packet in a finite lattice of size $N$ is described, as usual, by $\psi_n(t)=e^{-iHt}\psi_n(0)$. The wave dynamics is equivalent to that in a larger-size lattice, e.g., $2N$, before the wave packet touches the boundary at $x=N$, with another
wave evolution equation $\tilde\psi_n(t)=e^{-i\tilde Ht}\tilde\psi_n(0)$. Here $\tilde\psi_n(0),\tilde\psi_n(t)$ and $\tilde{H}$ are the wavefunctions and Hamiltonian that are the same as these in the original lattice but with a larger size $2N$.

We now perform a similarity transformation \cite{yao2018edge,li2022dynamic} to obtain the equivalent wave equation \begin{align}
\tilde\psi_n(t)=\tilde S\tilde S^{-1}e^{-i{\tilde{H}}t}\tilde S\tilde S^{-1}\tilde\psi_n(0)=
\tilde Se^{-i\bar{\tilde{H}}t}\tilde\varphi_n(0).
\end{align}
Here $\tilde S=\mathrm{diag}\{r,r^2,r^3,...r^{2N}\}$ with $r=\sqrt{J_R/J_L}$, and $\tilde\varphi_n(0)=\tilde S^{-1}\tilde\psi_n(0)=Ae^{(\sigma^2h^2+hn_0)}e^{-\frac{(n-n_1)^2}{4\sigma^2}+ik_0(n-n_0)}$ with $A=\frac{1}{(2\pi\sigma^2)^{1/4}}$, $h=-\mathrm{ln}r$ and $n_1=n_0+2\sigma^2h$, which is also a Gaussian wave packet but with a shifted center at $n_1$ and an amplitude adjustment $e^{(\sigma^2h^2+hn_0)}$ \cite{li2022dynamic} [see Fig.~\ref{fig3}(b)]. The evolution of the wave packet $\tilde\psi_n(t)$ can be obtained by evolving a wave packet $\tilde \varphi_n(0)$ in time under the Hermitian Hamiltonian $\bar{\tilde H}=\tilde{S}^{-1}\tilde H\tilde{S}$, and then multiplying it by a similarity matrix $\tilde S$. 

We now discuss the wave dynamics at the boundary. It is known that the reflected wave of the incident wave packet $\tilde \varphi_n(0)$, with the initial central momentum $k_0$ under the Hermitian lattice with $\bar{\tilde H}$, can be regarded as the transmitted wave of an auxiliary wave packet with an opposite central momentum $-k_0$ at the mirror-reflection point {$x=n_1'=2x_c-n_1$, in the form $\tilde \varphi_n'(0)=-Ae^{(\sigma^2h^2+hn_0)}e^{-\frac{(n-n_1')^2}{4\sigma^2}-ik_0(n-n_0')}$ with $x_c=N+1$ [see Fig.~\ref{fig3}(b)]. Note the minus in front of $A$ is due to the $\pi$ phase shift of reflection.} Therefore, the evolution equation for the wave dynamics is equivalent to 
\begin{align} \label{psi_SS}
\tilde\psi_n(t)=\tilde Se^{-i\bar{\tilde{H}}t}\left[\tilde\varphi_n(0)+\tilde\varphi_n'(0)\right]=e^{-i\tilde{H}t}\left[\tilde\psi_n(0)+\tilde\psi_n'(0)\right],
\end{align}
where 
$\tilde\psi_n'(0)=\tilde{S}\tilde\varphi_n'(0)=-Ae^{[4\sigma^2h^2+2h(n_0-x_c)]}\times e^{-\frac{(n-n_2)^2}{4\sigma^2}-ik_0(n-n_0')}$ centered at $n_2=2x_c-n_0-4\sigma^2h$. For the Hermitian case with $h=0$, the auxiliary wave packet changes back to $\tilde\psi_n'(0)=-Ae^{-\frac{(n-n_0')^2}{4\sigma^2}-ik_0(n-n_0')}$, located at the mirror-symmetric point of the initial wave packet $n_0'=2x_c-n_0$ [Fig.~\ref{fig3}(b)].

The wave dynamics at the boundary $x=N$ is changed to a bulk dynamics with the initial wave packet $\tilde\psi_n(0)$ (red curve) and an additional auxiliary wave packet $\tilde\psi_n'(0)$ located at $n_2$ (blue curve) [Fig.~\ref{fig3}(b)]. The initial wave function in momentum space is then $
\tilde\psi_k(0)=Ce^{-ikn_0}e^{-\sigma^2(k-k_0)^2}-Ce^{[4\sigma^2h^2+2h(n_0-x_c)]}e^{-i(k+k_0)n_2}e^{ik_0n_0'}e^{-\sigma^2(k+k_0)^2}$ with two central momenta $\pm k_0$ [see Fig.~\ref{fig3}(c)]. The evolution of the both wave packets in momentum space is then $\tilde{\psi}_k(t)=c_k(t)e^{-iE_Rt}$, with $c_k(t)=\tilde{\psi}_k(0)e^{E_I(k)t}$ shown in Fig.~\ref{fig3}(d).\\

\textit{Appendix C: Frequency shifts with the TGHSs}---Due to the momentum shift induced by non-Hermiticity, the reflection at the boundary no longer satisfies energy conservation, accompanied by a frequency shift. For the case $J_L>J_R$ in Fig.~\ref{fig4}(a), the momenta of the two initial wave packets at $k_0$ (red dotted circle) and $-k_0$ (blue dotted circle) move along the $+k$ direction toward the states highlighted by the red and blue solid circles, where the reflection occurs. Therefore, the incident state (red solid circle) before the boundary with a higher frequency projects to the reflected state at the boundary (blue solid circle) with a lower frequency, as indicated by the black solid arrow with a frequency red shift. Consequently, this projection does not satisfy the energy (frequency) conservation in Hermitian case denoted by the horizontal projection from the red dotted circle to the blue dotted circle.

Accordingly, for the case of $J_L<J_R$, the states in the band move along the $-k$ direction due to the negative slope of $d E_I(k)/d k$ around $k_0$. Therefore, the initial state projects from the red solid circle with a lower frequency to the blue solid circle with a higher frequency, indicating a frequency blue shift, as shown in Fig.~\ref{fig4}(d). During the projection process with the reflection, the group velocity of the reflected wave is unequal to that of the incident wave due to the evolution of the momentum, which can be calculated based on Eq.~(\ref{v_g}) \cite{supp}.




\clearpage
\newpage

\onecolumngrid
\section*{Supplementary materials for "Anomalous Wave-Packet Dynamics in One-Dimensional Non-Hermitian Lattices"}

\section{1. Derivation of the evolution of $k_{\mathrm{max}}(t)$}

{Here we derive the time evolution of $k_\mathrm{max}(t)$ given in Eq.~(3) of the main text.}
{As we wrote in the main text, we consider the wave packet whose real-space profile is $\psi_n (t = 0) = \frac{1}{(2\pi \sigma^2)^{1/4}}e^{-\frac{(n-n_0)^2}{4\sigma^2}}e^{ik_0(n-n_0)}$. If the wave packet is well spread in real space, the wavefunction in momentum space is}
\begin{align}
    \psi_k(0)=\frac{1}{\sqrt{2\pi}}{\sum_{n = -\infty}^\infty} \psi_n(0)e^{-ikn}
    \approx
    \frac{1}{\sqrt{2\pi}}\int_{-\infty}^\infty \psi_n(0)e^{-ikn}dn
    =
    \left(\frac{2\sigma^2}{\pi}\right)^{1/4}e^{-\sigma^2(k-k_0)^2}e^{-ikn_0}.
\end{align}
{Each momentum-component of the wave packet then evolves in time as}
\begin{align}
    \psi_k(t)=\psi_k(0)e^{-iE(k)t}=
    Ce^{-ikn_0}e^{-iE_R(k)t}e^{-\sigma^2(k-k_0)^2+E_I(k)t},
\end{align}
where $C=\left(\frac{2\sigma^2}{\pi}\right)^{1/4}$. The amplitude of $\psi_k(t)$ is
\begin{align}
    |\psi_k(t)|=Ce^{-\sigma^2(k-k_0)^2+E_I(k)t}=Ce^{X(t)},
\end{align}
with $X(t)=-\sigma^2(k-k_0)^2+E_I(k)t$. {The maximum $k_{\mathrm{max}}$ at time $t$ is determined by solving}
\begin{align}
    \frac{\partial X(t)}{\partial k}=-2\sigma^2(k-k_0)+\frac{d E_I}{d k}t=0,
\end{align}
which gives 
\begin{align}
    k_{\mathrm{max}}(t)=k_0+\frac{1}{2\sigma^2}\frac{d E_I(k)}{dk}\bigg |_{k=k_{\mathrm{max}}}t.\label{kmax}
\end{align}

{This expression implies that the momentum is not conserved in non-Hermitian systems, even when there is a translational symmetry. Generally, in non-Hermitian systems, even when an operator $\hat{A}$ commutes with the Hamiltonian, the expectation value of $\hat{A}$ can change in time. For completeness, we provide its proof here. Consider a non-Hermitian system obeying the Schr\"odinger equation $i\frac{d}{dt}|\psi(t)\rangle = \hat{H}|\psi(t)\rangle$ with a non-Hermitian $\hat{H}$. Then, the time evolution of the expectation value of an operator $\hat{A}$,
\begin{align}
    \langle \hat{A}(t)\rangle \equiv \frac{\langle\psi(t)|\hat{A}|\psi(t)\rangle}{\langle\psi(t)|\psi(t)\rangle},
\end{align}
changes in time as
\begin{align}
    \frac{d}{dt}\langle \hat{A}(t)\rangle
    =
    i\langle \hat{H}^\dagger \hat{A} - \hat{A}\hat{H} \rangle - i \langle \hat{H}^\dagger - \hat{H} \rangle \langle \hat{A}\rangle.
\end{align}
The right hand side is zero when the Hamiltonian is Hermitian ($\hat{H}^\dagger = \hat{H}$) and $\hat{H}$ commutes with $\hat{A}$, namely $\hat{H}\hat{A} - \hat{A}\hat{H} = 0$. However, in general for non-Hermitian Hamiltonians, the right hand side is nonzero even when $\hat{H}$ and $\hat{A}$ commute. Thus the expectation value $\langle \hat{A}(t)\rangle$ can change in time. Physically, the time evolution of momentum is due to the momentum-dependent gain and loss in the complex energy dispersion.
}



\clearpage

\section{\textcolor{black}{2. Derivation of the center of mass $\bar{n}(t)$ 
 and group velocity $V_g(t)$}}

As we defined in Eq.~(4) of the main text, the center of mass of the wave packet is
\begin{align}
    \bar{n}(t)=\frac{\sum_n n |\psi_n(t)|^2}{\sum_n|\psi_n(t)|^2}=\frac{P(t)}{D(t)},
\end{align}
{where we defined the numerator and the denominator as $P(t)$ and $D(t)$, respectively. The wave function in real space at time $t$ is}
\begin{align}
    \psi_n(t)=\frac{1}{\sqrt{2\pi}}\int_{-\pi}^\pi \psi_k(t)e^{ikn}dk=\frac{1}{\sqrt{2\pi}}\int_{-\pi}^\pi \psi_k(0)e^{-iE(k)t}e^{ikn}dk.
\end{align}
We {can then rewrite} $P(t)$ as
\begin{align}
    P(t)=\sum_n n |\psi_n(t)|^2=\frac{1}{2\pi}\int_{-\pi}^{\pi}\int_{-\pi}^{\pi}c_kc_{k'}^{*}e^{-iE(k)t}e^{iE^{*}(k')t}\sum_nne^{i(k-k')n}dk dk', \label{eq:zero}
\end{align}
where we set $c_k=\psi_k(0)$. Since $k\in(-\pi,\pi]$, the Dirac comb function gives
\begin{align}
   \sum_ne^{i(k-k')n}= 2\pi \delta (k'-k).
\end{align}
Then we have 
\begin{align}
\frac{\partial}{\partial k'}\sum_ne^{in(k-k')}=-\sum_nine^{in(k-k')}=2\pi \frac{\partial\delta(k'-k)}{\partial k'},
\end{align}
which gives
\begin{align}
    \sum_n ne^{in(k-k')}=i2\pi \frac{\partial \delta (k'-k)}{\partial k'}. \label{eq:first}
\end{align}
{Similarly, one can also obtain
\begin{align}
    \sum_n ne^{in(k-k')}=-i2\pi \frac{\partial \delta (k-k')}{\partial k}. \label{eq:second}
\end{align}
}
{Substituting Eq.~(\ref{eq:first}) into Eq.~(\ref{eq:zero})}, we obtain
\begin{align}
    P(t)=i\int_{-\pi}^{\pi}\int_{-\pi}^{\pi}c_kc_{k'}^{*}e^{-iE(k)t}e^{iE^{*}(k')t} \frac{\partial \delta (k'-k)}{\partial k'}dk dk'.
\end{align}
On the other hand, using the property of Dirac function,
\begin{align}
    \int f(k') \frac{\partial \delta(k'-k)}{\partial k'}dk'=-\frac{df(k')}{d k'}\bigg|_{k'=k},
\end{align}
with $f(k')=c_{k'}^*e^{iE^*(k')t}$, we have
\begin{align}
    \frac{d f(k')}{d k'}\bigg|_{k'=k}=\left[\frac{\partial c_{k}^{*}}{\partial k}e^{iE^*(k)t}+c_{k}^*it \frac{d E^*(k)}{d k}e^{iE^*(k)t}\right].
\end{align}
Then 
\begin{align}
    P(t)&=-i\int_{-\pi}^{\pi}c_ke^{-iE(k)t}\left[\frac{\partial c_{k}^{*}}{\partial k}e^{iE^*(k)t}+c_{k}^*it \frac{d E^*(k)}{dk}e^{iE^*(k)t}\right]dk\nonumber \\
    &=\int_{-\pi}^{\pi}e^{2E_I(k)t}\left[-ic_k \frac{\partial c_k^*}{\partial k}+|c_k|^2t \frac{d E^*(k)}{d k}\right]dk. \label{eq:pt1}
\end{align}
{Similarly, substituting Eq.~(\ref{eq:second}) into Eq.~(\ref{eq:zero}), we obtain
\begin{align}
    P(t)
    &=\int_{-\pi}^{\pi}e^{2E_I(k)t}\left[ic_k^* \frac{\partial c_k}{\partial k}+|c_k|^2t \frac{d E(k)}{d k}\right]dk. \label{eq:pt2}
\end{align}
Adding Eq.~(\ref{eq:pt1}) and Eq.~(\ref{eq:pt2}) and dividing by two, we obtain
\begin{align}
    P(t)
    &=\int_{-\pi}^{\pi}e^{2E_I(k)t}\left[\mathrm{Re}\left\{-ic_k \frac{\partial c_k^*}{\partial k}\right\}+|c_k|^2t \frac{d E_R(k)}{d k}\right]dk
    \notag \\
    &=\int_{-\pi}^{\pi}|c_k|^2 e^{2E_I(k)t}\left[n_0+ t \frac{d E_R(k)}{d k}\right]dk
    \notag \\
    &\equiv\int_{-\pi}^{\pi}w(k,t) \left[n_0+ t \frac{d E_R(k)}{d k}\right]dk,
\end{align}
where we have defined $w(k,t) \equiv |c_k|^2e^{2E_I(k)t}$.}

In the same way, one can also confirm
\begin{align}  D(t)&=\sum_n|\psi_n(t)|^2=\int_{-\pi}^{\pi}\int_{-\pi}^{\pi}c_kc_{k'}^{*}e^{-iE(k)t}e^{iE^{*}(k')t}\delta(k-k')dk dk'=\int_{-\pi}^{\pi}w(k,t)dk.
\end{align}
The center of mass $\bar{n}(t)$ is hence 
\begin{align}
    \bar{n}=\frac{P(t)}{D(t)}&=\frac{\int_{-\pi}^{\pi}w(k,t)[t(d E_R/dk)+n_0]dk}{\int_{-\pi}^{\pi}w(k,t)dk}
    \notag \\
    &=\frac{\int_{-\pi}^{\pi}w(k,t)[(d E_R/d k)]dk}{\int_{-\pi}^{\pi}w(k,t)dk}t+n_0=\bar{v}_g t+n_0,\label{nbar}
\end{align}
where $\bar{v}_g=\langle v_g\rangle=\langle d E_R/d k\rangle$ with $\langle f(k) \rangle =\frac{\int_{-\pi}^{\pi}w(k,t)f(k)dk}{\int_{-\pi}^{\pi}w(k,t)dk}=\frac{\int_{-\pi}^{\pi}|\psi_k(t)|^2f(k)dk}{\int_{-\pi}^{\pi}|\psi_k(t)|^2dk}$. 

Now we derive the group velocity of the wave packet, which can be expressed as
\begin{align}
    V_g&=\frac{d\bar{n}(t)}{dt}=\frac{\dot{P}D-{P}\dot{D}}{D^2}\nonumber\\
    &=\frac{\int w(2E_IE_R't+2E_In_0+E_R')dk\int wdk-\int w (E_R't+n_0)dk\int 2E_I w dk}{\int wdk\int wdk}\nonumber\\
    &=\frac{\int w(2E_IE_R't+ E_R'+2E_In_0)dk}{\int wdk}\nonumber\\
    &-\frac{\int w E_R'tdk \int w 2E_I dk+\int wdk\int w2n_0E_I dk}{\int wdk\int wdk}\nonumber\\
    &=\langle E_R'\rangle+2t[\langle E_IE_R'\rangle-\langle E_R'\rangle \langle E_I\rangle]\nonumber\\
    &=\bar{v}_g(t)+2t[\langle E_I(k)v_g\rangle-\langle v_g\rangle \langle E_I(k)\rangle], \label{V}
\end{align}
where $E_R'=dE_R/dk=v_g$. On the other hand, from Eq.~(\ref{nbar}), one can also see
\begin{align}
    V_g=\frac{d\bar{n}(t)}{dt}=\bar{v}_g(t)+\frac{d\bar{v}_g(t)}{dt}t. \label{V2}
\end{align}
{Comparing Eq.~(\ref{V}) and Eq.~(\ref{V2}), we arrive at}
\begin{align}
   \frac{d\bar{v}_g(t)}{dt}t=2t[\langle E_I(k)v_g\rangle- \bar{v}_g \langle E_I(k)\rangle],
\end{align}
{which is the equation we had below Eq.~(5) in the main text.}

\textcolor{black}{The expression of the group velocity Eqs.~(\ref{V}-
\ref{V2}) is valid in one-dimensional general non-Hermitian systems with complex energy bands.}
\textcolor{black}{We note that we did not assume specific forms for the initial wave packet; the expression is valid for general forms of localized initial wave packets.}

\clearpage

\section{\textcolor{black}{3. Influence of different parameters on the self-induced Bloch oscillation}}

{Here we discuss how the self-induced Bloch oscillation is influenced by the choice of parameters.}

\begin{figure}[h!]
\centering
\includegraphics[width=15 cm]{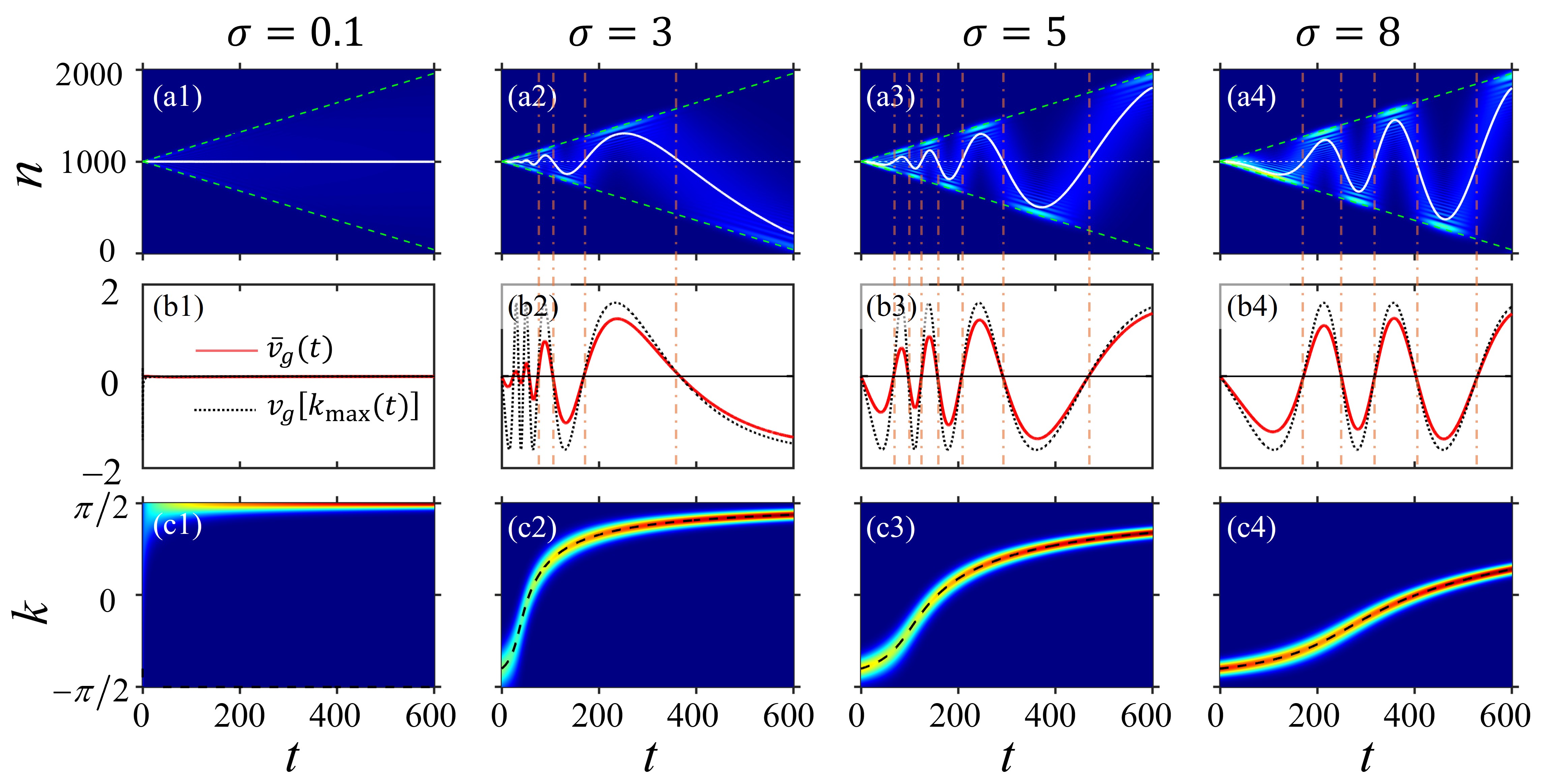}
\caption{ Self-induced Bloch oscillation with different widths of the wave packet. $\sigma=0.1$ (a1-c1), $\sigma=3$ (a2-c2), $\sigma=5$ (a3-c3), and $\sigma=8$ (a4-c4). Other parameters are the same as those used in Fig.~1(b) of the main text. The red dash-dotted lines indicate the correspondence between $\bar{n}(t)=n_0$ and $\bar{v}_g=0$. 
\label{Fig_BO}}
\end{figure}

{We first discuss the influence of $\sigma$, namely the width of the wave packet.} In Fig.~\ref{Fig_BO}, we plot the wave-packet dynamics in real and momentum spaces, as well as the group velocities $\bar{v}_g(t)$ and $v_g[k_{\mathrm{max}}(t)]$ . We see that for very small width with $\sigma=0.1$, which corresponds to {essentially} the single-site excitation, all the momentum states on the band are excited at $t=0$. These states evolve with different group velocities, with the  maximum and minimum determined by the extreme value $dE_R/dk$ on the band [see the green dashed lines in Fig.~\ref{Fig_BO}]. With $t\ne0$, only the state at the momentum $k^*$ remains, due to its highest gain [Fig.~\ref{Fig_BO}(c1)]. When $\sigma$ is increased to $\sigma=3$, the wave packet excites a range of momentum states around $k_0$, and hence the wave function in momentum space moves quickly to the state at $k^*$, as shown in the momentum-space evolution in Fig.~\ref{Fig_BO}(c2). For the wider $\sigma$, the width of the wave function in momentum space becomes narrower, so it moves slowly [Figs.~\ref{Fig_BO}(c3,~c4), see also Eq.~(\ref{kmax})].  The envelop amplitude of the oscillation of the wave packet lies within the range determined by the maximum and minimum group velocities $v_{g}$, as shown by the green dashed lines. The time when the wave packet comes back to the initial point $n_0$ is consistent with $\bar{v}_g(t)=0$, as shown in Figs.~\ref{Fig_BO}(b2-b4). When the width $\sigma$ increases, the width in momentum space decreases, and hence the group velocity $\bar{v}_g(t)$ approaches {$v_g[k_{\mathrm{max}}(t)]$}. Because of the non-linear change of the momentum [see Figs.~\ref{Fig_BO}(c2-c4)], the oscillation period is not equal with increasing time.

\begin{figure}[h!]
\centering
\includegraphics[width=15 cm]{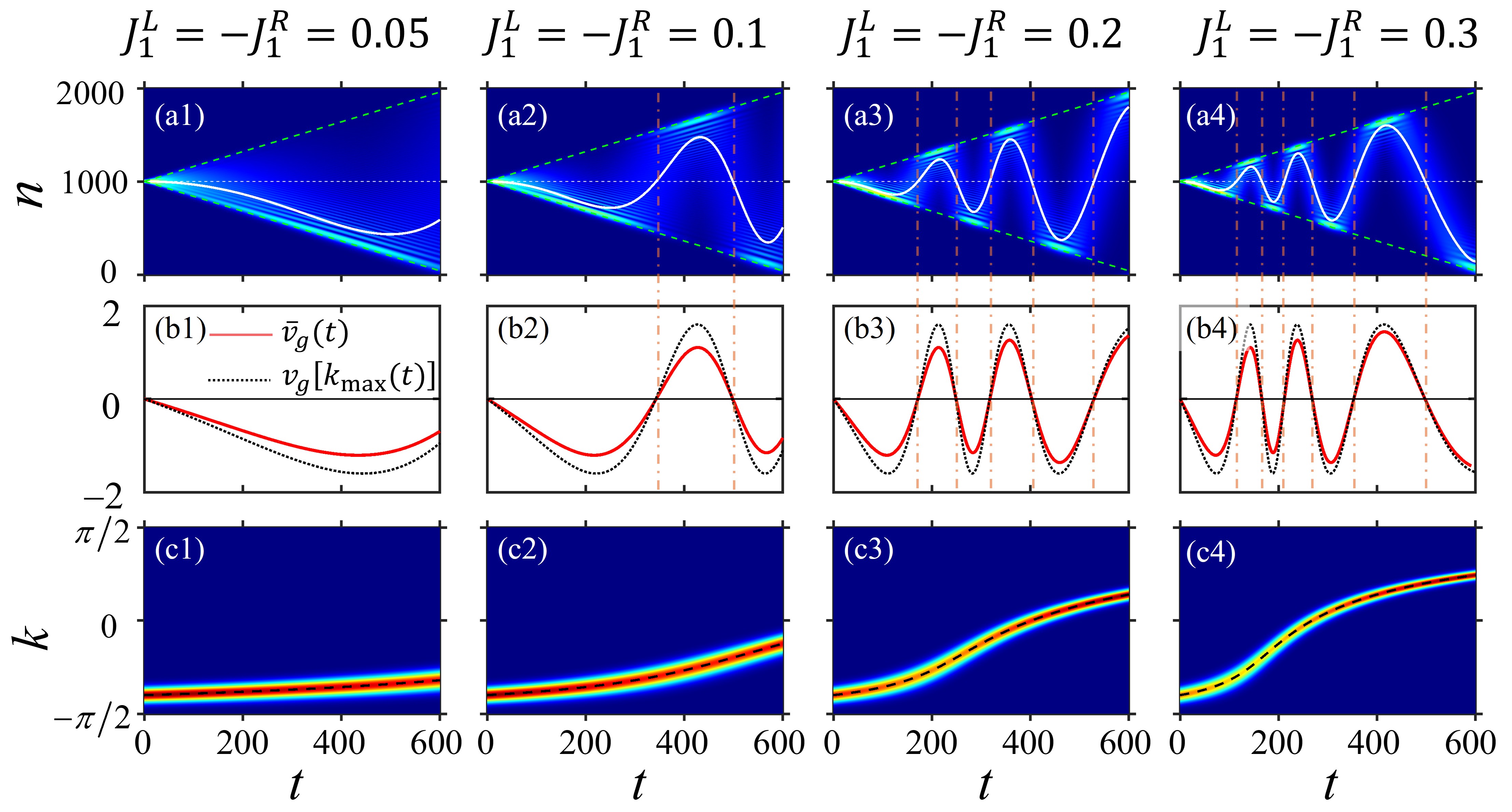}
\caption{Self-induced Bloch oscillation with different strength of the non-Hermiticity. $J_1^L=-J_1^R=0.05$ (a1-c1), $J_1^L=-J_1^R=0.1$ (a2-c2), $J_1^L=-J_1^R=0.2$ (a3-c3), and $J_1^L=-J_1^R=0.3$ (a4-c4). $\sigma=8$. Other parameters are the same as those in Fig.~1(b) of the main text. The red dash-dotted lines indicate the correspondence between $\bar{n}(t)=n_0$ and $\bar{v}_g=0$. 
\label{Fig_BO_2}}
\end{figure}
We next discuss the influence of the strength of non-Hermiticity on the self-induced Bloch oscillation. In Fig.~\ref{Fig_BO_2}, we show the self-induced Bloch oscillations for $J_1^L=~-J_1^R=0.05,~0.1,~0.2$, and $0.3$, which increases the strength of the non-Hermiticity due to the increase of $dE_I/dk$. From Eq.~(\ref{kmax}), we see that the evolution of $k_{\mathrm{
max}}(t)$ depends on $\sigma$ and $dE_I/dk$. Decreasing (increasing) $\sigma$ corresponds to increasing (decreasing) of $d E_I/dk$. Thus, the strength of  non-Hermiticity plays an opposite role with respect to $\sigma$ in the evolution of $k_\mathrm{max}(t)$, and hence the resulting self-induced Bloch oscillation. We see that the shift of the momentum increases with the increase of non-Hermiticiy [see Figs.~\ref{Fig_BO_2}(c1-c4)], which hence induces the decrease of the period of the oscillation [see Figs.~\ref{Fig_BO_2}(a1-a4)].

\begin{figure}[h!]
\centering
\includegraphics[width=15 cm]{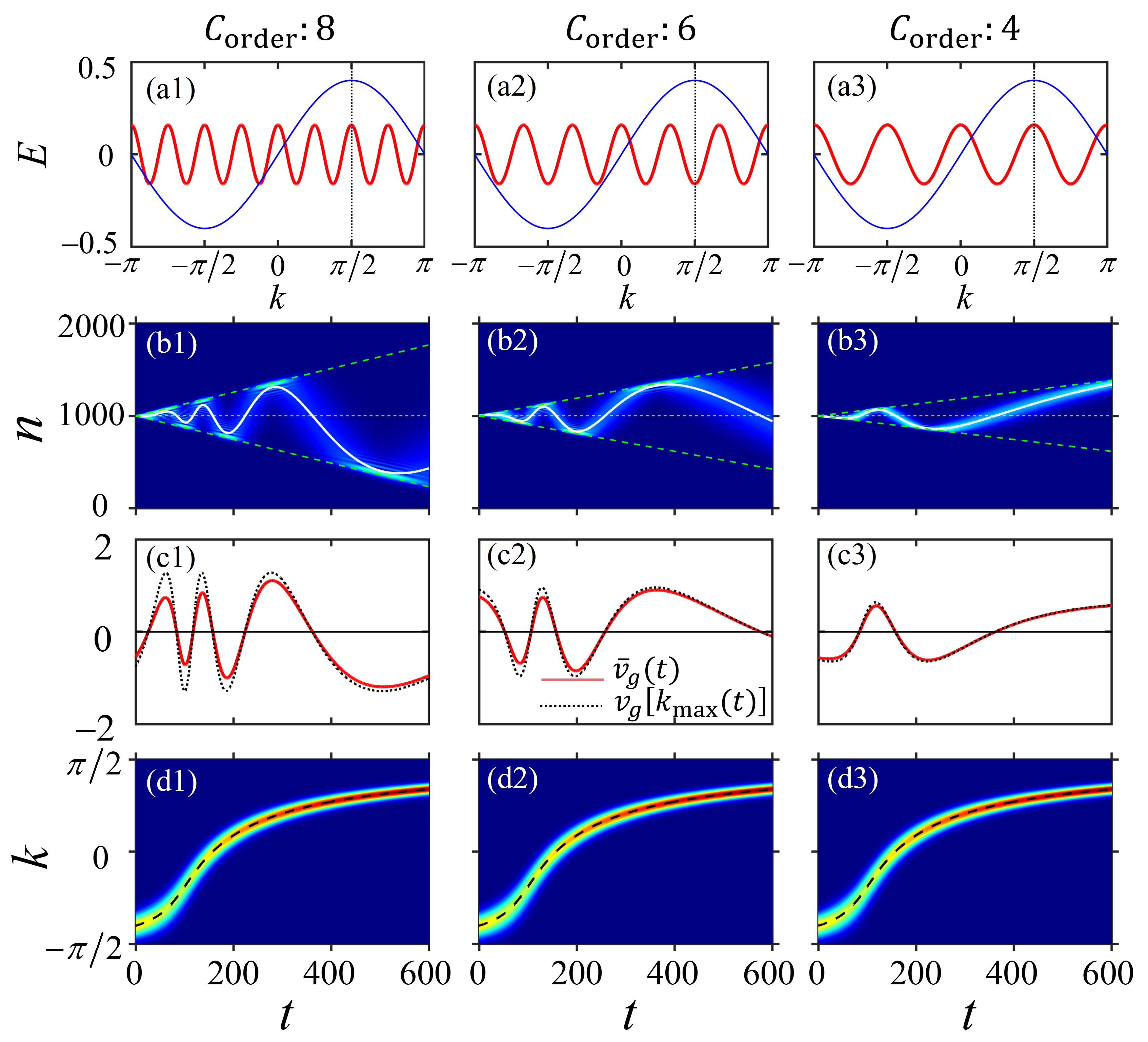}
\caption{\textcolor{black}{Self-induced Bloch oscillation with different coupling orders. Other parameters are the same as those in Fig.~1(b) of the main text. The red dash-dotted lines indicate the correspondence between $\bar{n}(t)=n_0$ and $\bar{v}_g=0$. }
\label{Fig_BO_order}}
\end{figure}

\textcolor{black}{We then discuss the effect of the long-range coupling order $C_{\text{order}}$ on the self-induced Bloch oscillation. In the main text, we consider the next-nearest neighbor coupling with $J_1^L=-J_1^R$ and long-range coupling $J_{10}^L=J_{10}^R\ne0$ with $J_{m\ne10}^L=J_{m\ne10}^R=0$, which corresponds to $C_{\text{order}}=10$. In Fig.~\ref{Fig_BO_order}, we show the self-induced Bloch oscillations for different coupling orders, while keeping the other parameters the same. One can see that, with decreasing the coupling order, the envelop amplitude of the oscillation decreases and the oscillation period increases, which is due to the decrease of the maximum and minimum group velocities governed by $dE_R/dk$ and the increase of the period of $E_R(k)$. The evolution of $k_{\text{max}}(t)$ remains the same for different coupling orders. This is because the coupling order only changes the period of $E_R(k)$, but does not affect $E_I(k)$, which governs the evolution of the momentum.} 

\clearpage

\begin{figure}[h!]
\centering
\includegraphics[width=15 cm]{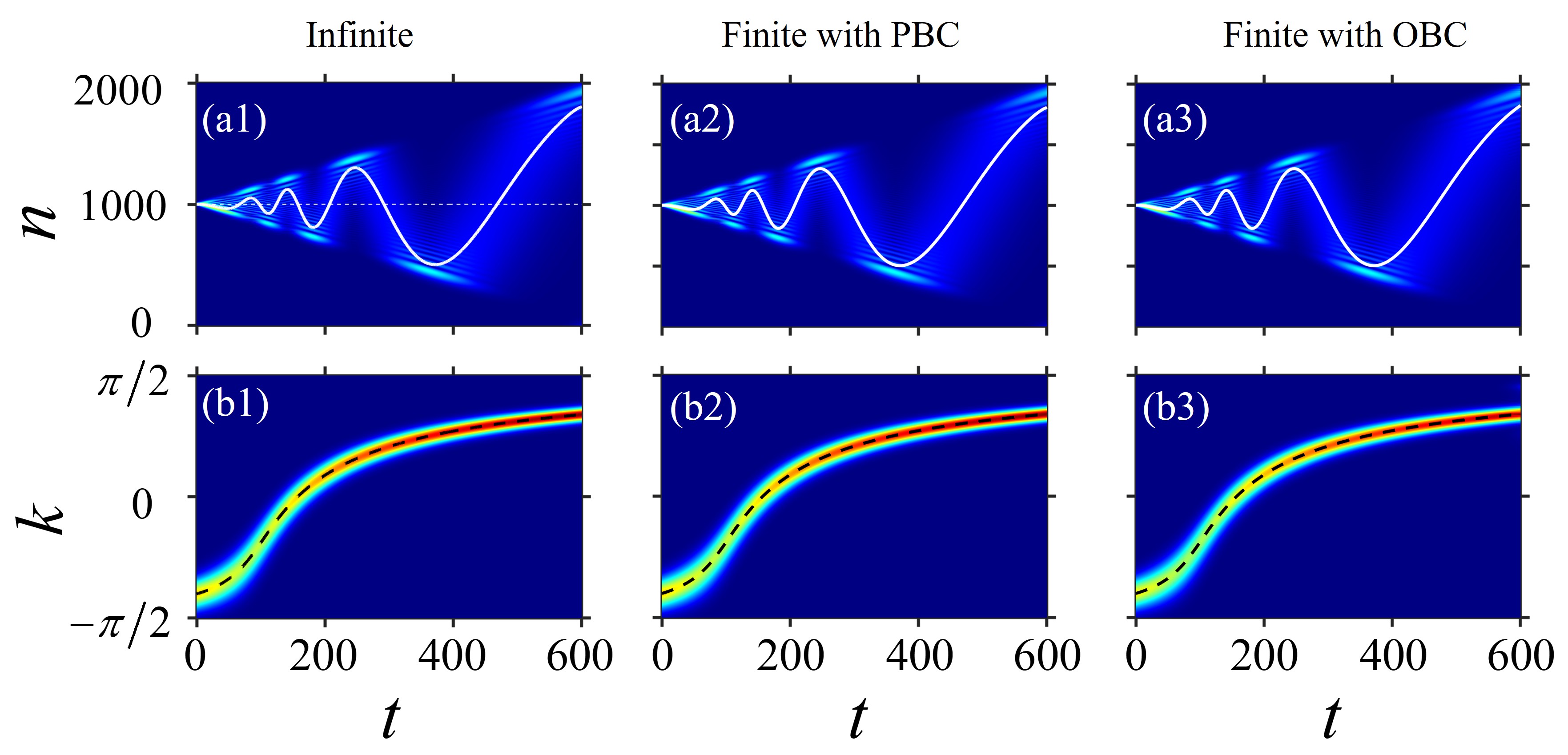}
\caption{\textcolor{black}{Self-induced Bloch oscillation for the infinite size (a1,~b1), finite size with PBC (a2,~b2), and finite size with OBC (a3,~b3). Other parameters are the same as those in Fig.~1(b) of the main text. The white lines in (a1-a3) represents the center of mass, while the black dashed lines in (b1-b3) denote the theoretical $k_{\text{max}}$ according to Eq.~(3) of the main text. 
\label{Fig_BO_size_2000}}}
\end{figure}

\textcolor{black}{We now discuss the effect of the lattice size on the self-induced Bloch oscillation. In the main text, we assume an infinite lattice, which is an idealized limit and cannot be realized experimentally. Here we compare the results of the infinite case and the finite cases with PBC and OBC \textcolor{black}{for the lattice size of $N = 2000$ and $\sigma = 5$} in Fig.~\ref{Fig_BO_size_2000}. We see that the self-induced Bloch oscillation remains the same for three cases, indicating that the finite-size effect does not affect the self-induced Bloch oscillation when the lattice size is large enough compared to the width of the wave packet.} 

\textcolor{black}{
When the size is reduced, the wave packet dynamics is affected with the existence of NH wave-packet jumps. We show the results with PBC and OBC for $N=1000$ and $500$, and see that the self-induced Bloch oscillation can also exist for the PBC case, while it breaks for the OBC with the existence of the NH wave-packet jump. }

\textcolor{black}{We can choose suitable parameters to avoid the NH wave-packet jumps even for a small lattice size, which is more accessible experimentally. From Fig.~\ref{Fig_BO_order}, one sees that the envelop amplitude decreases for a low coupling order, which can hence avoid the NH wave-packet jumps caused by the finite-size effect. In Fig.~\ref{Fig_BO_size_250}, we change the long-range coupling order to $4$, and observe the complete self-induced Bloch oscillation during the time $t\in (0-150)$.}

\begin{figure}[h!]
\centering
\includegraphics[width=12 cm]{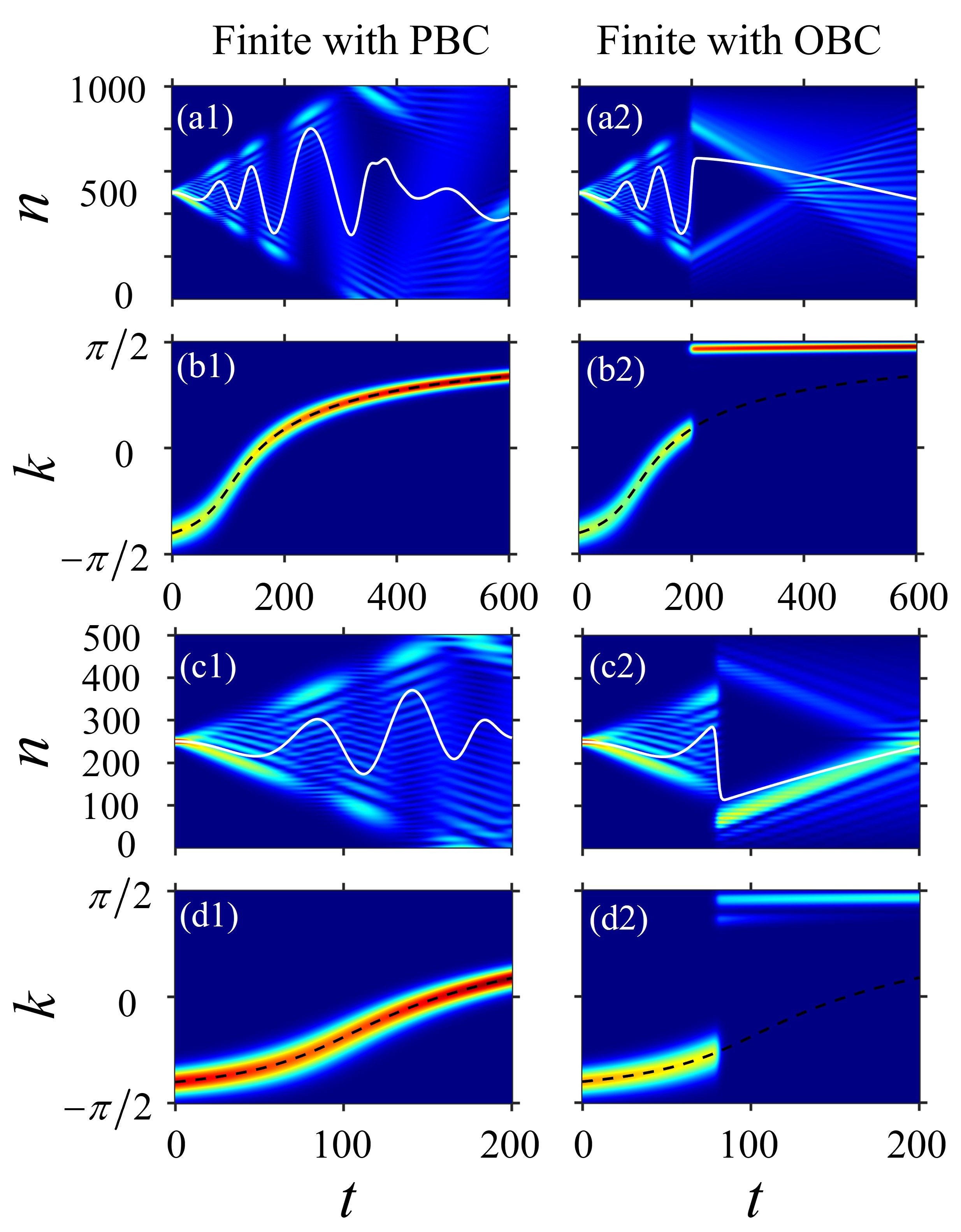}
\caption{\textcolor{black}{Self-induced Bloch oscillation for the finite size with PBC (a1-d1), and finite size with OBC (a2-d2). Other parameters are the same as those in Fig.~1(b) of the main text. 
\label{Fig_BO_size_1000}}}
\end{figure}

\begin{figure}[h!]
\centering
\includegraphics[width=15 cm]{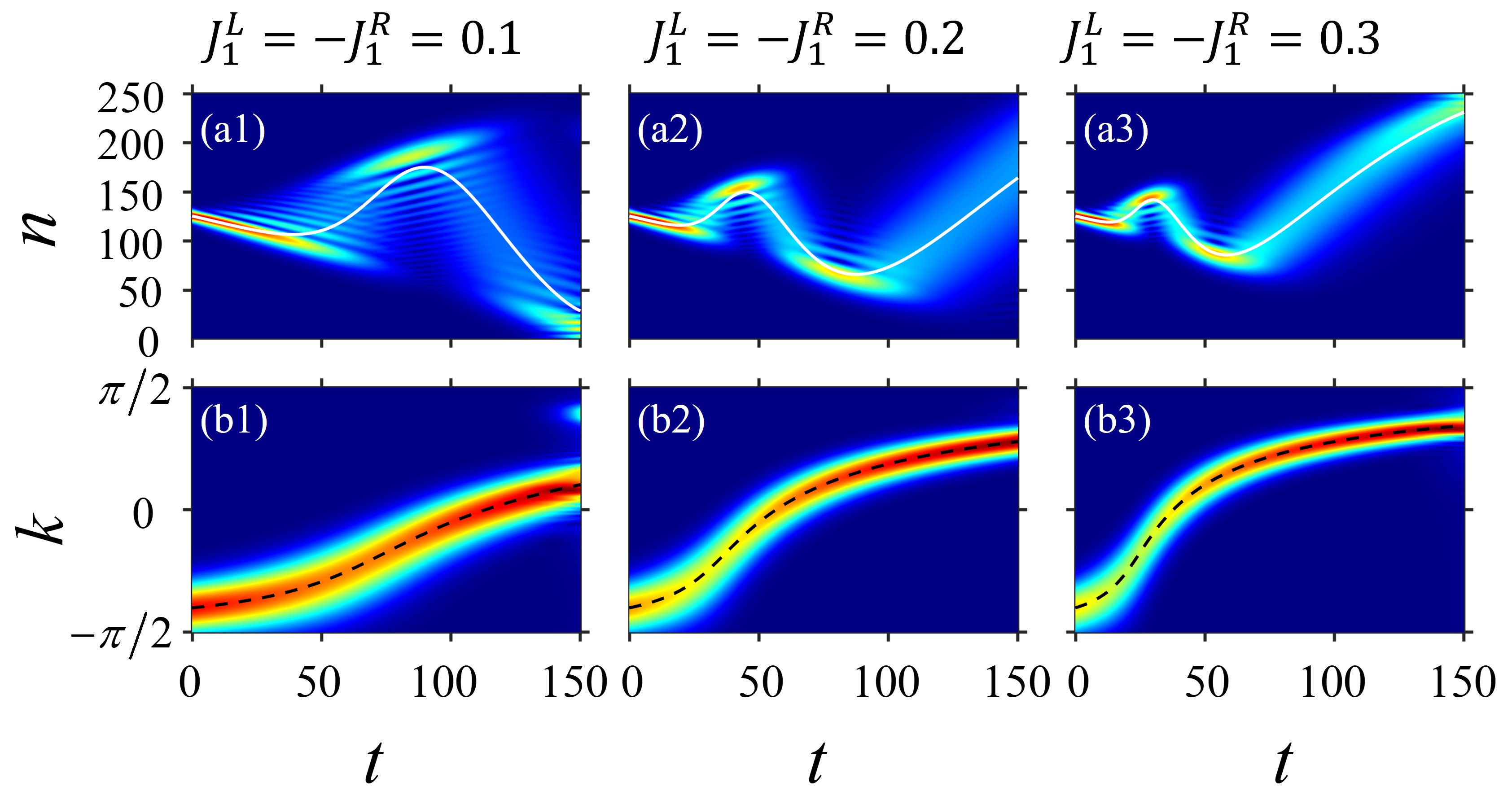}
\caption{\textcolor{black}{Self-induced Bloch oscillation with different $J_1^{L,R}$ under the OBC. The other parameters are $\sigma=3$ and $J_4^L=J_4^R=0.1$ with $J_{m\ne1,4}^{L,R}=0$. 
\label{Fig_BO_size_250}}}
\end{figure}

\clearpage

\section{\textcolor{black}{4. Comparison between self-induced Bloch oscillation and Bloch oscillation}}

\begin{figure}[h!]
\centering
\includegraphics[width=15 cm]{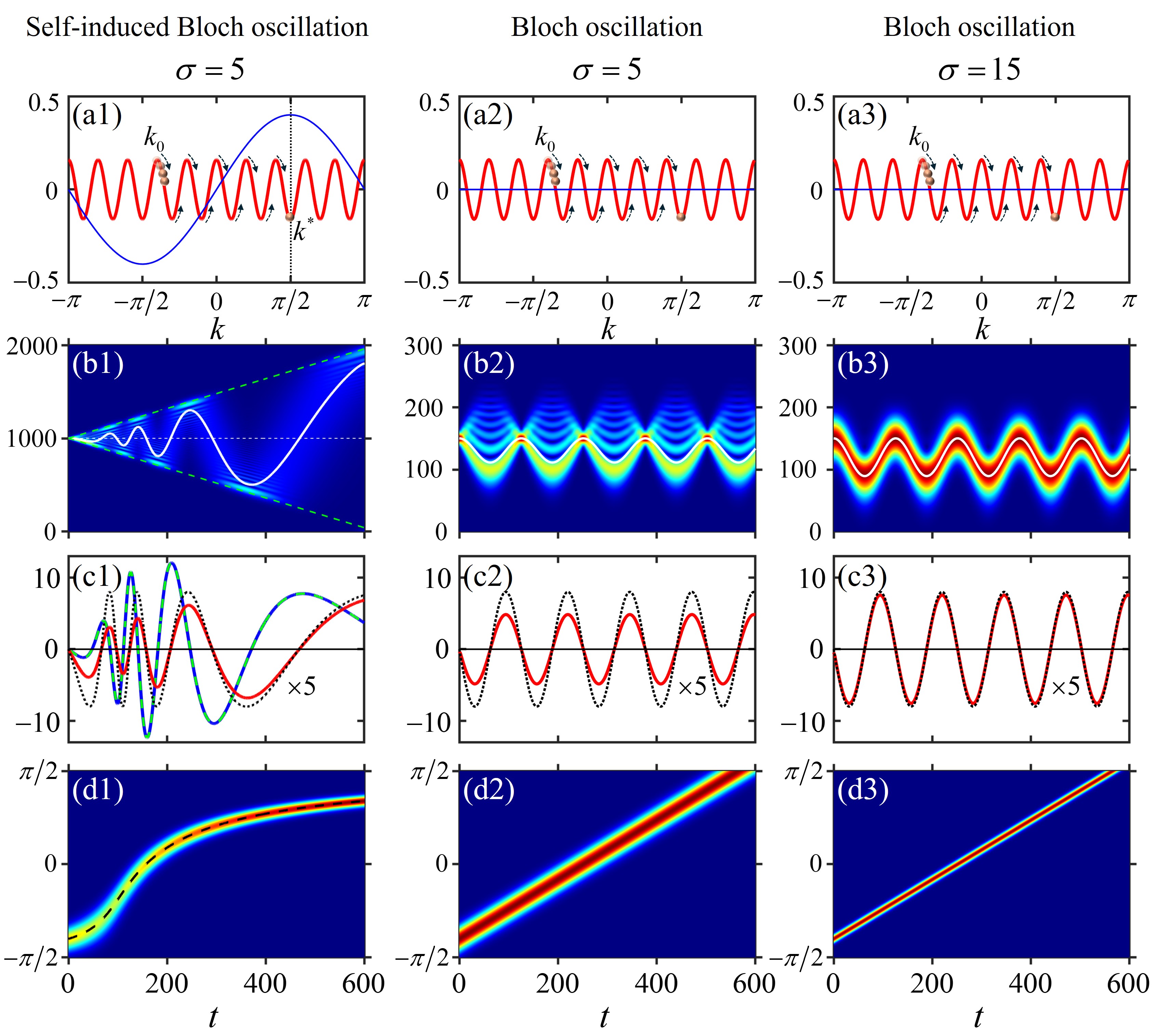}
\caption{\textcolor{black}{(a1-a3) Band structures, $E_R(k)$ (red lines) and $E_I(k)$ (blue lines). (b1-b3) Evolution of a wave packet in real space. (c1-c3) The evolution of group velocities $\bar{v}_g(t)$ (red lines), $v_g[k_\mathrm{max}(t)]$ (black dotted lines), the numerical $V_g(t)$ (blue solid lines), and the predicted $V_g(t)$ (green dashed lines) according to Eq.~(5) in the main text. (d1-d3) Corresponding evolution of the wave packets in momentum space. The parameters in (a1-d1) are the same as those used in Fig.~(a2-d2) of the main text. The parameters in (a2-d2) are $F=-0.005$, $J_{10}^L=J_{10}^R=0.08, J_{m\ne10}^{L,R}=0$, and $\sigma=5$. (a3-d3), same as (a2-d2), except $\sigma=15$.} \label{Fig_BO_compare} }
\end{figure}

\textcolor{black}{In this section, we compare the self-induced Bloch oscillation in NH lattices with the Bloch oscillation in Hermitian lattices. We keep the long-range coupling as in Fig.~1(a2-d2) [Fig.~\ref{Fig_BO_compare}(a1-d1)], $J_{10}^L=J_{10}^R=0.08$, and set the next-nearest coupling as zero, where only the real band structure remains, as shown in Fig.~\ref{Fig_BO_compare}(a2). We consider a nonzero force $F=-0.005$, which induces a linear evolution of momentum $k_\text{max}(t)=-Ft$ in Fig.~\ref{Fig_BO_compare}(d2). The Bloch oscillation of the wave dynamics in real space is shown in Fig.~\ref{Fig_BO_compare}(b2). The corresponding group velocity nearly matches the $v_g[k_\text{max}(t)]$, where the difference is due to the smaller width $\sigma$. When we increase $\sigma=15$, a narrower width in momentum space is excited, and hence the simulated group velocity matches well with $v_g[k_\text{max}(t)]$, as shown in Fig.~\ref{Fig_BO_compare}(c3). Compared with the Bloch oscillation with a static force [Fig.~\ref{Fig_BO_compare}(a3-d3)], the self-induced Bloch oscillation exhibits several different properties.} 

\textcolor{black}{(i) The momentum changes non-linearly and approaches $k^*$ in the long-time limit, which is different from the Bloch oscillation with a force where the momentum changes linearly with time without saturation. This can be explained from Eq.~(\ref{kmax}), i.e., $k_{\mathrm{max}}(t)=k_0+\frac{1}{2\sigma^2}\frac{d E_I(k)}{dk}\big |_{k=k_{\mathrm{max}}}t$, where $k_{\mathrm{max}}$ increases with the gradient $dE_I/dk>0$ and does not change for $dE_I/dk=0$, which hence makes $k_{\mathrm{max}}$ saturated to the momentum at $k^*$ with the largest $E_I$. Since the momentum changes nonlinearly, the period of the oscillation changes with time, which is different from Hermitian case where the period of oscillation remains the same.} 

\textcolor{black}{(ii) The envelop amplitude changes linearly for self-induced Bloch oscillation, while it stays the same for Bloch oscillation with a force. The envelop amplitude of the oscillation lies within the range determined by the maximum and minimum group velocities $v_{g}$ of the real band structure, as shown by the green dashed lines. This aspect has been explained in Section 3.}

\textcolor{black}{(iii) The group velocity of the self-induced Bloch oscillation originates from the \textcolor{black}{conventional (Hermitian)} contribution $\bar{v}_g(t)$ that is due to the evolution of momentum, and the other component of anomalous group velocity $d\bar{v}_g(t)/dt$, which is absent in Hermitian Bloch oscillation. Therefore, the group velocity changes more in non-Hermitian systems compared to the Hermitian case.}

\textcolor{black}{The self-induced Bloch oscillation originates from the evolution of momentum induced by the non-Hermiticity and the anomalous group velocity. Both contributions give rise to such a self-induced Bloch oscillation with distinct properties form its counterpart in Hermitian case with a force. }

\clearpage

\section{\textcolor{black}{5. Self-induced Bloch oscillation beyond Gaussian wave packets}}

\begin{figure}[h!]
\centering
\includegraphics[width=15 cm]{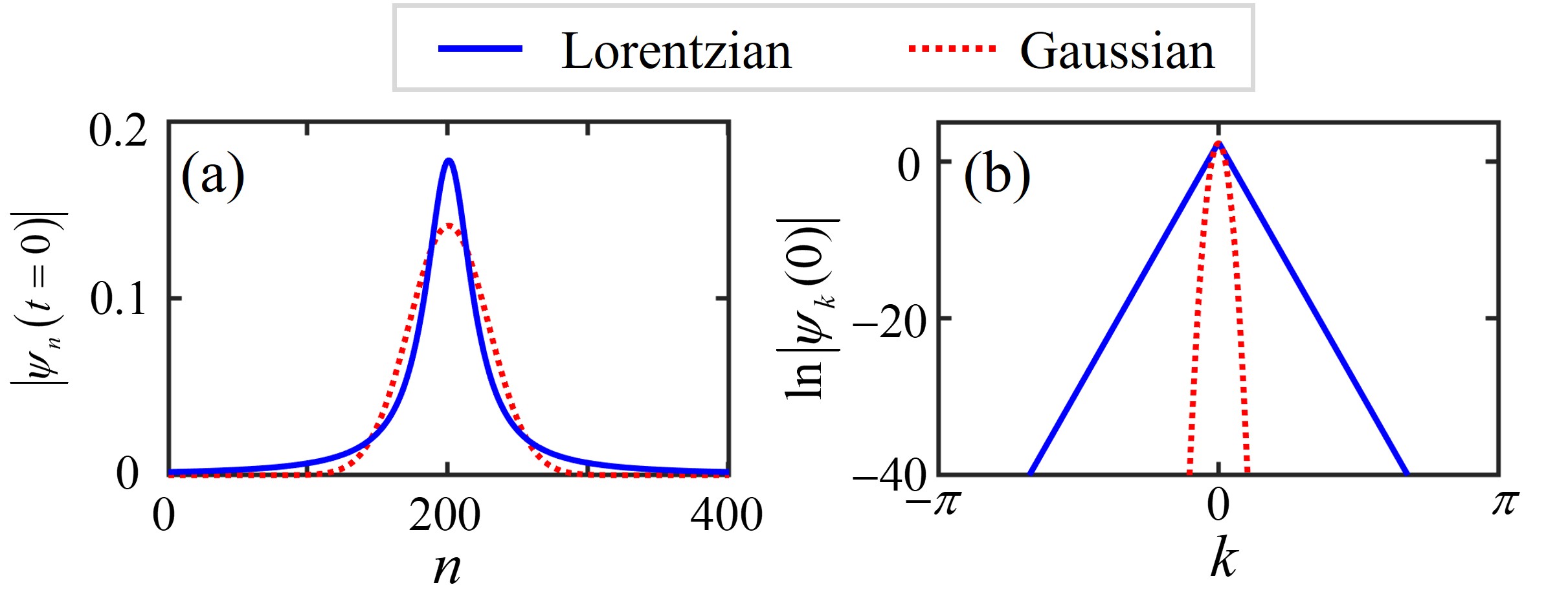}
\caption{\textcolor{black}{ Typical normalized Gaussian (red dashed lines) and Lorentzian wave packet (blue solid lines) in (a) real and (b) momentum spaces. } \label{Fig_Lorentz_Gaussian} }
\end{figure}

\textcolor{black}{We discuss the wave-packet dynamics beyond the Gaussian wave packets.} \textcolor{black}{For concreteness, we consider a wave packet which is of Lorentzian form, which in real space is}
\textcolor{black}{
\begin{equation}
    \psi_n(t=0)=\sqrt{\frac{2\sigma^3}{\pi}} \frac{e^{ik_0(n-n_0)}}{(n-n_0)^2+\sigma^2}.
\end{equation}
The corresponding wave packet in momentum space in an infinite lattice has the form of 
\begin{equation}
    \psi_k(t)=C{e^{-\sigma|k-k_0|}}e^{-i(k-k_0)n_0},
\end{equation}
where $C$ is a normalized coefficient. We compare the Lorentzian and Gaussian wave packets in real and momentum spaces, as shown in Fig.~\ref{Fig_Lorentz_Gaussian}.}

\begin{figure}[h!]
\centering
\includegraphics[width=13 cm]{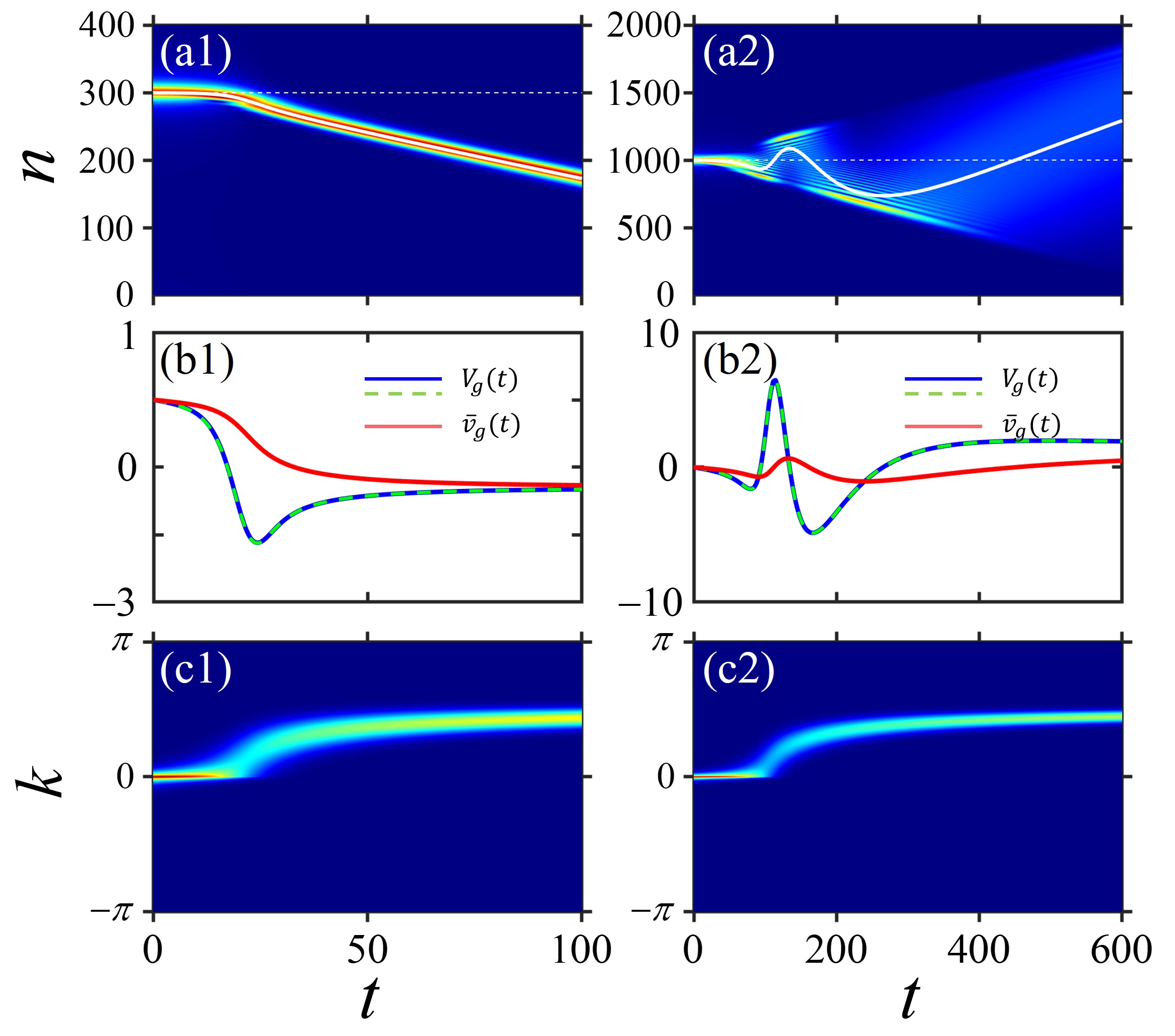}
\caption{
\textcolor{black}{(a1,~a2) Time evolution of a Lorentzian wave packet, where the white solid lines denote the predicted center of mass according to Eq.~(4) of the main text. (b1,~b2) The evolution of group velocities $\bar{v}_g(t)$ (red lines) and numerical (blue solid lines) and predicted (green dashed lines ) $V_g(t)$. (c1,~c2) The evolution of the wave packet in momentum space. The parameters are $J_1^L=0.9,~J_1^R=0.4$, $J_{m>1}^{L,R}=0$, $\sigma=10$, and $k_0=0$ in (a1-c1), while $J_1^L=0.1,~J_1^R=-0.1$, $J_{7}^{L}=J_7^R=0.1$, $\sigma=20$, and $k_0=0$ in (a2-c2). }\label{Fig_Lorentz_BO} }
\end{figure}

\textcolor{black}{
We plot the time evolution of a Lorentzian wave packet in real and momentum spaces, as well as the corresponding group velocities $\bar{v}_g(t)$ and $V_g(t)$ in Fig.~\ref{Fig_Lorentz_BO}(a1-c1). We see that the group velocity of the Lorentzian wave packet changes in time, which agrees well with the analytical expression $V_g(t)$ in Eq.~(5) of the main text. This agreement \textcolor{black}{is expected because the theoretical expression of group velocity $V_g(t)$ was derived without assuming a specific form for a wave packet}. The momentum also changes in time, as shown in Fig.~\ref{Fig_Lorentz_BO}(c1). }

\textcolor{black}{The combination effects of the evolution of momentum and the anomalous group velocity can give rise to a self-induced Bloch oscillation. We illustrate this effect by introducing non-Hermitian nearest-neighbor coupling, $J_1^L=0.1,~J_1^R=-0.1$, as well as Hermitian long-range coupling, $J_{7}^{L}=J_7^R=0.1$. The evolution of a Lorentzian wave packet in real space is shown in Fig.~\ref{Fig_Lorentz_BO}(a2). We see that the center of mass oscillates in time, indicating the self-induced Bloch oscillation. }

\clearpage

\section{6. Coincidence of NH wave-packet jumps with real spectra under the OBC and PBC}

\begin{figure}[h!]
\centering
\includegraphics[width=10 cm]{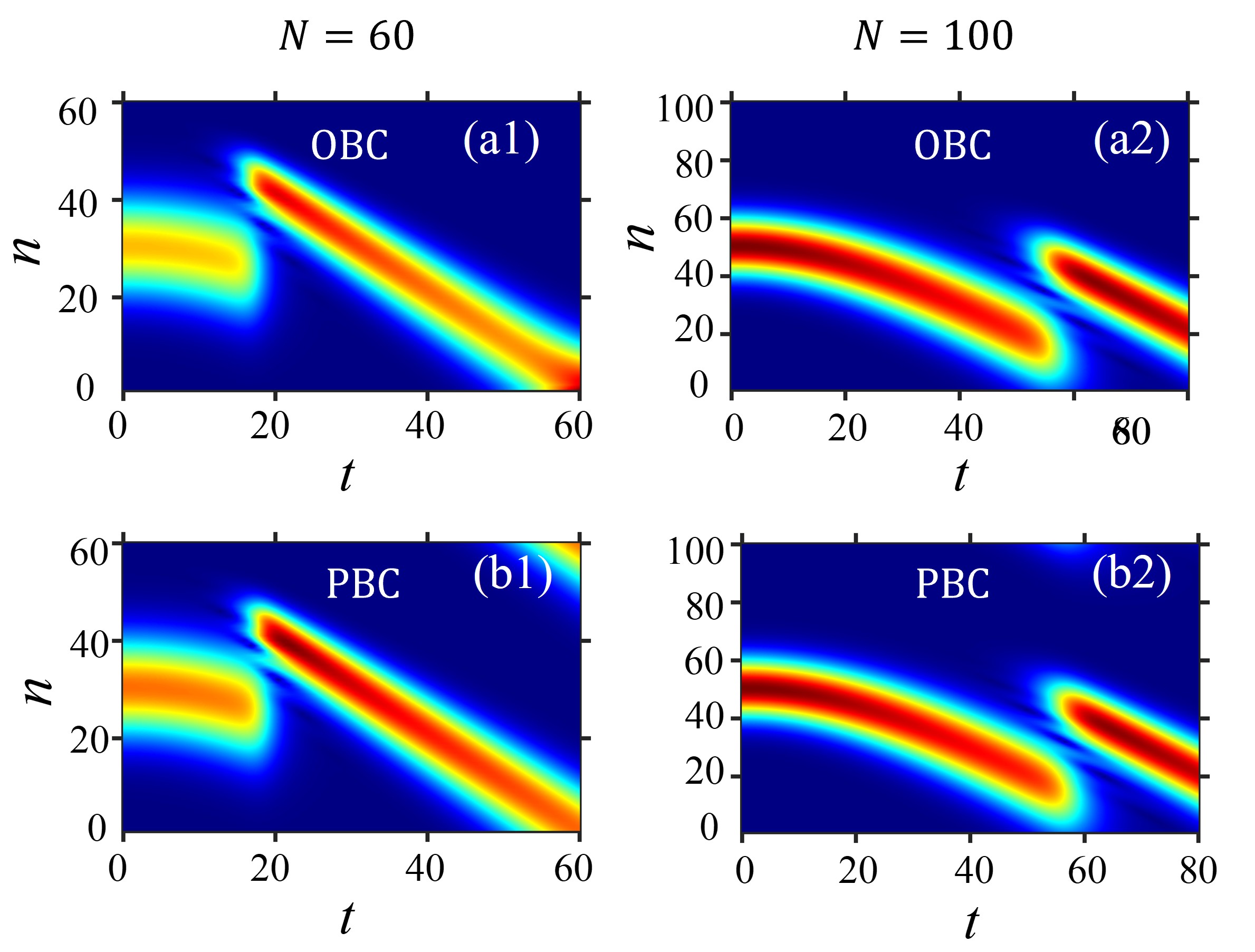
}
\caption{Coincidence of NH wave-packet jumps under the PBC and OBC. The parameters are the same as those used in Fig.~2 of the main text.
\label{Fig_PBC_OBC}}
\end{figure}

In the section of \textit{NH wave-packet jumps with real spectra} in the main text, we discuss NH wave-packet jump in a finite lattice under the OBC with an entirely real spectra. Since the wave propagates in the bulk far away from the edge, the dynamics can be explained from the PBC spectra, which is complex. Different momentum states with different gain or loss compete with each other and the state at $k^*$ with the highest gain dominates at $t_c$, leading to the NH wave-packet jumps. The wave dynamics should be same for the OBC and PBC if the wave packet does not approach the edge. 
{To confirm this expectation, in Fig.~\ref{Fig_PBC_OBC}, we plot the time evolution of a wave packet under both OBC and PBC for $N=60$ and $N=100$}. We see that the wave dynamics is exactly same under the OBC and PBC before the wave packet reaches the edge. Thus, the NH wave-packet jumps persist regardless of whether the boundary condition is open (OBC) or closed (PBC).

\clearpage

\section{7. Identify the position of the NH wave-packet jumps with real spectra}

{In the section of \textit{NH wave-packet jumps with real spectra} in the main text, we discuss NH wave-packet jumps which occur when the amplitude of the state at $k^*=\pi/2$ (highest gain) exceed other states and dominate the dynamics. Here, we now analyze the position where the jump occurs.}

\begin{figure}[h!]
\centering
\includegraphics[width=15 cm]{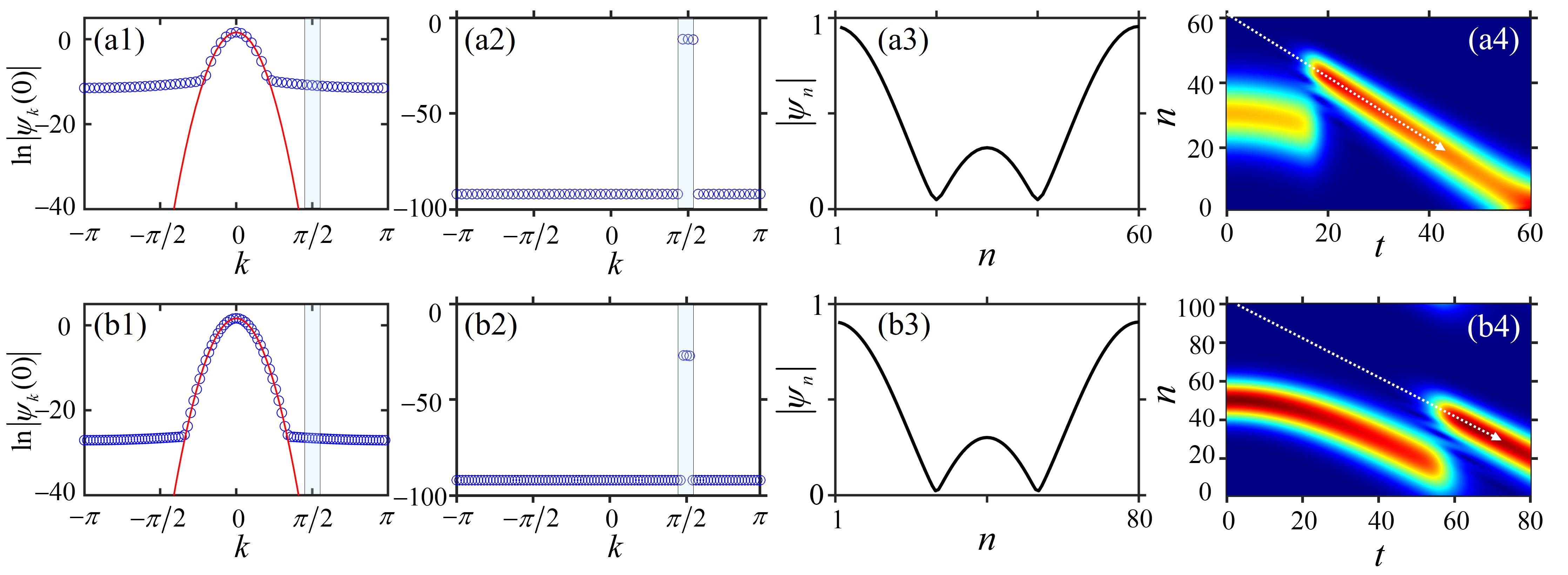}
\caption{ (a1,~b1) Initial wave packet in momentum space with finite size under the PBC (blue circles) and infinite lattice (red lines). (a2,~b2) The distribution $\mathrm{ln}|\psi_k(0)|$ around $k=k^*=\pi/2$. (a3,~b3) The wave function in real space corresponding to the wave function in (a2,~b2). 
The
evolution of the wave packet in real space (a4,~b4), where the white dotted lines denote trajectory of the wave packet with $k$ around $\pi/2$. The sizes $N=60$ (a1-a4) and $N=100$ (b1-b4). Other parameters are the same as these in Fig.~2 of the main text. 
\label{Fig_PBC}}
\end{figure}

Since the jump occurs when the component of the wave packet with $k$ around $k^*=\pi/2$ starts to dominate the evolution, we extract the momentum-space component of the wave packet around $k^*$ and set other momenta as extreme small values, as shown in Figs.~\ref{Fig_PBC}(a2,~b2). We then take the Inverse Fourier Transform to obtain the corresponding wave function of theses momenta in real space, as shown in Figs.~\ref{Fig_PBC}(a3,~b3), from which we see that the wave function mainly locates near the edges of the lattice. Therefore, the wave packet with the momentum near $k^*$ propagates from the edge and dominates the evolution {after $t = t_c$. The position at which the NH wave-packet jump occurs is determined by the trajectory of this component of the wave near $k \approx k^*$ which propagates from the edge, as we indicated by the white dotted lines in Figs.~\ref{Fig_PBC}(a4,~b4).}

\clearpage
\section{8. Influence of different parameters on the NH wave-packet jumps with real spectra}

\subsection{8.1 Influence of the strength of non-Hermiticity on the NH wave-packet jumps with real spectra}

\begin{figure}[h!]
\centering
\includegraphics[width=15 cm]{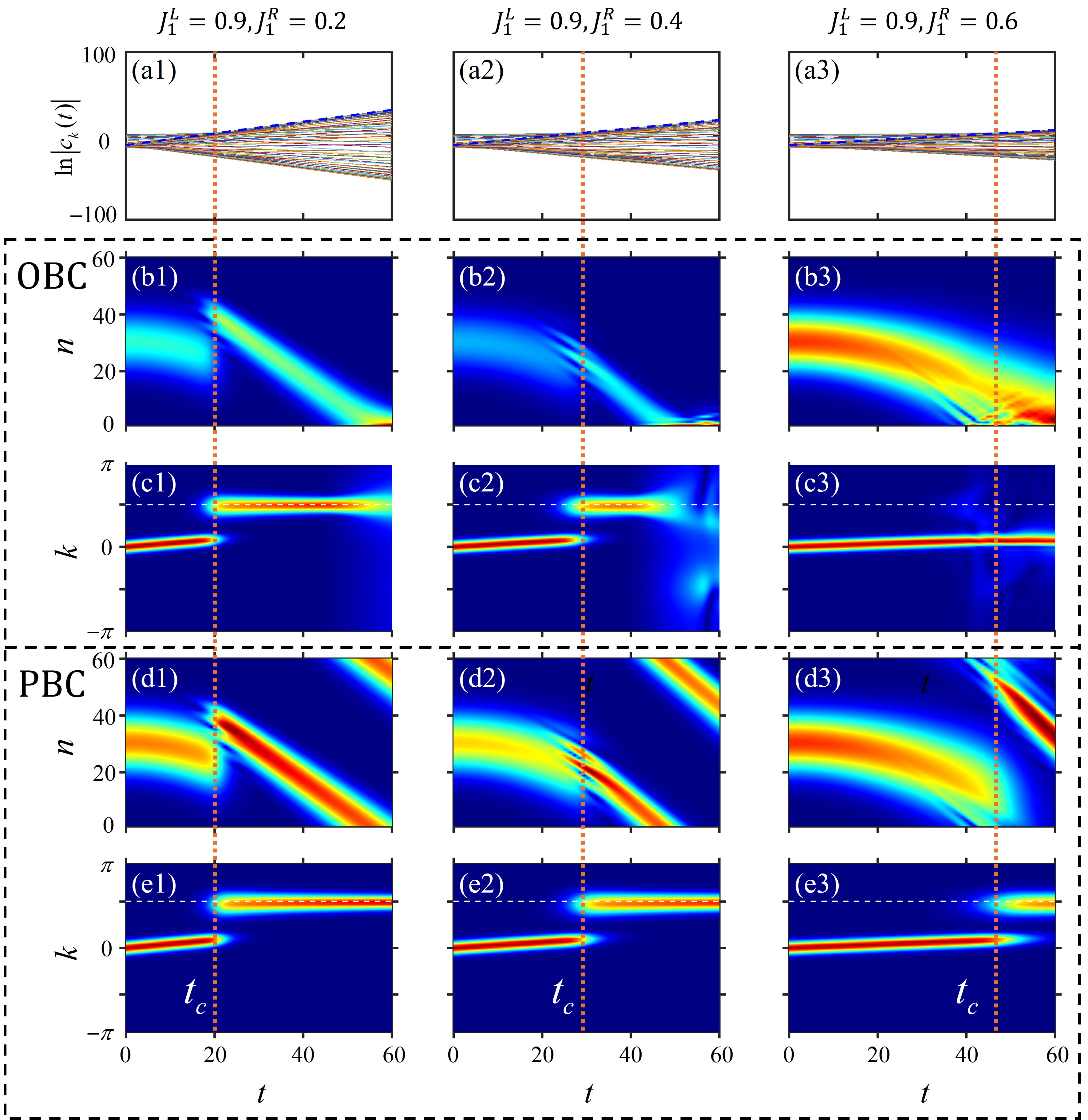}
\caption{ (a1-a3) The evolution of $c_k(t)$, where $c_{k=k^*=\pi/2}(t)$ is highlighted by the blue dashed lines. The corresponding wave function in real space under the OBC (b1-b3) and PBC (d1-d3). The evolution of the wave packet in momentum space under the OBC (c1-c3) and PBC (e1-e3). The white dashed lines denote $k=\pi/2$. The red dotted lines denote the jump time $t_c$. The size is $N=60$.  Other parameters are the same as these in Fig.~2 of the main text. 
\label{Fig_PBC_J}}
\end{figure}

We first analyze the influence of the strength of non-Hermiticity on the jump time. We fix $J_1^L=0.9$ and reduce the strength of non-Hermiticity by changing $J_1^R=0.2,0.4$, and $0.6$. The corresponding wave dynamics in real and momentum spaces under the OBC is shown in Figs.~\ref{Fig_PBC_J}(b1-b3) and~\ref{Fig_PBC_J}(c1-c3), respectively. For comparison, we also plot the results under the PBC in Figs.~\ref{Fig_PBC_J}(d1-d3) and~\ref{Fig_PBC_J}(e1-e3). When the non-Hermiticity is large with $J_1^R=0.2$, the wave packet propagates in the bulk and the NH wave-packet jump occurs at $t_c\approx20$ under the OBC, which is the same for the PBC case. When the strength of non-Hermiticity is reduced to $J_1^R=0.4$, the jump time increases to $t_c\approx28$ for both the OBC and PBC cases. For the small strength of non-Hermiticiy with $J_1^R=0.6$, the wave pack reaches the edge, making the results under OBC and PBC different. For the OBC case, the wave packet piles up at the edge and there is no jump. For the PBC case, since the wave packet can still propagate, {NH wave-packet jump occurs at a later time at} $t_c\approx47$. The jump time for all the cases are consistent with {the condition that the state at $k^*$ exceeds all other components}, as shown in Figs.~\ref{Fig_PBC_J}(a1-a3).

\clearpage

\subsection{\textcolor{black}{8.2 Influence of the lattice size on the NH wave-packet jumps with real spectra}}

\textcolor{black}{In this section, we examine in details the lattice-size dependence of the NH wave-packet jump. For a finite lattice with the size $N$, the momentum-space wave packet is $\psi_k'=\frac{1}{\sqrt{2\pi}}{\sum_{n = 1}^N} \psi_n(t=0)e^{-ikn}$. For a large $N$, the wavefunction approaches its continuous counterpart with 
\begin{align}
    \psi_k'
    &\approx\frac{1}{\sqrt{2\pi}}\int_{1}^N \psi_n(0)e^{-ikn}dn\nonumber\\
    &=
    \frac{1}{\sqrt{2\pi}}\int_{1}^N \frac{1}{(2\pi \sigma^2)^{1/4}}e^{-\frac{(n-n_0)^2}{4\sigma^2}}e^{ik_0(n-n_0)}e^{-ikn}dn\nonumber\\
     &=\frac{1}{\sqrt{2\pi}}\frac{1}{(2\pi \sigma^2)^{1/4}}e^{-ik_0n_0}\int_{1}^N e^{-\frac{(n-n_0)^2}{4\sigma^2}}e^{-i(k-k_0)n}dn.\label{psi_k_finite}
\end{align}
Let $m=n-n_0\in[1-n_0,N-n_0]$, we have
\begin{align}
    \psi_k'&\approx\frac{1}{\sqrt{2\pi}}\frac{1}{(2\pi \sigma^2)^{1/4}}e^{-ik_0n_0}\int_{1-n_0}^{N-n_0} e^{-\frac{m^2}{4\sigma^2}}e^{-i(k-k_0)(m+n_0)}dm\nonumber\\
    &=\frac{1}{\sqrt{2\pi}}\frac{1}{(2\pi \sigma^2)^{1/4}}e^{-ikn_0}\int_{1-n_0}^{N-n_0} e^{-\frac{m^2}{4\sigma^2}}e^{-i(k-k_0)m}dm
    \nonumber\\
    &=\frac{1}{\sqrt{2\pi}}\frac{e^{-ikn_0-\sigma^2 (k-k_0)^2}}{(2\pi \sigma^2)^{1/4}}\int_{1-n_0}^{N-n_0} e^{-\frac{(m+2i\sigma^2 (k-k_0))^2}{4\sigma^2}}dm
\end{align}}
\textcolor{black}{We now rewrite this integral in terms of error functions. We change the variable of integral to $u \equiv (m+2i\sigma^2 (k-k_0))/(2\sigma)$. We then have
\begin{align}
    \psi_k'
    =
    \frac{2\sigma}{\sqrt{2\pi}}\frac{e^{-ikn_0-\sigma^2 (k-k_0)}}{(2\pi \sigma^2)^{1/4}}\int_{\frac{1-n_0}{2\sigma}+i\sigma (k-k_0)}^{\frac{N-n_0}{2\sigma} + i\sigma (k-k_0)} e^{-u^2}du.
\end{align}
Note that now the integral is a complex integral, but the result does not depend on the specific path one chooses in the complex plain because $e^{-u^2}$ is holomorphic in the entire complex plane.
Using the error function
\begin{align}
    \text{erf}(z)=\frac{2}{\sqrt{\pi}}\int_0^z e^{-u^2}du,
\end{align}
we can rewrite as
\begin{align}
    \psi_k'
    &=
    \left( \frac{2\sigma^2}{\pi} \right)^{1/4} e^{-ikn_0-\sigma^2 (k-k_0)}
    \frac{1}{2}
    \left\{\text{erf}\left( \frac{N-n_0}{2\sigma} + i\sigma (k-k_0)\right)
    -
    \text{erf}\left( \frac{1-n_0}{2\sigma}+i\sigma (k-k_0)\right)
    \right\}
    \notag \\
    &=
    \psi_k
    \frac{1}{2}
    \left\{\text{erf}\left( \frac{N-n_0}{2\sigma} + i\sigma (k-k_0)\right)
    -
    \text{erf}\left( \frac{1-n_0}{2\sigma}+i\sigma (k-k_0)\right)
    \right\},
\end{align}
where $\psi_k$ is the momentum-space wave packet on an infinite lattice. 
The finite size correction is thus due to the terms with error functions. We now assume that $n_0$ is at the middle of the lattice so that $N-n_0 = N/2$ and $1-n_0 = 1-N/2 \approx -N/2$.
The momentum-space wave packet on a finite lattice is then
\begin{align}
    \psi_k'
    &=
    \psi_k
    \frac{1}{2}
    \left\{\text{erf}\left( \frac{N}{4\sigma} + i\sigma (k-k_0)\right)
    -
    \text{erf}\left( -\frac{N}{4\sigma}+i\sigma (k-k_0)\right)
    \right\}.
\end{align}
For notational convenience, we write $a \equiv \frac{N}{4\sigma}$ and $b \equiv \sigma (k-k_0)$.
Since the error function is an odd function, $\text{erf}(-z) = -\text{erf}(z)$, we can rewrite as
\begin{align}
    \psi_k'
    &=
    \psi_k
    \frac{1}{2}
    \left\{\text{erf}\left( a + bi\right)
    +
    \text{erf}\left( a-bi\right)
    \right\}. \label{psi_k1}
\end{align}
Let us first check that when the lattice is large, $N \to \infty$, the finite result $\psi_k'$ coincides with the infinite result $\psi_k$. To see this, we note that when $\mathrm{Re}(z) > 0$ and $|z|$ is large, the error function asymptotically becomes $\text{erf}(z) \approx 1 - \frac{e^{-z^2}}{z\sqrt{\pi}}$.
Then, when $|a + bi|$ is large,
\begin{align}
    \psi_k' &\approx \psi_k
    \frac{1}{2}
    \left\{2 - \frac{e^{-(a+bi)^2}}{(a+bi)\sqrt{\pi}} -\frac{e^{-(a-bi)^2}}{(a-bi)\sqrt{\pi}}
    \right\}
    \notag \\
    &=
    \psi_k \left\{ 1 - \frac{e^{b^2 - a^2}}{\sqrt{\pi}(a^2 + b^2)}\left( a \cos (2ab) - b \sin (2ab)\right)\right\}. \label{eq:asymp}
\end{align}
In the limit of $N\to \infty$, which means $a \to \infty$, we thus see that $\psi_k' \to \psi_k$, which is expected.}

\begin{figure}[h!]
\centering
\includegraphics[width=15 cm]{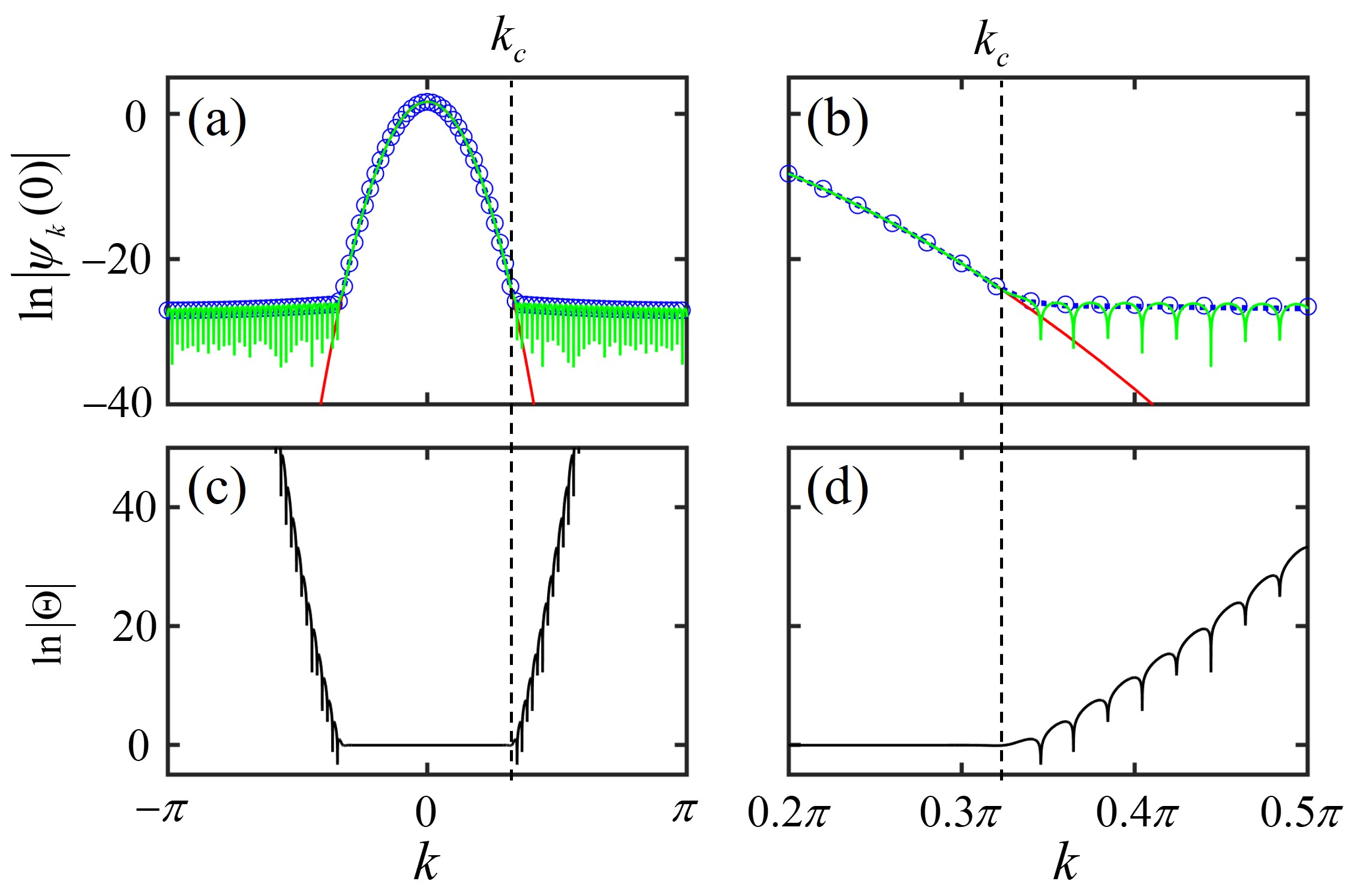}
\caption{ \textcolor{black}{(a) The wave packet in momentum space under the PBC. The blue circles represent $\psi_k'$ for the finite discrete lattice. The red solid line denotes $\psi_k$ for the infinite case. The blue dashed line denotes $\psi_k'$ for a finite discrete lattice based on Eq.~(\ref{psi_k1}), while the green solid line denotes $\psi_k'$ for the finite continuous case based on Eq.~(\ref{psi_k1}). (b) Enlarged area of (a) for $k\in[0.2\pi,~0.5\pi]$. (c) The function $\ln{|\Theta(k)|}$. (d) Enlarged area of (c) for $k\in[0.2\pi,~0.5\pi]$.}
\label{Fig_PBC_critical_N}}
\end{figure}

\textcolor{black}{In Fig.~\ref{Fig_PBC_critical_N}, we compare the distribution of the wave packet in momentum space for a finite case $\psi_k'$ and an infinite case $\psi_k$. One can see that $\ln|\psi_k'|$ and $\ln|\psi_k|$ agree very well near $k\approx 0$, but they deviate at large $|k|$. This deviation is due to the error function contribution, $\Theta (k) \equiv \frac{1}{2} \left\{\text{erf}\left( a + bi\right) + \text{erf}\left( a-bi\right)\right\}$, which is plotted in Fig.~\ref{Fig_PBC_critical_N}(c,~d). There is a point $k = k_c$, which we call the inflection point, beyond which the derivation occurs.}


\textcolor{black}{We can estimate the value of $k_c$ using the asymptotic expansion of the error function. In the asymptotic expression Eq.~(\ref{eq:asymp}), making $k$ larger implies making $b$ larger. As $b$ is made larger, the factor $e^{b^2 - a^2}$ grows exponentially. The correction term thus becomes dominant when $e^{b^2 - a^2} \ge 1$, which implies $b \ge a$. This condition is equivalent to $|k-k_0| \ge N/(4\sigma^2)$.}

\textcolor{black}{Now we explain the relation between the inflection point and the existence of NH wave-packet jumps. We should keep in mind that the NH wave-packet jump cannot exist for an infinite lattice, as the central momentum changes continuously to $k^*$, giving a continuous wave evolution, as seen in Fig.~1 of the main text. Having a finite size lattice can introduce deviations in the initial wave packet distribution in momentum space relative to its infinite counterpart beyond $k_c$. When $k_c<k^*$, which corresponds to $\ln|\psi'_{k=k^*}(0)|>\ln|\psi_{k=k^*}(0)|$, the amplitude of the momentum state at $k=k^*$ with the largest increasing rate will exceed other components and dominate the evolution at a later time, indicating the existence of NH wave-packet jumps. When $k_c>k^*$, which corresponds to $\ln|\psi'_{k=k^*}(0)|=\ln|\psi_{k=k^*}(0)|$, the momentum state will evolve from $k_0$ to $k^*$ continuously, giving a continuously wave evolution in real and momentum space, so the NH wave-packet jumps disappear. Therefore, the critical point for the existence of NH wave-packet jumps is $k_c=k^*$.} 

\textcolor{black}{ For the Hatano-Nelson model with $k^*=\pi/2,~k_0=0$, the critical size $N$ where the NH wave-packet jumps can occur is given by $k_c=k^*=\pi/2$, i.e., 
\begin{align}
    N_c= 4\sigma^2 |k-k_0| = 2\pi \sigma^2.
\end{align}
For $\sigma=5$ used in Fig.~2 of the main text, the critical lattice size is $N_c=2\pi\sigma^2\approx157$.}
\begin{figure}[h!]
\centering
\includegraphics[width=14 cm]{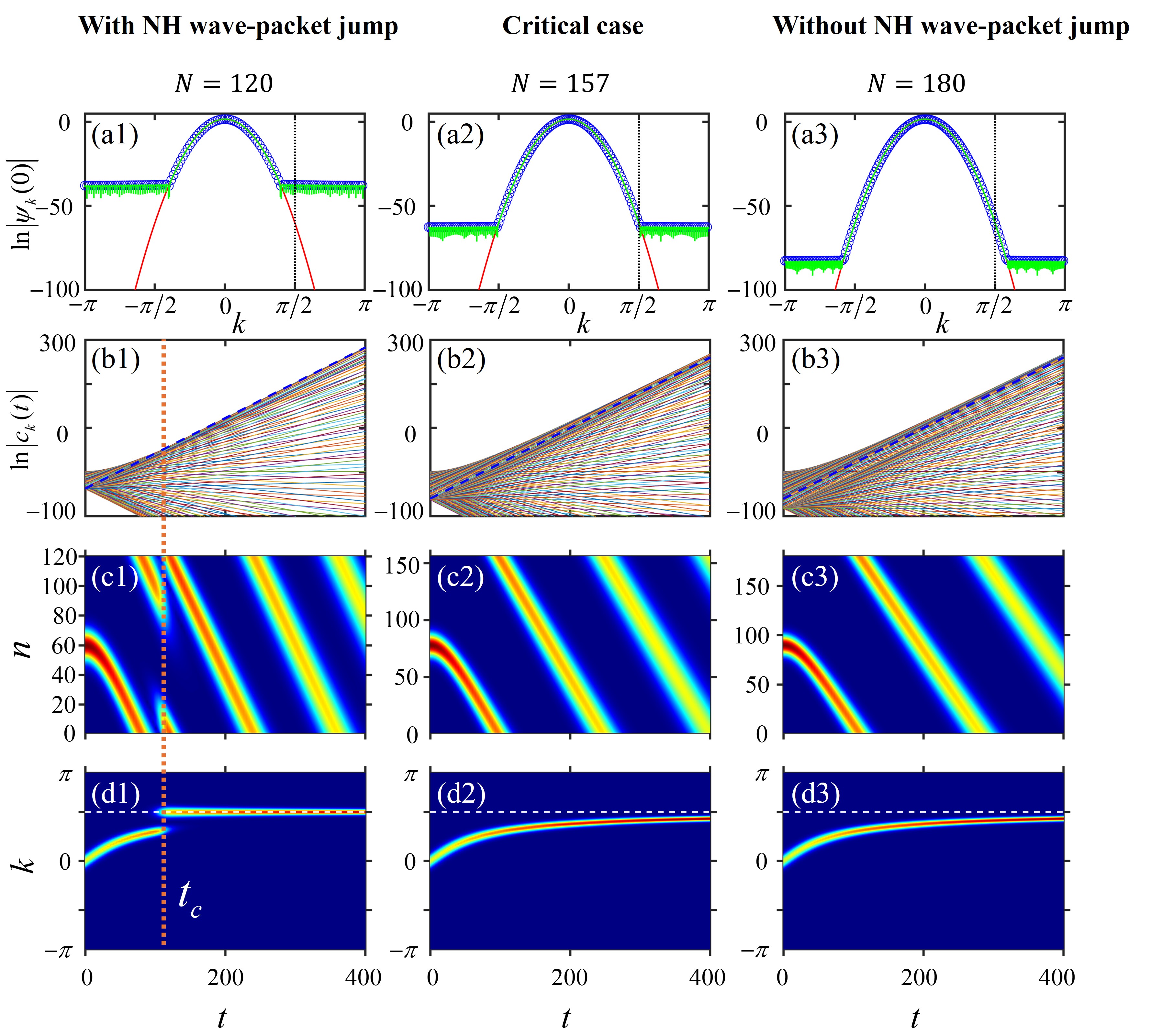}
\caption{\textcolor{black}{(a1-a3) Initial wave packet in momentum space for a finite discrete (blue circles), continuous (green solid lines), and  infinite case (red solid lines). (b1-b3) The evolution of $c_k(t)$, where $c_{k=k^*=\pi/2}(t)$ is highlighted by the blue dashed lines. (c1-c3) The  evolution of the wave packet  in real space under the PBC. (d1-d3) The evolution of the wave packet in momentum space under the PBC. The white dashed lines denote $k=k^*=\pi/2$. Other parameters are the same as those in Fig.~2 of the main text.} 
\label{Fig_PBC_N}}
\end{figure}

{\textcolor{black}{To demonstrate this point, we show the time evolution of a wave packet for different lattice sizes $N=120, 157$, and $180$ under the PBC in Fig.~\ref{Fig_PBC_N}, where we keep the other parameters the same as those in Fig.~2 of the main text. We see that there is a NH wave-packet jump for $N=120$, while NH wave-packet jumps disappear for $N=157$ and $180$.}
{\textcolor{black}{Thus, the NH wave-packet jump cannot exist when the lattice size is larger than a critical value, i.e., $N>N_c$. Since this value is related to the width of the wave packet, we can increase the width $\sigma$ to achieve NH wave-packet jumps at a larger lattice.} 

{\textcolor{black}{To demonstrate this point, in Fig.~\ref{Fig_PBC_sigma}, we keep the lattice size as $N=180$, and increase the width of the wave packet to $\sigma=7,~8$, and $10$, which corresponds $N_c=2\pi\sigma^2\approx308,~402$, and $628$. Since $N=180<N_c$, the NH wave-packet jumps reappear. This is because when the width of the wave packet increases, the amplitude of momentum at $k^*$ for the finite size increases and becomes larger than that of the infinite size, i.e., $\ln|\psi'_{k=k^*}(0)|>\ln|\psi_{k=k^*}(0)|$, as shown by the blue dots in Fig.~\ref{Fig_PBC_sigma}(a1-a3), and hence the amplitude of the wave packet at $k^*$ will exceed other components at some certain time, indicating the existence of the NH wave-packet jumps.}

\begin{figure}[h!]
\centering
\includegraphics[width=14 cm]{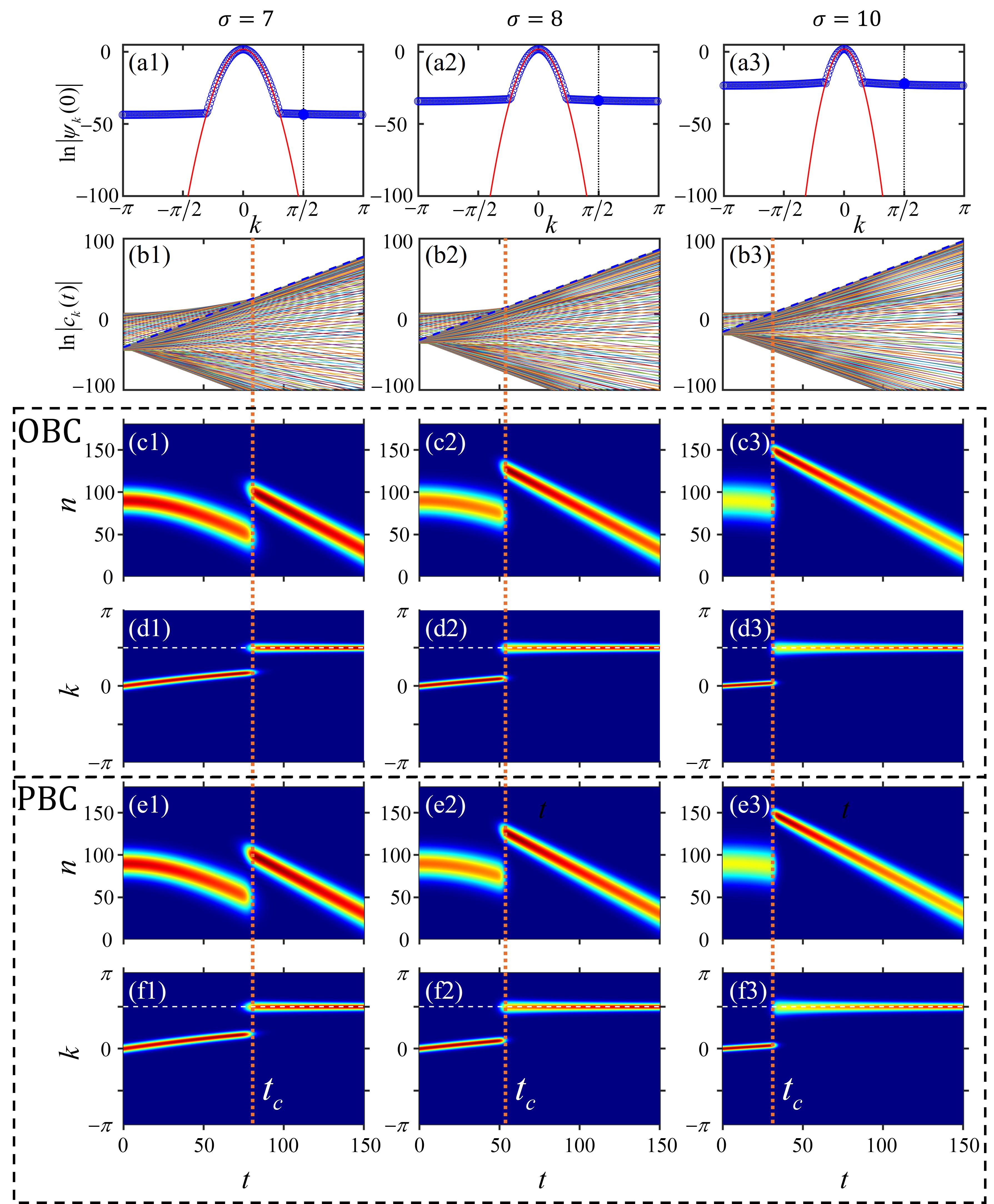}
\caption{\textcolor{black}{(a1-a3) Initial wave packet in momentum space for a finite (blue circles) and infinite cases (red lines). The blue dots indicate the values for $k=k^*=\pi/2$. (b1-b3) The evolution of $c_k(t)$, where $c_{k=k^*=\pi/2}(t)$ is highlighted by the blue dashed lines. The corresponding wave evolution in real space under the OBC (c1-c3) and PBC (e1-e3). The evolution of the wave packet in momentum space under the OBC (d1-d3) and PBC (f1-f3). The white dashed lines denote $k=\pi/2$. The lattice size is $N=180$. Other parameters are the same as those in Fig.~2 of the main text. }
\label{Fig_PBC_sigma}}
\end{figure}

\clearpage

\subsection{8.3 Influence of the initial central momentum on the NH wave-packet jumps with real spectra}

{Now, we analyze NH wave-packet jumps for incident wave packets with a nonzero initial momentum, $k_0\ne0$.} The PBC requires that the incident momentum $k_0$ and the size  $N$ satisfy the relation 
\begin{align}
    e^{ik_0}=e^{ik_0(N+1)},
\end{align}
which gives $e^{ik_0N=1}$, and hence $k_0=2\pi m/N$, with $m=0,\pm 1,\pm 2...$. {If we consider $N=80$, the allowed momenta are, for example,} $k_0=-0.25\pi$, $k_0=-0.4\pi$, and $k_0=-0.8\pi$. In Fig.~\ref{Fig_PBC_k0}, we plot the wave dynamics in real and momentum spaces under the OBC and PBC, respectively. {We see that NH wave-packet jumps can occur when $k_0$ takes different values.} The NH wave-packet jump time $t_c$ is nearly independent of the choices of $k_0$.

\begin{figure}[h!]
\centering
\includegraphics[width=15 cm]{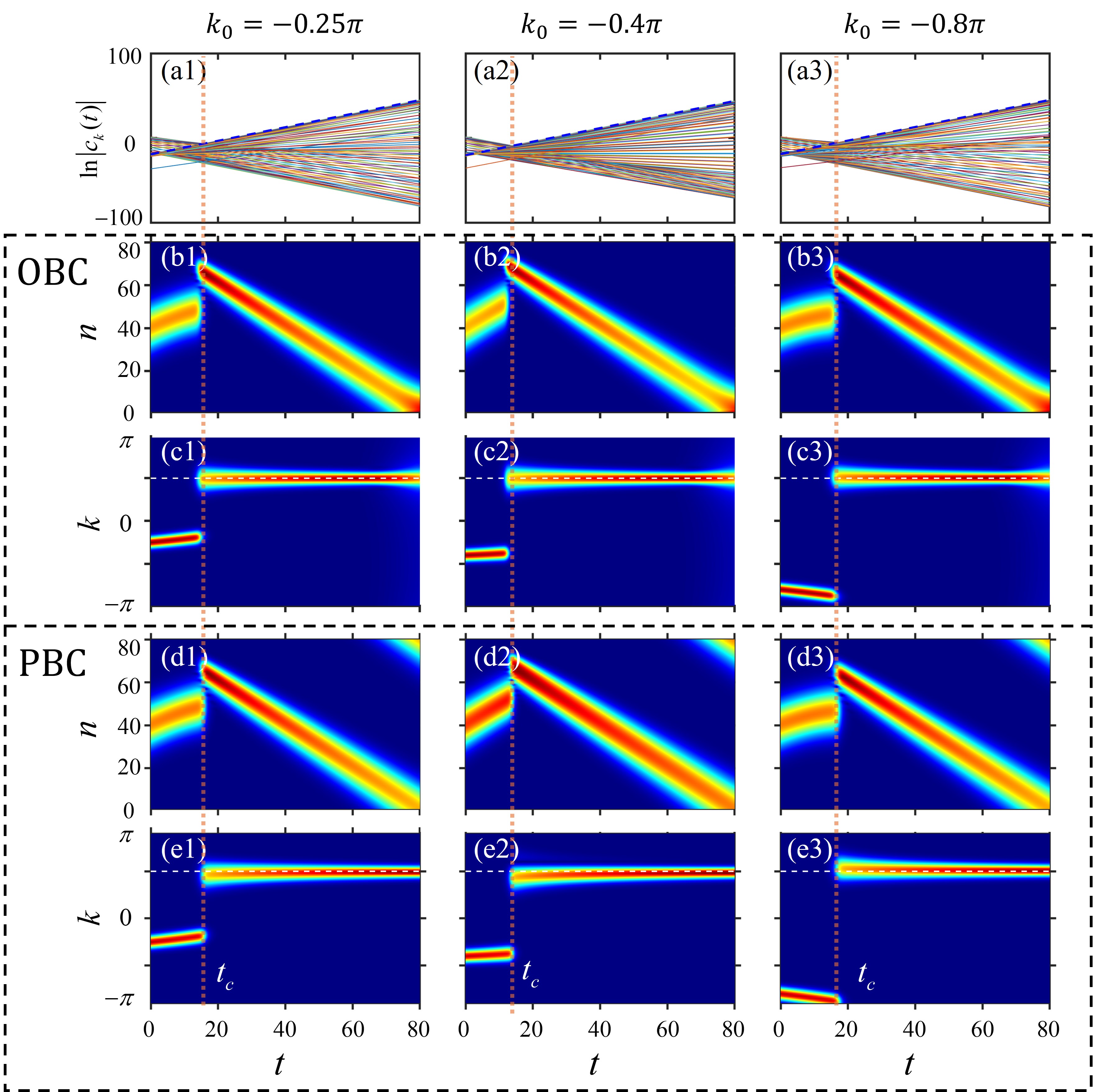}
\caption{ (a1-a3) The evolution of $c_k(t)$, where $c_{k=k^*=\pi/2}(t)$ is highlighted by the blue dashed lines. The corresponding wave function in real space under the OBC (b1-b3) and PBC (d1-d3). The evolution of the wave packet in momentum space under the OBC (c1-c3) and PBC (e1-e3), where the white dashed lines denote $k=\pi/2$. The red dotted lines denote the jump time $t_c$. The size is $N=80$. The initial central momentum $k_0=-0.25\pi$ (a1-e1), $k_0=-0.4\pi$ (a2-e2), and $k_0=-0.8\pi$ (a3-e3). Other parameters are the same as these in Fig.~2 of the main text.
\label{Fig_PBC_k0}}
\end{figure}

\clearpage

\section{\textcolor{black}{9. NH wave-packet jumps with real spectra beyond the Gaussian wave packets}}

\textcolor{black}{We now demonstrate that the NH wave-packet jumps with real spectra can exist also for Lorentzian wave packets. In Fig.~\ref{Fig_Lorentz_OBC}, we plot the evolution of the wave packet on the Hatano-Nelson lattice for size $N=50$ and $100$. The NH wave-packet jumps in real and momentum spaces are demonstrated. The wave packet jumps from the initial one to that at $k^*$ with the largest $E_I$. Such NH wave-packet jumps are also size-dependent.}

\begin{figure}[h!]
\centering
\includegraphics[width=14 cm]{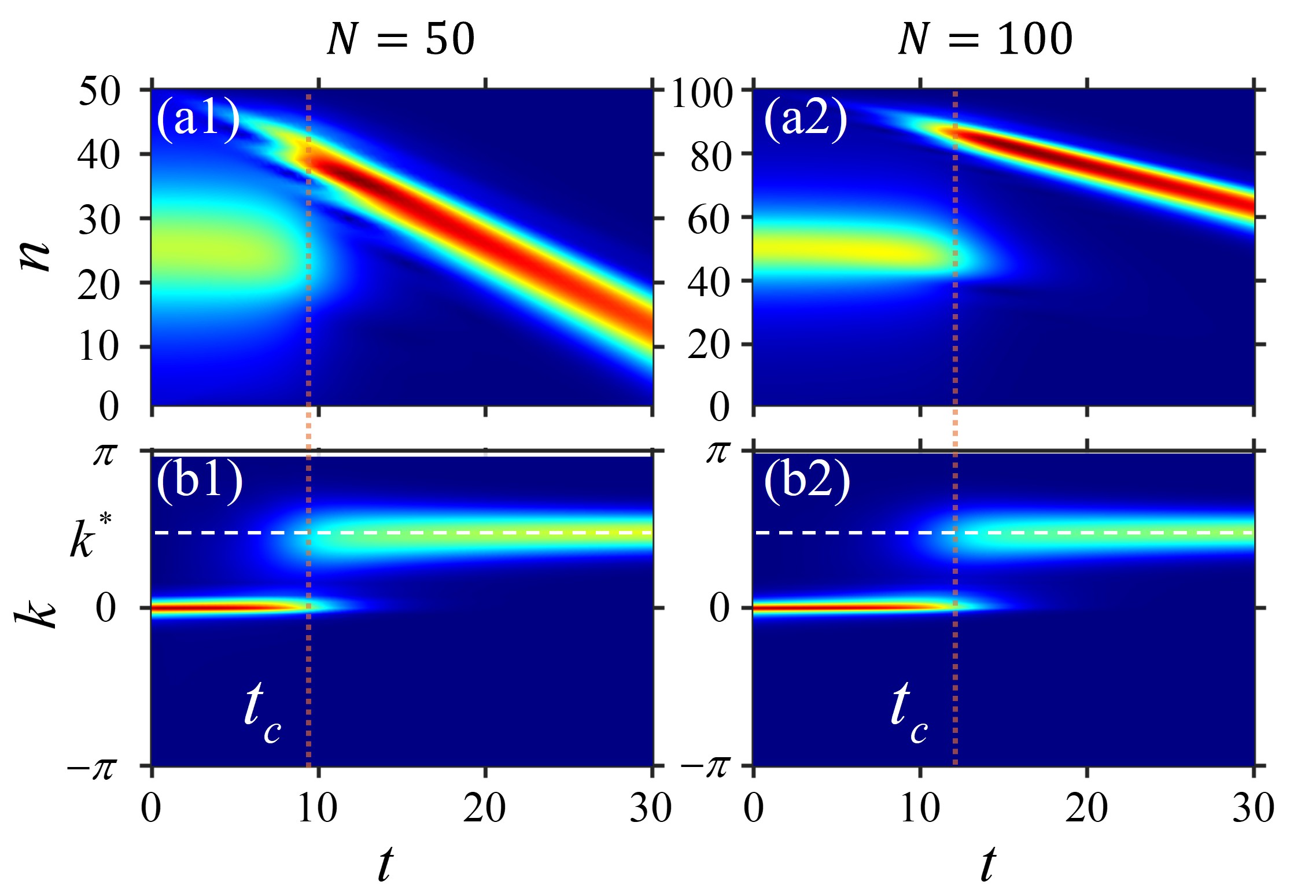}
\caption{\textcolor{black}{
(a1,~a2) Time evolution of a Lorentzian wave packet in (a1,~a2) real and (b1,~b2) momentum spaces. The vertical red dotted lines highlight the time $t_c$ when the NH wave-packet jumps occur. The sizes in (a,~b1) and (a2,~b2) are $N=50$ and $N=100$, respectively. Other parameters are $J_1^L=0.9,~J_1^R=0.1,~J_{m>1}^{L,R}=0,~k_0=0$,~and $\sigma=10$.} \label{Fig_Lorentz_OBC} }
\end{figure}

\clearpage

\section{10. NH wave-packet jumps with real spectra beyond the Hatano-Nelson model}

{In the section of \textit{NH wave-packet jumps with real spectra} in the main text, we discuss the NH wave-packet jumps with models where the momentum jumps to $k=k^*=\pi/2$, which shows the highest gain. Here we consider other models with $k^*\ne\pi/2$ and show that NH wave-packet jumps always occurs at $k=k^*$.}

In Fig.~\ref{Fig_PBC_km}, we show the cases with complex couplings $J_1^L, J_1^R$ with different $k^*=0.33\pi$, $~0.42\pi$, and $0.76\pi$, and see that the NH wave-packet jump can also occur. The momentum jumps from the initial $k_0=0$ to $k^*$ [see Figs.~\ref{Fig_PBC_km}(c1-c3) and~\ref{Fig_PBC_km}(e1-e3)].
The results under the PBC are exactly the same to those of OBC.   

\begin{figure}[h!]
\centering
\includegraphics[width=15 cm]{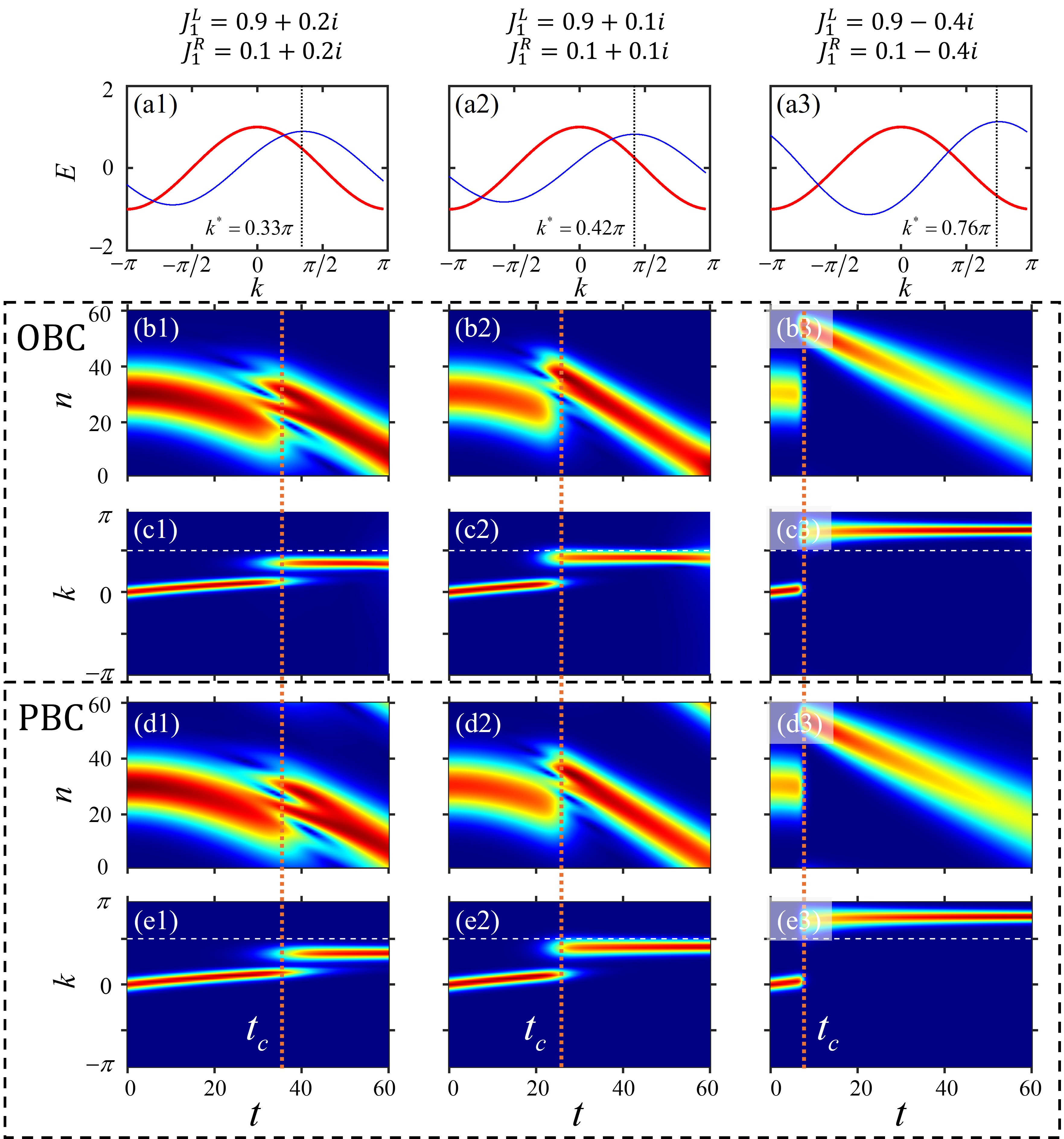}
\caption{ (a1-a3) Band structures, $E_R(k)$ (red lines) and $E_I(k)$ (blue lines). The black dotted lines represent $k^*$, where $E_I$ has its maximum. The evolution of the wave wavepacket in real space under the OBC (b1-b3) and PBC (d1-d3). The evolution of the wave packet in momentum space under the OBC (c1-c3) and PBC (e1-e3), where the white dashed lines denote $k=\pi/2$. The red dotted lines denote the jump time $t_c$. The parameters are $J_1^L=0.9+0.2i,~J_1^R=0.1+0.2i$ (a1-e1), $J_1^L=0.9+0.1i, J_1^R=0.1+0.1i$ (a2-e2), and $J_1^L=0.9-0.4i, J_1^R=0.1-0.4i$ (a3-e3). $N=60,~\sigma=5$, and $k_0=0$.
\label{Fig_PBC_km}}
\end{figure}

{The physical reason for the NH wave-packet jump is the competition among different momentum states.} The state with the highest gain at $k^*$ increases at the maximum rate, and hence dominates the evolution, indicating the NH wave-packet jump. This phenomena can exist in any general NH lattices {regardless of the existence of asymmetric coupling and skin effects}. To demonstrate this statement, we consider the symmetric coupling strengths with $J_1^L=J_1^R=0.5+0.3i$, where the skin effect is absent. As shown in Fig.~\ref{Fig_PBC_Beyond_HN}(a), the band structure is symmetric with respect to $k=0$, and $k^*=0$. The PBC spectra in the complex energy plane show a line, which hence denotes the absence of the skin effect [see Fig.~\ref{Fig_PBC_Beyond_HN}(b)]. The wave dynamics in real and momentum spaces under the OBC is shown in Figs.~\ref{Fig_PBC_Beyond_HN}(c,~d), where we see a NH wave-packet jump at $t_c\approx20$.

\begin{figure}[h!]
\centering
\includegraphics[width=10 cm]{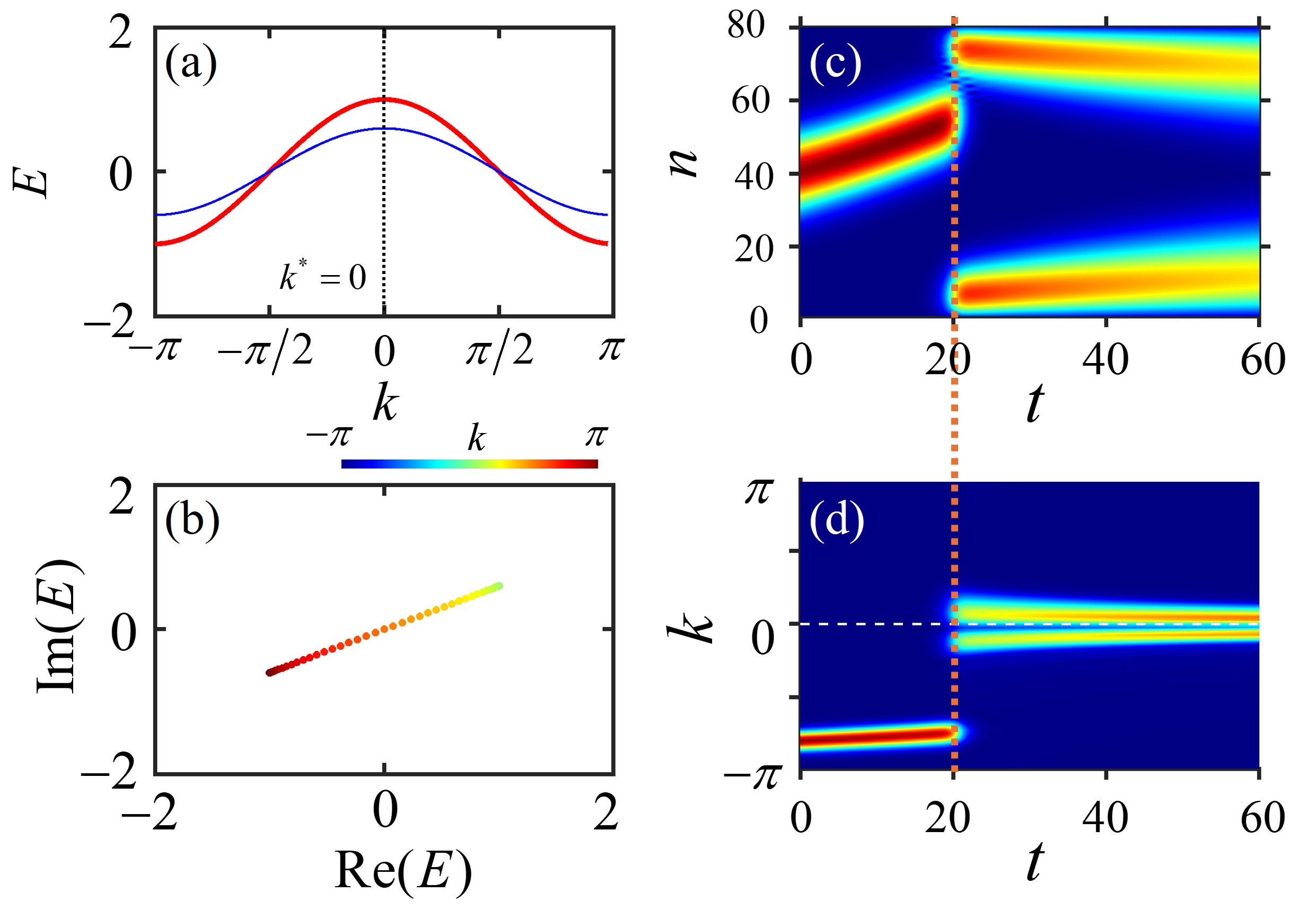}
\caption{ (a) Band structure, $E_R(k)$ (red lines) and $E_I(k)$ (blue lines). The black dotted lines represent $k^*$, where $E_I$ has its maximum. (b) The PBC spectrum in the complex energy plane. (c) The corresponding evolution of wave function in real space under the OBC. (d) The evolution of the wave packet in momentum space, where the white dashed lines denote $k=0$. The parameters are $J_1^L=J_1^R=0.5+0.3i$, $J_{m>1}^{L,R}=0$ $N=80,~\sigma=5$, and $k_0=-0.8\pi$.
\label{Fig_PBC_Beyond_HN}}
\end{figure}

\clearpage

\section{11. The correspondence between the wave-packet dynamics under OBCs and that under PBCs with the auxiliary wave packet}

In this section, we show the correspondence between the wave-packet dynamics in finite lattices under OBC and the extended lattices under PBC with the auxiliary wave packet. The edge effect under the OBC is equivalent to the {combined dynamics} of the initial wave packet and the auxiliary wave packet. 

\begin{figure}[h!]
\centering
\includegraphics[width=15 cm]{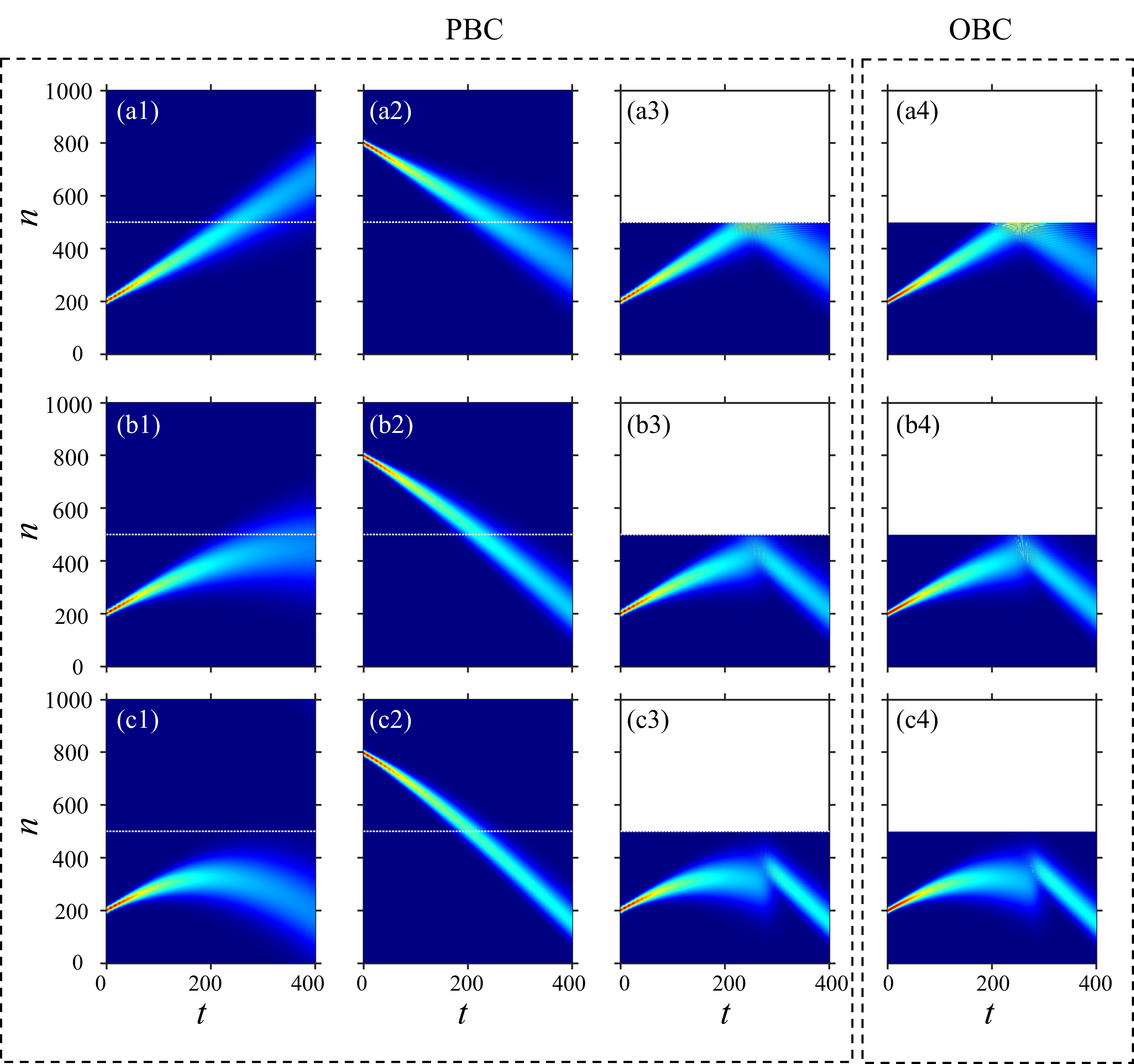}
\caption{ (a1-c1) Wave dynamics of the initial wave packet in the lattice under the PBC.  (a2-c2) Wave dynamics of the auxiliary wave packet in the lattice under the PBC. (a3-c3) Wave dynamics of the initial and auxiliary wave packets in the lattice under the PBC.  (a4-c4) Wave dynamics of the initial wave packet in the lattice under the OBC. The parameters are $J_L=J_R=1$ (a1-a4), $J_L=1,~J_R=0.96$ (b1-b4), and $J_L=1,~J_R=0.92$ (c1-c4). $\sigma=5$ and $k_0=-0.2\pi$. 
\label{Fig_OBC_PBC_1}}
\end{figure}

We first consider the wave reflection in the Hermitian case. For the Hermitian case, the reflection of the initial wave packet centered at $n_0$ can be considered as the transmission of the auxiliary wave packet at the mirror-reflection point $n_0'=2x_c-n_0$ with an opposite momentum $-k_0$. These two wave packets have opposite group velocities $\pm v_g=\frac{dE}{dk}\big|_{k=k_0}$, and hence propagate toward the edge. They reach the edge simultaneously at time $t=\frac{x_c-n_0}{|v_g|}=\frac{n_0'-x_c}{|v_g|}$. The auxiliary wave packet transmits across the edge and become the reflected wave packet of the initial wave packet, which is the reflection process in Hermitian case. The wave dynamics for the initial and auxiliary wave packet is shown in Figs.~\ref{Fig_OBC_PBC_1}(a1) and ~\ref{Fig_OBC_PBC_1}(a2), respectively. The wave dynamics of both wave packets is plotted in Fig.~\ref{Fig_OBC_PBC_1}(a3), where only the wave dynamics in the original lattice with size $N$ is shown. We see that the wave dynamics fully agrees with that in the original lattice under the OBC [Fig.~\ref{Fig_OBC_PBC_1}(a4)], thus demonstrating that the reflection at the edge is equivalent to the {combined dynamics of} the initial and auxiliary wave packets.

We then consider the NH case. For the NH lattice, the auxiliary wave packet locates at $n_2=2x_c-n_0-4\sigma^2h$ with the opposite momentum $-k_0$ and counter-propagates toward the edge with respect to the initial wave packet. Different from the Hermitian case, the group velocities and momenta of these two wave packets change with time. For the weak non-Hermiticity, the auxiliary wave packet reaches the edge earlier than the initial wave packet, as {the group velocity of the auxiliary (initial) wave packet increases (decreases) with time}. In this case, these two wave packets meet near but before the edge [see Fig.~\ref{Fig_OBC_PBC_1}(b3)]. The wave dynamics also agrees with that under the OBC [Fig.~\ref{Fig_OBC_PBC_1}(b4)]. For larger non-Hermiticiy, since the changes of the group velocities become faster, and the initial wave packet first experiences loss, while the auxiliary wave packet experiences gain. At time $t_c$ when the amplitude of the auxiliary wave packet exceed that of the initial wave packet, NH wave-packet jump occurs where the wave packet jumps from the initial one to the auxiliary one. The bulk dynamics in the larger lattice under the PBC with the two wave packets totally agrees with the single initial wave-packet dynamics in the original lattice under the OBC [see Figs.~\ref{Fig_OBC_PBC_1}(c3,~c4)].

\begin{figure}[h!]
\centering
\includegraphics[width=15 cm]{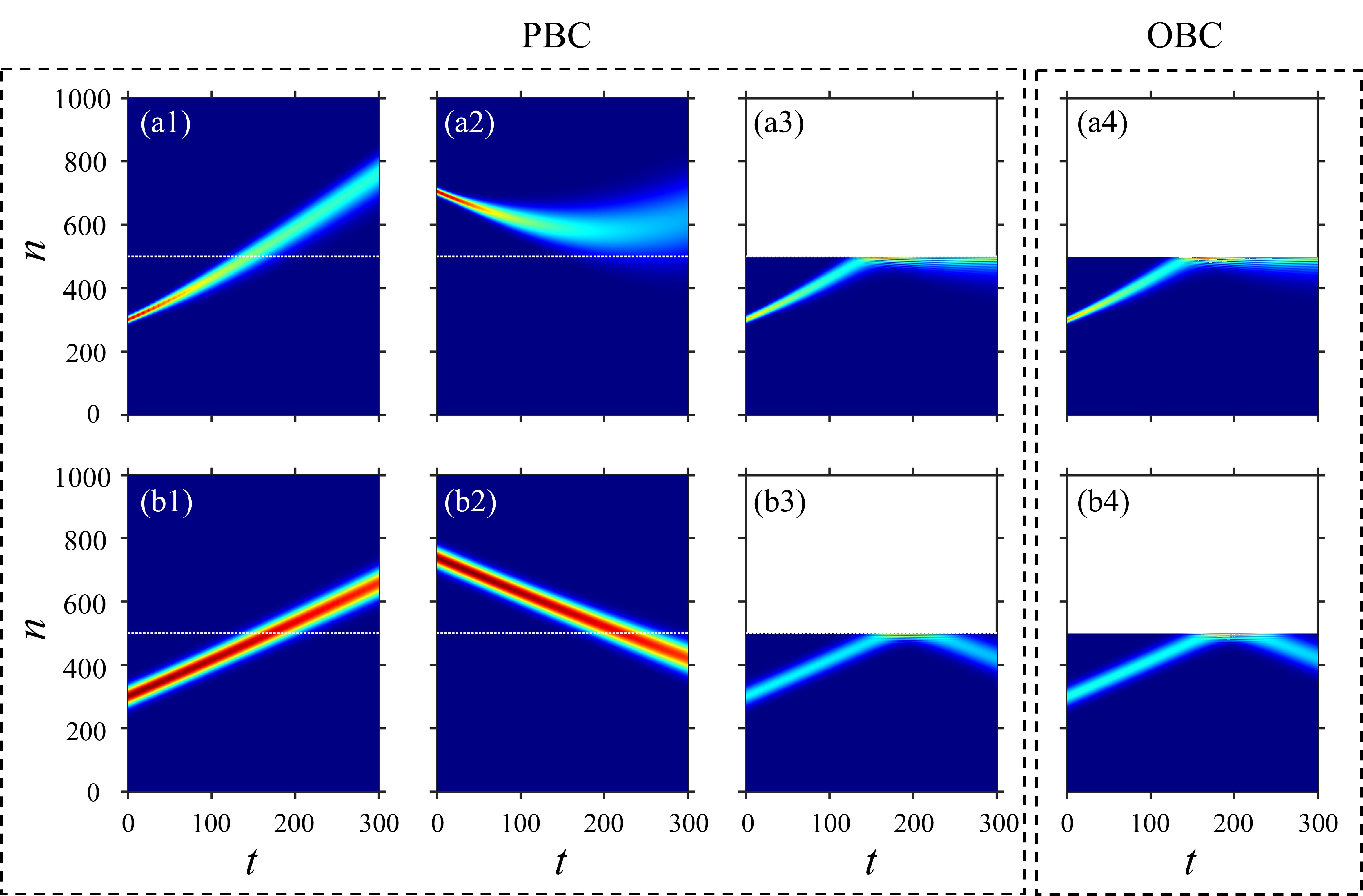}
\caption{ (a1,~b1) Wave dynamics of the initial wave packet in the lattice under the PBC.  (a2,~b2) Wave dynamics of the auxiliary wave packet in the lattice under the PBC. (a3,~c3) Wave dynamics of the initial and auxiliary wave packets in the lattice under the PBC. (a4,~b4) Wave dynamics of the initial wave packet in the lattice under the OBC. The width are $\sigma=5$ (a1-a4). $\sigma=15$ (b1-b4). $J_L=0.92,~J_R=1$, and $k_0=-0.2\pi$. 
\label{Fig_OBC_PBC_2}}
\end{figure}
{So far we have considered $J_R / J_L < 1$. We now consider the opposite case of $J_R/J_L > 1$ where the direction of the non-Hermiticity (skin effect) is flipped.}
For small width of the initial wave packet $\sigma=5$, the momentum of the auxiliary wave packet changes quickly, which hence causes the group velocity changes quickly accordingly. Therefore, the auxiliary wave packet cannot transmit across the edge, and hence there is no reflection [see Figs.~\ref{Fig_OBC_PBC_2}(a2,~a3)], which is called the dynamic skin effect in previous studies. The bulk wave dynamics with both the wave packets under the PBC agree well with that under the OBC [see Fig.~\ref{Fig_OBC_PBC_2}(a4)]. For larger width $\sigma=15$, the group velocity changes slowly due to the slowly-changed momentum, and hence the auxiliary wave packet transmits across the edge, becoming the reflected wave of the initial wave packet. Since the unequal central position of the initial wave packet, i.e., $n_2\ne -n_0$, there is a positive TGHS at the edge [see Fig.~\ref{Fig_OBC_PBC_2}(b3)]. The results under the PBC agree with that of a single initial wave packet under the OBC [Fig.~\ref{Fig_OBC_PBC_2}(b4)].

Therefore, the wave dynamics in the finite lattice under the OBC is equivalent to that in a larger lattice with both the initial and auxiliary wave packets. The edge effect corresponds to the interplay between the initial and auxiliary wave packets.

\begin{figure}[h!]
\centering
\includegraphics[width=10 cm]{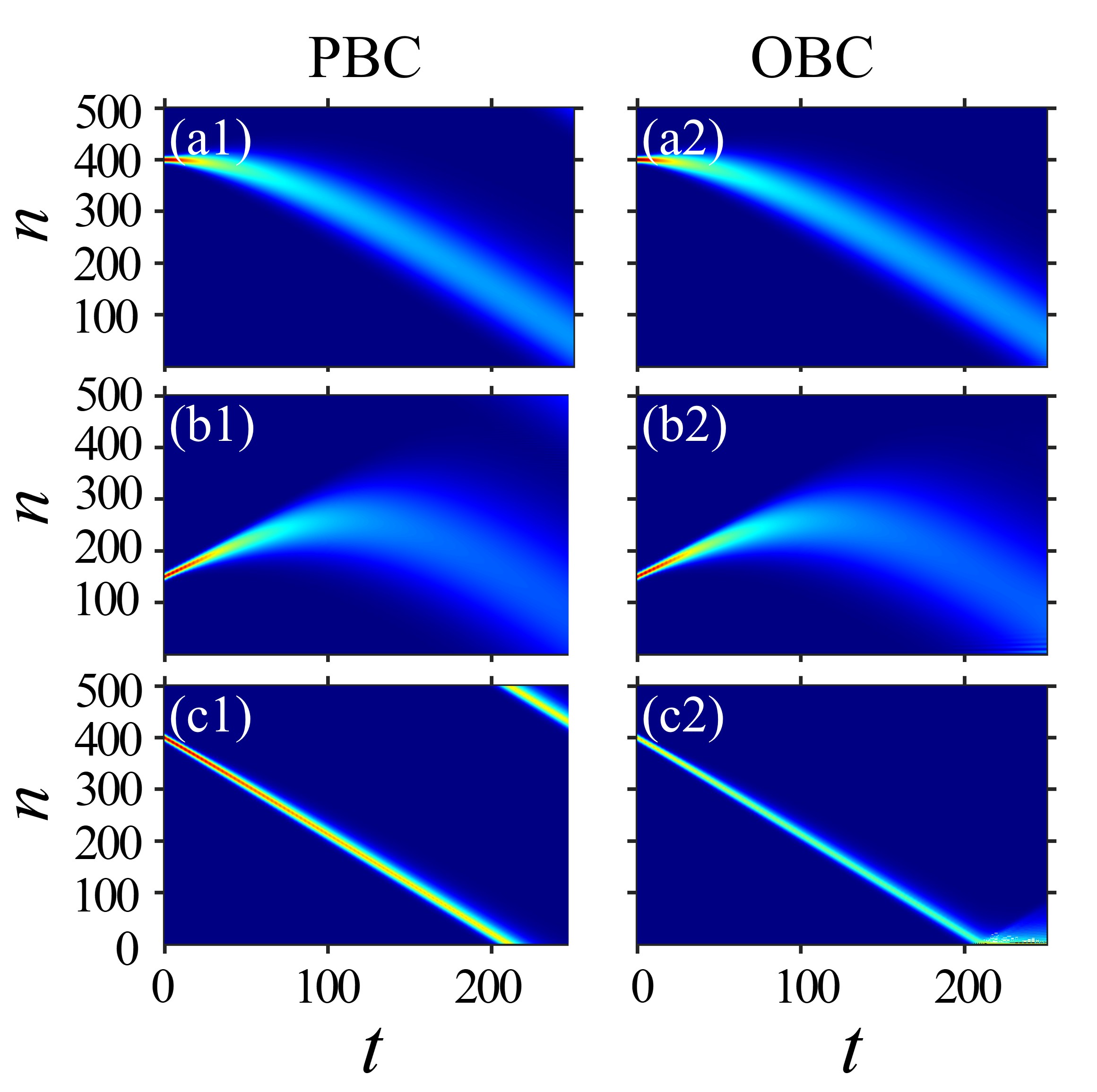}
\caption{ Wave dynamics of the initial wave packet in the lattice under the PBC (a1-c1) and OBC (a2-c2). The parameters are $k_0=0$ (a1,~a2), $k_0=-0.3\pi$ (b1,~b2), and $k_0=0.4\pi$ (c1,~c2). $\sigma=3$, $J_L=1$, and $J_R=0.92$.
\label{Fig_OBC_PBC_3}}
\end{figure}

{Finally, we show that, in some cases, the wave dynamics is the same under OBC and the not-extended PBC lattice, when the influence of the auxiliary wave packet can be ignored.} In Fig.~\ref{Fig_OBC_PBC_3}, we show the wave-packet dynamics in real space under the PBC and OBC for different initial central momentum $k_0$. We find that the wave dynamics is the same under different boundary conditions when the wave packet propagates in the bulk. The auxiliary wave packet does not affect the wave dynamics since its amplitude is always lower than that of the initial wave packet, so the initial wave packet dominates the whole evolution. For the case of initial wave packet experiencing gain at $k_0=0.4\pi$, the wave dynamics under both boundary conditions is also the same before the wave packet reaches the edge [Figs.~\ref{Fig_OBC_PBC_3}(c1,~c2)].

\clearpage

\section{12. The influence of different parameters on NH wave-packet jumps close to the edge}

In this section, we discuss the influence of different parameters including the strength of non-Hermiticify, the width $\sigma$ of the wave packet, the size $N$ of the lattice on the NH wave-packet jump time. 

\subsection{12.1 The influence of the strength of non-Hermiticity on the jump time}

We first consider the influence of the strength of non-Hermiticity on the NH wave-packet jump time. We fix $J_L=1$ and change $J_R$ from $0.89$ to $0.92$ and then $0.95$, corresponding the decrease of the non-Hermiticity. The wave dynamics in real and momentum spaces is shown in Fig.~\ref{Fig_OBC_J}. We see that the NH wave-packet jump time $t_c$ decreases with the decrease of the non-Hermiticity, and finally becomes the reflection at the edge [see Fig.~\ref{Fig_OBC_J}(b3)]. This can be demonstrated from the evolution of $c_k$ in Figs.~\ref{Fig_OBC_J}(a1-a3), where we see that when the amplitude of the initial (auxiliary) wave packet gradually decreases (increases) and NH wave-packet jump occurs when the amplitudes of these two wave packets are equal. For the smaller non-Hermiticity, the amplitudes of these two wave packets nearly do not change, indicating the absences of 
the NH wave-packet jump [see Fig.~\ref{Fig_OBC_J}(b3)]. In the Hermitian limit, the NH wave-packet jump changes to the reflection at the edge, with no shift of the momentum.

\begin{figure}[h!]
\centering
\includegraphics[width=15 cm]{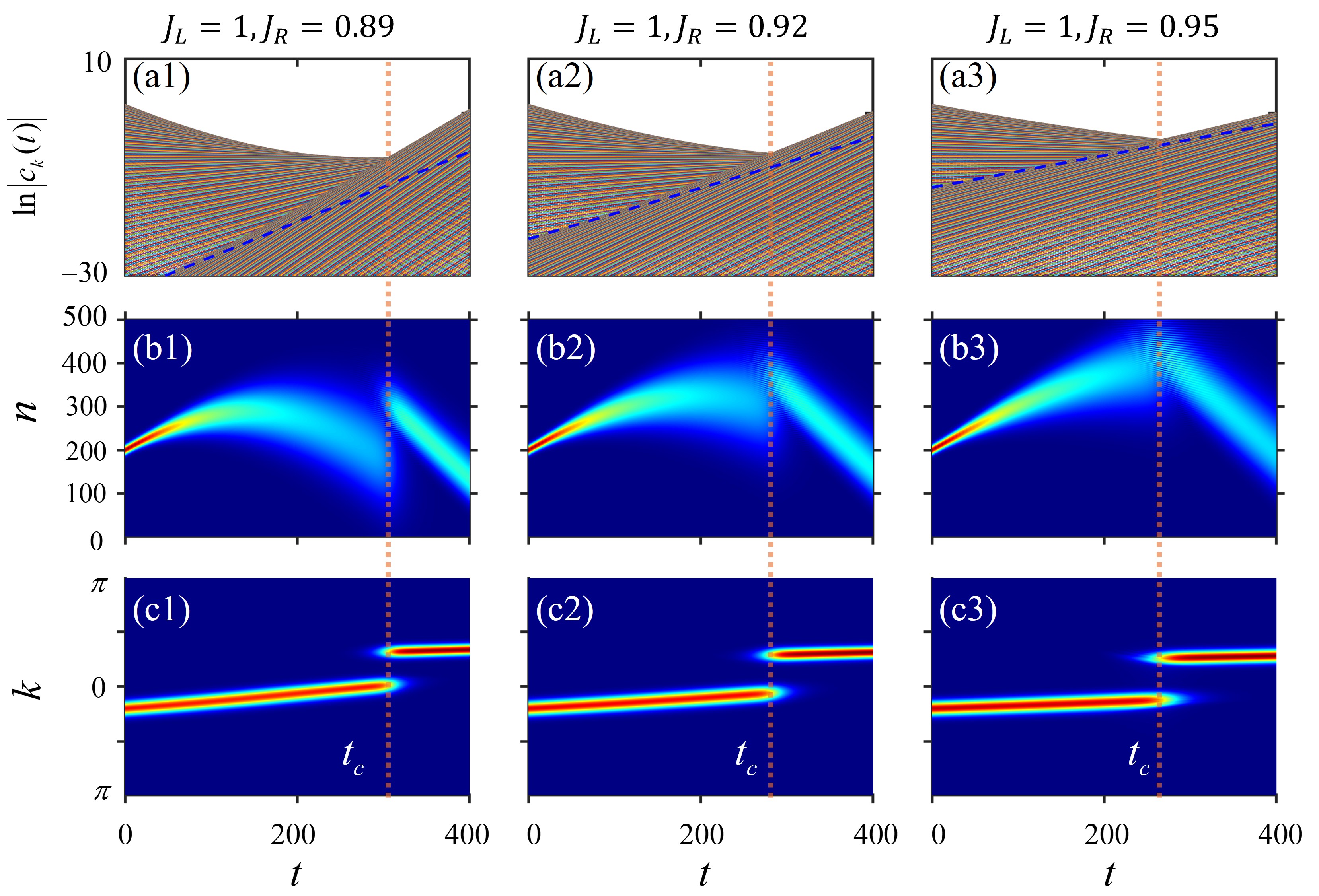}
\caption{
(a1-a3) The evolution of $c_k(t)$, where $c_{k=-k_0}(t)$ is highlighted by the blue dashed lines. (b1-b3) The evolution of the wave packet in real space. (c1-c3) The evolution of the wave packet in momentum space. The red dotted lines denote the jump time $t_c$. The size is $N=500$. The parameters are $J_L=1, J_R=0.89$ (a1-c1), $J_L=1, J_R=0.92$ (a2-c2), and $J_L=1, J_R=0.95$ (a3-c3). Other parameters are the same as those in Fig.~3 of the main text.
\label{Fig_OBC_J}}
\end{figure}
\clearpage

\subsection{12.2 The influence of the width of the wave packet $\sigma$ on the jump time}

We next discuss the influence of the width of the wave packet $\sigma$ on the NH wave-packet jumps. In Fig.~\ref{Fig_OBC_sigma}, we plot the wave-packet dynamics in real and momentum spaces, as well as the evolution of $c_k(t)$. We see that, with increasing width $\sigma$, the jump time $t_c$ decreases. Meanwhile, we see that the shift of the momentum decreases, which hence corresponds to the {smaller change of the} group velocity. Therefore, the wave packet touches the edge and {the NH wave-packet jump becomes a simple reflection} [see Fig.~\ref{Fig_OBC_sigma}(b3)]. The width $\sigma$ plays an effect opposite to the strength of non-Hermiticity ($d E_I/dk$) on the wave-packet dynamics, which can be seen from the evolution $k_{\mathrm{max}}(t)=k_0+\frac{t}{2\sigma^2}\frac{d E_I(k)}{d k}\bigg|_{k=k_{\mathrm{max}}}$. The increase of $\sigma$ corresponds to the decrease of {the second term, weakening the effect of non-Hermiticity.}

\begin{figure}[h!]
\centering
\includegraphics[width=15 cm]{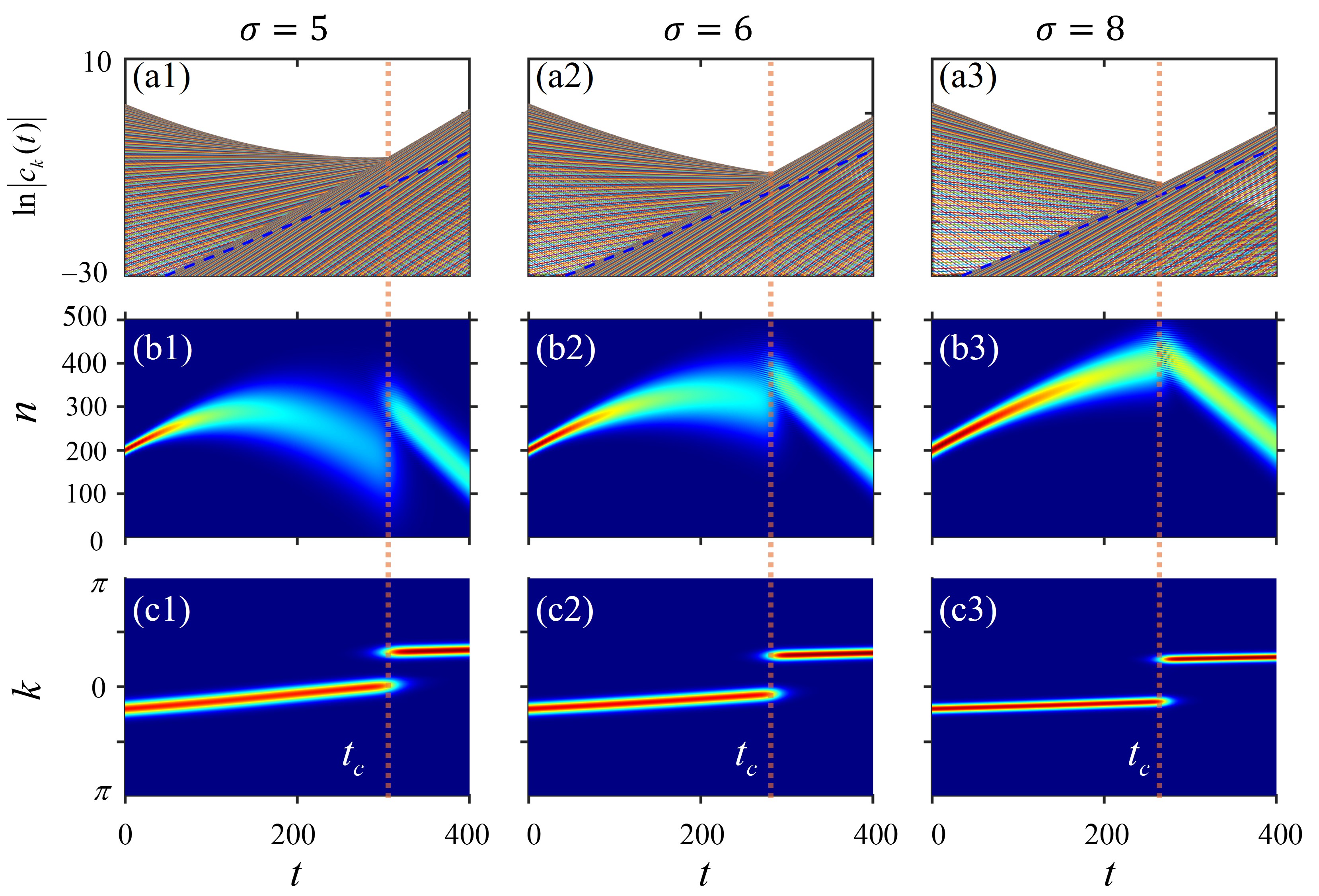}
\caption{
(a1-a3) The evolution of $c_k(t)$, where $c_{k=-k_0}(t)$ is highlighted by the blue dashed lines. (b1-b3) The evolution of the wave packet in real space. (c1-c3) The evolution of the wave packet in momentum space. The red dotted lines denote the jump time $t_c$. The width is $\sigma=5$ (a1-c1), $\sigma=6$ (a2-c2), and $\sigma=8$ (a3-c3). Other parameters are $N=500$. $J_L=1, J_R=0.89$, and $k_0=-0.2\pi$.
\label{Fig_OBC_sigma}}
\end{figure}

\clearpage
\subsection{12.3 The influence of the lattice size $N$ on the jump time}

We then discuss the influence of the lattice size on the jump time. {In Fig.~\ref{Fig_OBC_N}, we} show the wave-packet dynamics in real and momentum spaces for $N=450$,~$500$, and $N=550$, keeping the initial position $n_0=200$. In this case, the change of size $N$ corresponds the change of the distance between the initial wave packet to the edge of the lattice, which increases with the size $N$. As shown in Fig.~\ref{Fig_OBC_N}, we see that the jump time $t_c$ increases with the increase of $N$. This is because when $N$ increases, the corresponding distance between the auxiliary wave packet and the edge increases. Therefore, it needs more time to transmit to the edge where it induces the jump.

\begin{figure}[h!]
\centering
\includegraphics[width=15 cm]{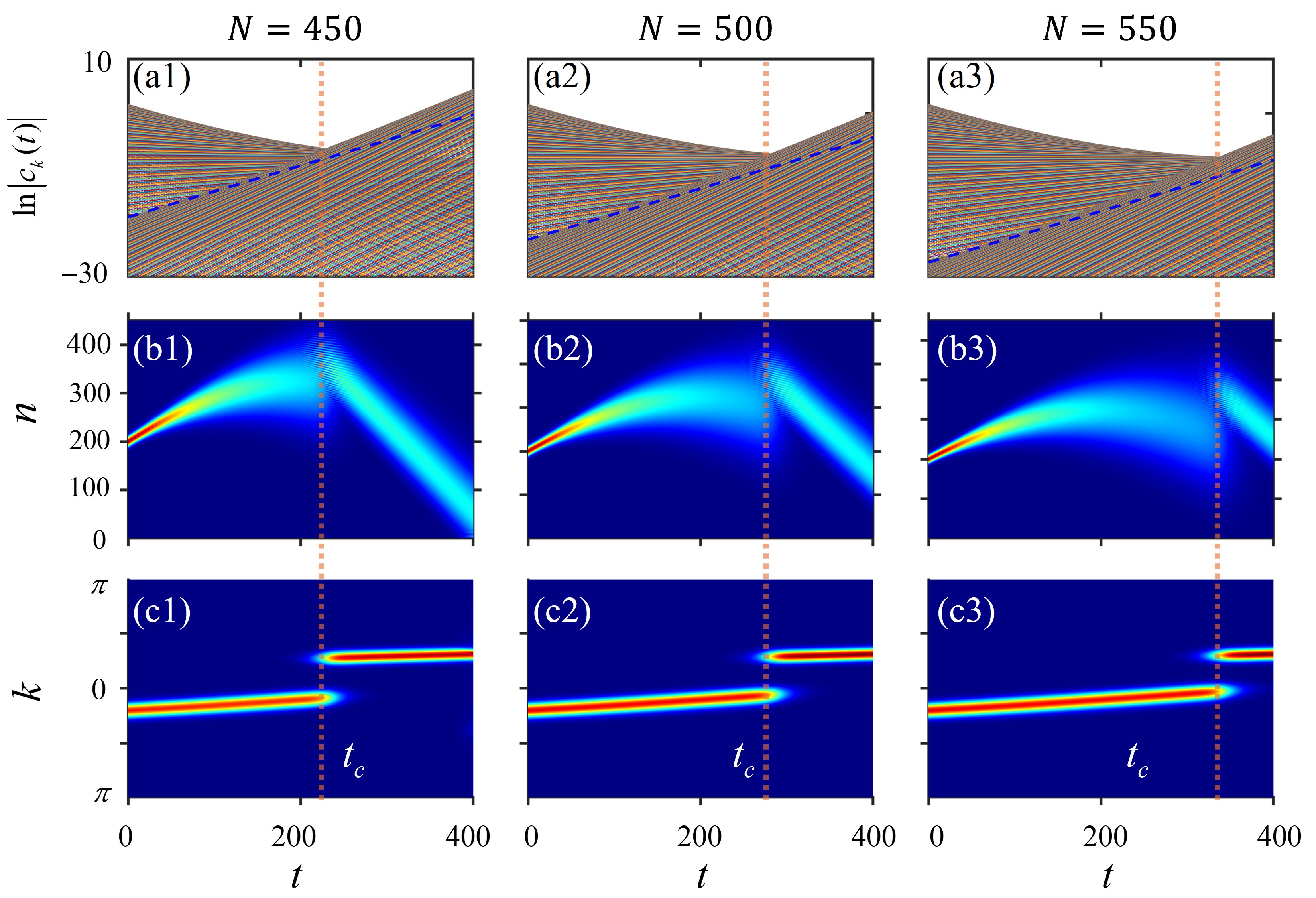}
\caption{
(a1-a3) The evolution of $c_k(t)$, where $c_{k=-k_0}(t)$ is highlighted by the blue dashed lines. (b1-b3) The evolution of the wave packet in real space. (c1-c3) The evolution of the wave packet in momentum space. The red dotted lines denote the jump time $t_c$. The lattice size is $N=450$ (a1-c1), $N=500$ (a2-c2), and $N=550$ (a3-c3). Other parameters are $k_0=-0.2\pi$. $J_L=1, J_R=0.92$, and $\sigma=5$.
\label{Fig_OBC_N}}
\end{figure}

\clearpage
\subsection{12.4 The influence of the initial momentum $k_0$ on the NH wave-packet jump time}
We then discuss the influence of the initial momentum $k_0$ on the jump time $t_c$. We show the results with $k_0=-0.2\pi,~-0.25\pi$, and $-0.3\pi$ in Fig.~\ref{Fig_OBC_k0}. One can see that the jump time $t_c$ decreases when $k_0$ changes from $-0.2\pi$ to $-0.3\pi$. This is consistent with the evolution of $c_k$ shown in Figs.~\ref{Fig_OBC_k0}(a1-a3), where we see that the component of the initial (auxiliary) wave packet decreases (increases) at a faster rate when $k_0$ decreases from $-0.2\pi$ to $-0.3\pi$ due to the larger loss (gain) of the initial (auxiliary) wave packet. Therefore, the time when the amplitude of the auxiliary wave packet exceed that of the initial wave packet becomes earlier, indicating the decrease of $t_c$.

\begin{figure}[h!]
\centering
\includegraphics[width=15 cm]{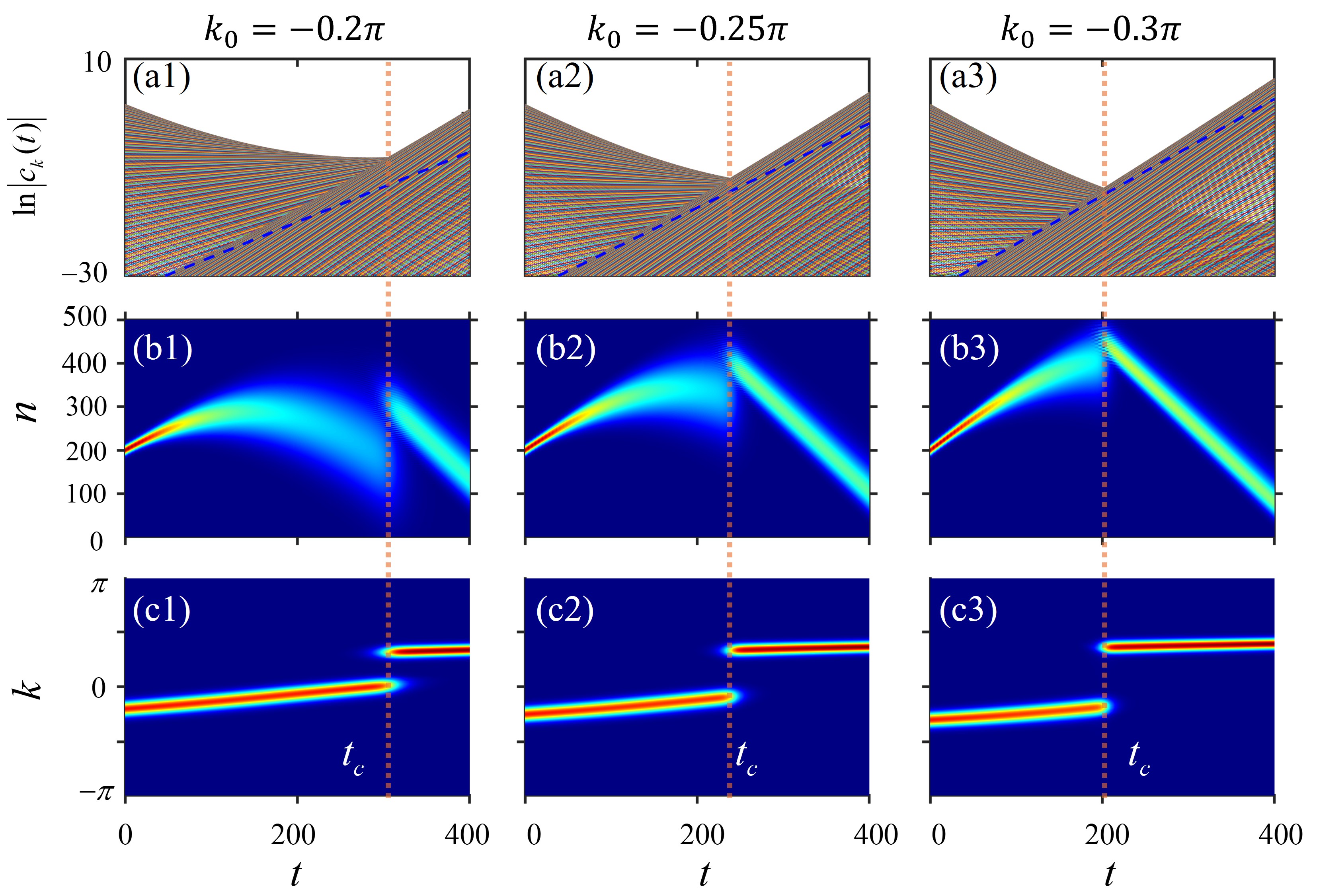}
\caption{
(a1-a3) The evolution of $c_k(t)$, where $c_{k=-k_0}(t)$ is highlighted by the blue dashed lines. (b1-b3) The evolution of the wave packet in real space. (c1-c3) The evolution of the wave packet in momentum space. The red dotted lines denote the jump time $t_c$. The initial momentum of the wave packet is $k_0=-0.2\pi$ (a1-c1), $k_0=-0.25\pi$ (a2-c2), $k_0=-0.3\pi$ (a3-c3). Other parameters are $N=500$. $J_L=1, J_R=0.89$, and $\sigma=5$.
\label{Fig_OBC_k0}}
\end{figure}

\clearpage

\subsection{\textcolor{black}{12.5 The influence of the global gain on the NH wave-packet jumps}}

\textcolor{black}{In Fig.~3 of the main text, we show the NH wave-packet jump near the edge, which is due to competition between the initial and auxiliary wave packets.  The initial wave packet experiences loss, while the auxiliary wave packet experiences gain. The NH wave-packet jump occurs at the time when the intensities of the two wave packets become equal. We plot the total intensity (probability) $I=\sum_n|\psi_n(t)|^2$} \textcolor{black}{as a function of time when we do not normalize the wavefuction at each time step in Fig.~\ref{fig_V_fig3}(d1). The minimum of the total intensity corresponds to the point where the NH wave-packet jump occurs, i.e. the jump from the initial wave packet to the auxiliary wave packet occurs}}

\begin{figure}[h!]
\centering
\includegraphics[width=13 cm]{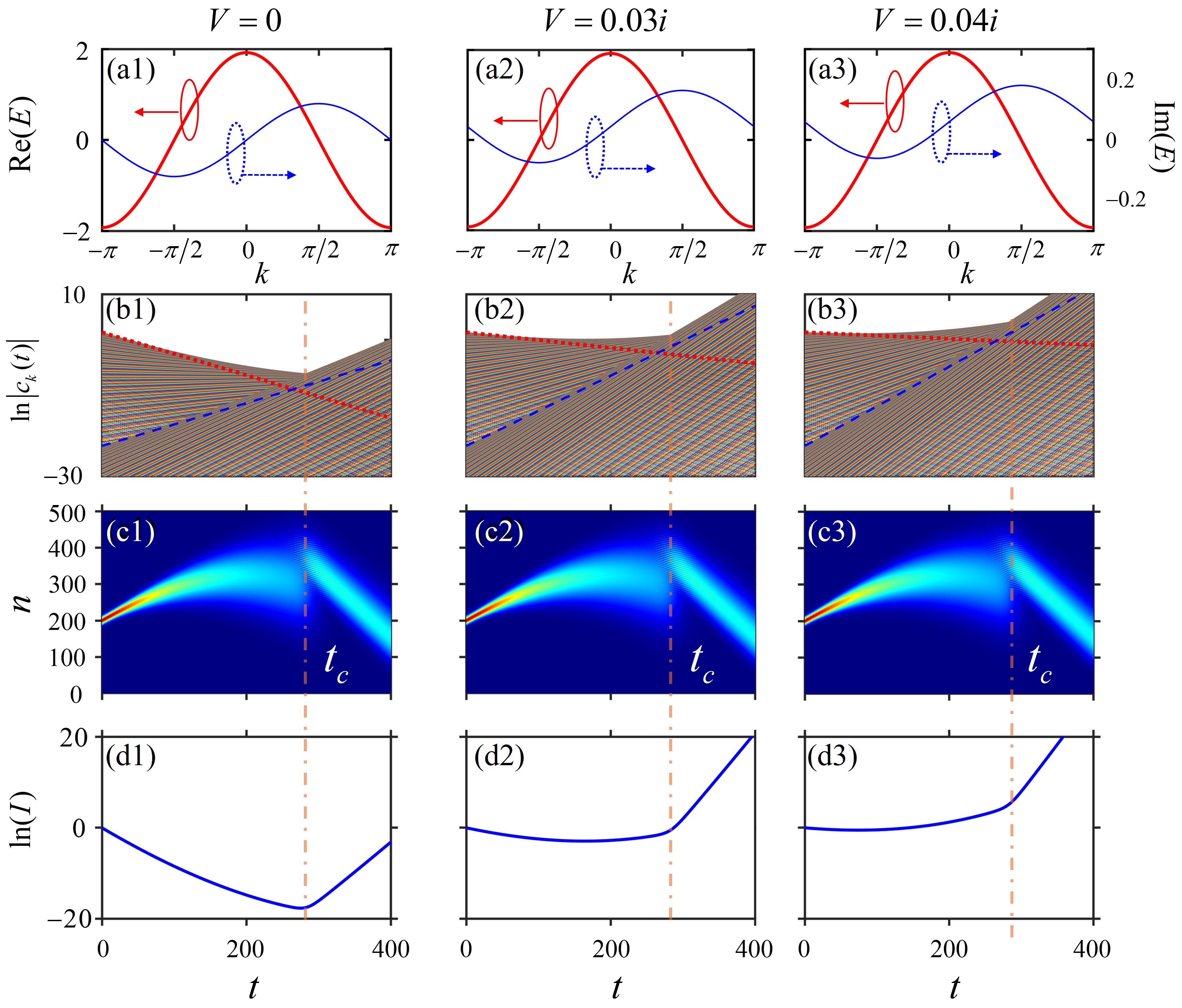}
\caption{\textcolor{black}{(a1-a3) Band structure with $\text{Re}(E)$ (red lines) and $\text{Im}(E)$ (blue lines). (b-b3) Evolution of the $c_k$, where $c_{k=k_0}(t)$ is highlighted by the red dotted lines, while $c_{k=-k_0}(t)$ is highlighted by the blue dashed lines. (c1-c3) Evolution of the wave packet with NH wave-packet jumps in real space. (d1-d3) The corresponding total intensity during the time evolution. }}
\label{fig_V_fig3}
\end{figure}

\textcolor{black}{The total intensity where the NH wave-packet jump occurs is quite small, $\sim e^{-20}$, with the parameters used here. In order to observe the NH wave-packet jump experimentally in realistic setups, it is desirable to work in a regime with a larger intensity. One way to achieve it is to add a global on-site gain, which does not affect the wave-packet dynamics apart from the change of the total intensity.}
\textcolor{black}{For instance, in Fig.~\ref{fig_V_fig3}(a2-d2), we consider an on-site gain potential $V=0.03i$, which adds $0.03i$ globally to $\text{Im}(E)$. The wave-packet dynamics shows a NH wave-packet jump at the time exactly equal to the original problem without the global gain. From Fig.~\ref{fig_V_fig3}(b2), we can see that the slopes of all modes increases, and so does the total intensity. We further consider $V=0.04i$ and find that the total intensity can increase further.}
\begin{figure}[h!]
\centering
\includegraphics[width=13 cm]{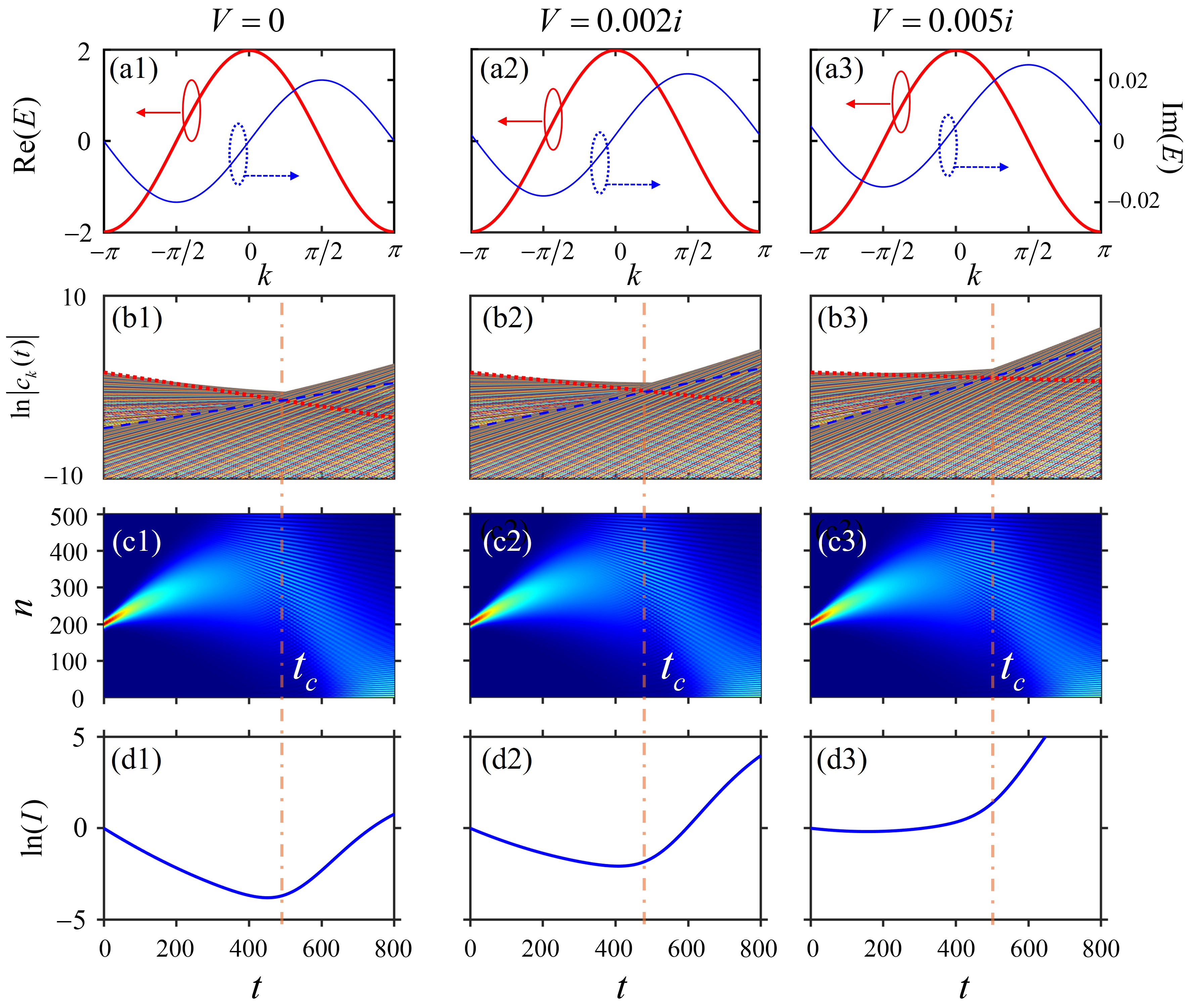}
\caption{\textcolor{black}{(a1-a3) Band structure with $\text{Re}(E)$ (red lines) and $\text{Im}(E)$ (blue lines). (b-b3) Evolution of the $c_k$, where $c_{k=k_0}(t)$ is highlighted by the red dotted lines, while $c_{k=-k_0}(t)$ is highlighted by the blue dashed lines. (c1-c3) Evolution of the wave packet with NH wave-packet jumps in real space. (d1-d3) The corresponding total intensity during the time evolution.}
\label{fig_V}}
\end{figure}

\textcolor{black}{We can also find parameters where the intensity at the NH wave-packet jump is not so small. By choosing smaller initial momentum and smaller non-Hermiticity, the intensity at the NH wave-packet jump can increase, with a trade-off of less sharper signal. In Fig.~\ref{fig_V}, we plot the wave dynamics for $J_L=1, J_R=0.98,~k_0=-0.1\pi$, where the intensity at the NH wave-packet jump is $\sim e^{-4}$. Also for this case, of course, the total intensity at the NH wave-packet jump can be increased when a global gain is added.}

\clearpage

\section{13. NH wave-packet jumps from edge to bulk}

In the main text, we have discussed the NH wave-packet jump {close to the edge} of lattices, which arises from the competition between the initial wave packet and the auxiliary one. For $J_R<J_L$, the amplitude of the initial wave packet decreases due to the loss for $k_0\in(-\pi,0)$, while the amplitude of the auxiliary wave packet increases since the gain for $-k_0\in(0,\pi)$. Therefore, the NH wave-packet jump occurs when the amplitude of the auxiliary wave packet exceeds the amplitude of that of the initial wave packet. Such a NH wave-packet jump can always occur for $k_0\in(-\pi,0)$ even the {initial wave packet touches and propagates along the edge}. To demonstrate this statement, we show the wave-packet dynamics in real and momentum spaces for different strength of the non-Hermiticity in Fig.~\ref{Fig_OBC_long_time}. We see that the NH wave-packet jumps can still occurs even when the wave packet propagates along the edge, thus demonstrating the NH wave-packet jumps from edge to bulk.

\begin{figure}[h!]
\centering
\includegraphics[width=15 cm]{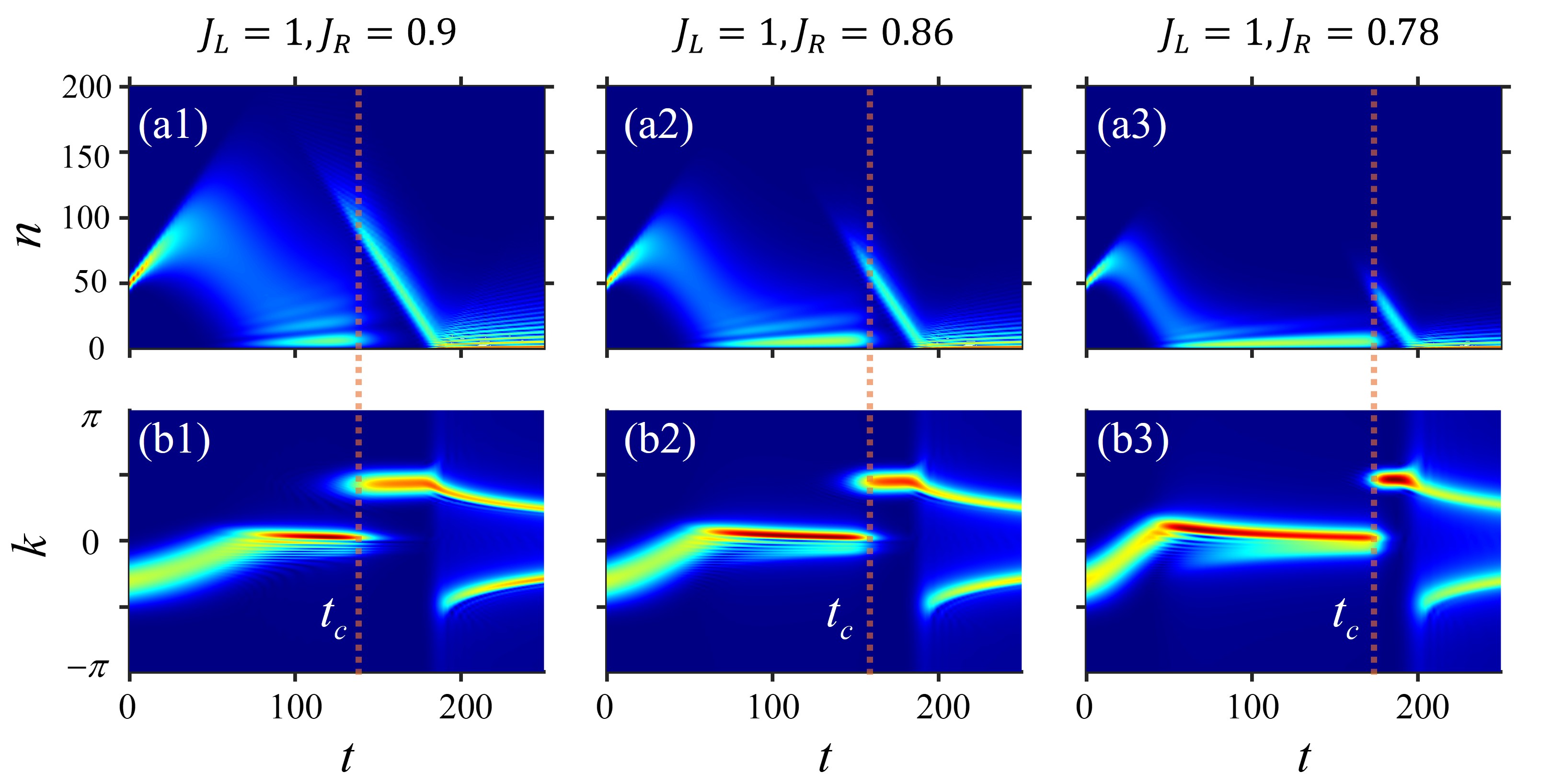}
\caption{
(a1-a3) The evolution of the wave packet in real space. (b1-b3) The evolution of the wave packet in momentum space. The red dotted lines denote the jump time $t_c$. The coupling strength is $J_L=1,~J_R=0.9$ (a1-c1), $J_L=1,~J_R=0.86$ (a2-c2), and $J_L=1,~J_R=0.78$ (a3-c3). Other parameters are $k_0=-0.2\pi$, $N=200$, and $\sigma=2$.
\label{Fig_OBC_long_time}}
\end{figure}

However, when the initial wave packet experiences no loss or gain with $k_0\in[0,~\pi]$, NH wave-packet jumps may disappear. This is because the initial wave packet feels continuous gain, while the auxiliary wave packet first undergoes loss and then gain, with its amplitude remaining lower than that of the initial wave packet. Thus, the initial wave packet dominates the whole evolution. We demonstrate this statement in Fig.~\ref{Fig_OBC_long_time_2}, where we show the wave dynamics in real and momentum spaces with different $k_0\in[0,\pi]$. We see that the the wave packet touches the edge, but no jumps occur.

\begin{figure}[h!]
\centering
\includegraphics[width=15 cm]{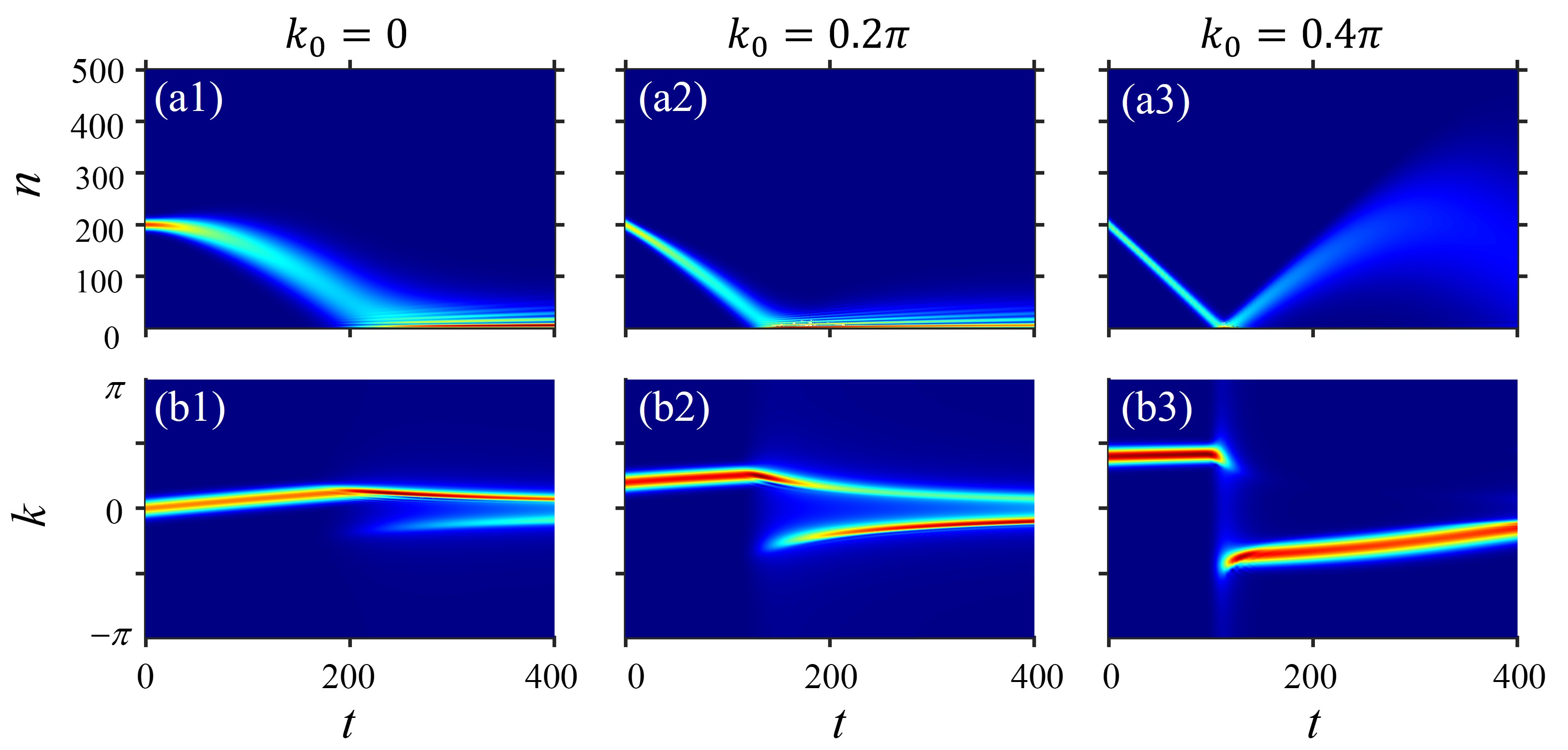}
\caption{
(a1-a3) The evolution of the wave packet in real space. (b1-b3) The evolution of the wave packet in momentum space. The coupling strength is $J_L=1,~J_R=0.89$. Other parameters are $N=500$, $J_L=1, J_R=0.89$, and $\sigma=5$.
\label{Fig_OBC_long_time_2}}
\end{figure}

\clearpage

\section{14. Identify the group velocities with TGHSs}
In this section, we discuss the changes of group velocities during the reflection. In Fig.~\ref{Fig_OBC_GH_1}, we show the case of $J_J>J_R$, which means the NH direction (direction of the skin effect) is opposite to the edge. We can see a negative TGHS in the real-space evolution [Figs.~\ref{Fig_OBC_GH_1}(a1)]. The momentum-space evolution shows a jump behavior, from $k_0$ to {nearly} $-k_0$. Because of the shift of the momentum, the group velocity before and after the reflection become different in {magnitude}.
We term the group velocities of the initial and auxiliary wave packets as $V_g$ and $V_g'$, respectively. As shown in  Fig.~\ref{Fig_OBC_GH_1}(c1), these two group velocities change with time according to Eq.~(\ref{V}) (green dashed lines), which agree with the extracted group velocities from the separated evolutions (blue solid lines). The initial group velocities of these two wave packets are highlighted by the red and blue dotted circles, while the $V_g(t_1)$ and $V_g'(t_2)$ where the reflection occurs are highlighted by the red and blue solid circles. The change in group velocities at the edge is denoted by the black arrow. We see that the {absolute value of the} group velocity after the reflection is larger than that before the reflection, i.e., $|V_g'(t_2)|>|V_g(t_1)|$. In the Hermitian limit, i.e., $J_L=J_R$, the two group velocities satisfy $V_g'(t)=-V_g(t)$ [Fig.~\ref{Fig_OBC_GH_1}(c2)]. For the case of $k_0=-0.5\pi$ with $dE_I/dk=0$, the group velocities also satisfy $V_g'(t)=-V_g(t)$ [Fig.~\ref{Fig_OBC_GH_1}(c3)], since the momentum does not change [Fig.~\ref{Fig_OBC_GH_1}(b3)].

\begin{figure}[h!]
\centering
\includegraphics[width=15 cm]{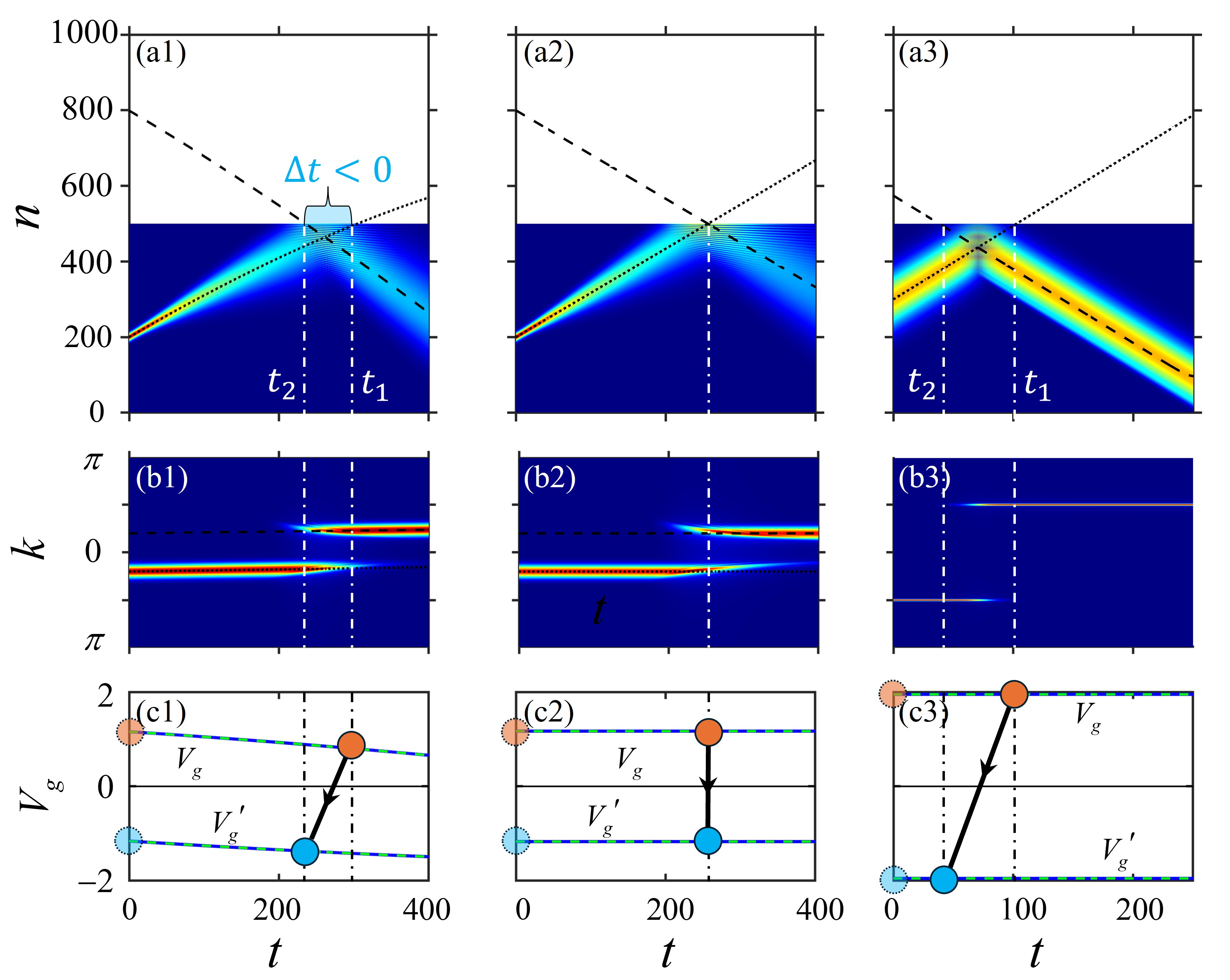}
\caption{
(a1-a3) Time evolution of the wave packet in the finite lattice of size $N = 500$ under the OBC. (b1-b3) The evolution of the wave packet in momentum space. (c1-c3) The evolution of group velocities of the initial (upper lines) and auxiliary (lower lines) wave packets. The red (blue) dotted circles denote the initial incident (reflected) state, which moves to the state of the red (blue) solid
circles when the reflection occurs.
The parameters for (a1-c1) and (a2-c2) are $J_L=1,~J_R=0.98$ and $J_L=J_R=1$.  $k_0=-0.2\pi$, and $\sigma=5$. The parameters in (a3-c3) are the same as those used in Fig.~4(h) of the main text.
\label{Fig_OBC_GH_1}}
\end{figure}

\clearpage

For the case of $J_L<J_R$, i.e., the NH direction toward the edge, the TGHS is positive with $\Delta t=t_2-t_1>0$. For the large non-Hermiticity, the group velocities of both the initial and auxiliary wave packets increase rapidly, as shown in Fig.~\ref{Fig_OBC_GH_2}(c1). During the reflection process, the group velocities satisfy $|V_g'(t_2)|<|V_g(t_1)|$ . When the non-Hermiticity decreases, the shift of momentum is small, and hence the change of group velocity is small. In this case, the group velocity of the reflected wave packet also satisfies  $|V_g'(t_2)|<|V_g(t_1)|$ [see Fig.~\ref{Fig_OBC_GH_2}(c2)]. For the case of $k_0=-0.5\pi$, the central momentum does not change, and hence the group velocities remain $V_g'(t)=-V_g(t)$, same to the Hermitian case, as shown in Fig.~\ref{Fig_OBC_GH_2}(c3). 

\begin{figure}[h!]
\centering
\includegraphics[width=15 cm]{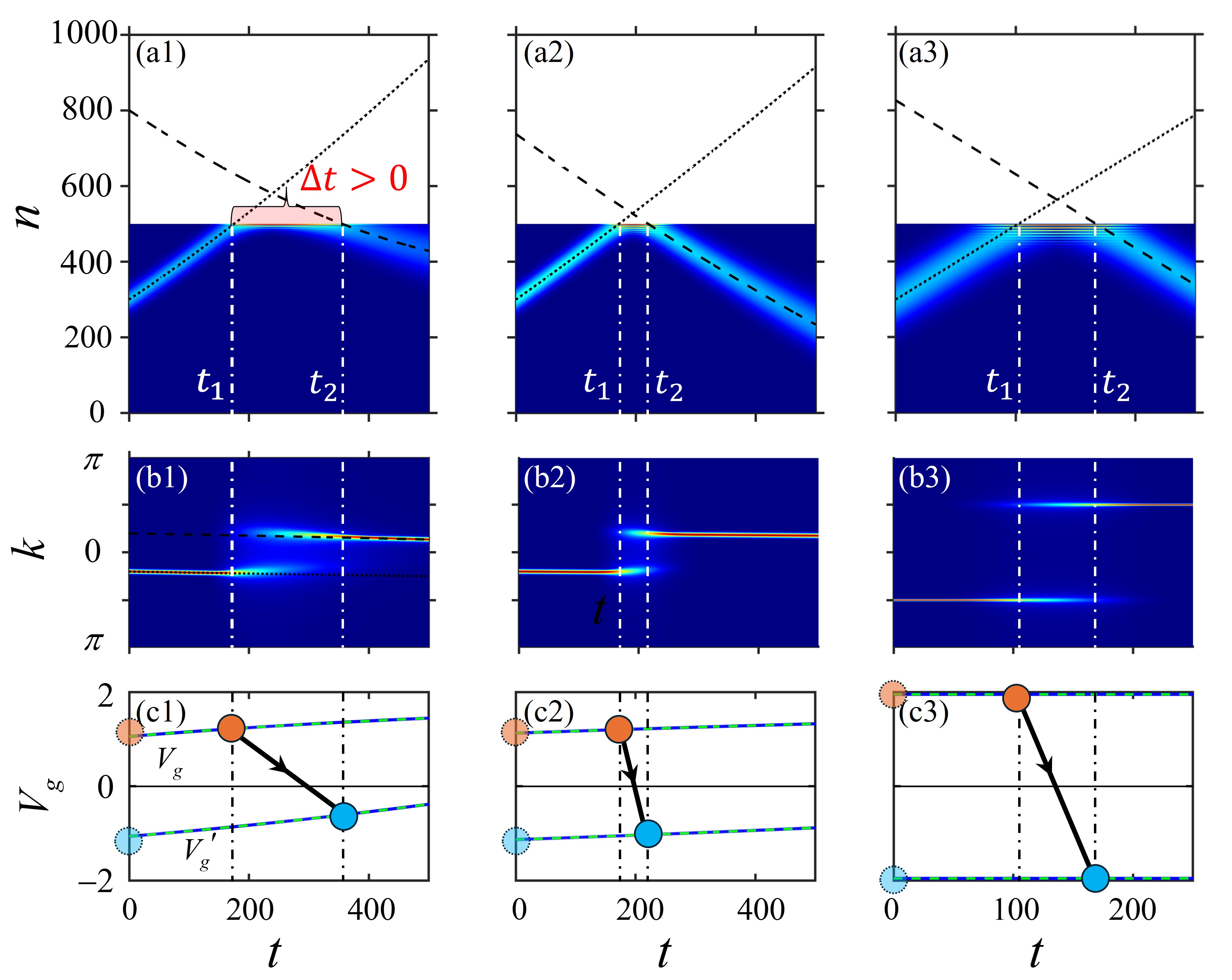}
\caption{
(a1-a3) Time evolution of the wave packet in the finite lattice of size $N = 500$ under the OBC. (b1-b3) The evolution of the wave packet in momentum space. (c1-c3) The evolution of group velocities  of the incident (upper lines) and reflected (lower lines) wave packets. The red (blue) dotted circles denote the initial incident (reflected) group velocity, which moves to that of the red (blue) solid
circles where the reflection occurs.
The parameters for (a1-c1) and (a3-c3) are the same as those in Figs.~4(f,~i) of the main text. The parameters for (a2-c2) are the same as those in (a1-c1) except $J_L=0.92,~J_R=1$.
\label{Fig_OBC_GH_2}}
\end{figure}

\clearpage

\section{\textcolor{black}{15. NH wave-packet jumps, dynamic skin effects, and TGHSs beyond the Hatano-Nelson model}}

\textcolor{black}{Our method {to understand the NH wave-packet jump close to the edge through the analysis of auxiliary wave packet} can be used for other NH models with real spectra beyond the Hatano-Nelson model} \textcolor{black}{which are called pseudo-Hermitian lattices. Pseudo-Hermitian Hamiltonians can be brought into Hermitian forms by similarity transformations, and hence they have the real spectra.}

\textcolor{black}{We take the \textcolor{black}{single-band non-Hermitian lattices with long-range couplings} as an example to show that the dynamic skin effects, NH wave-packet jumps, and positive/negative TGHSs can also occur. Let us assume the real-space Hamiltonian
\begin{align}
H=\begin{pmatrix}
\ddots & \vdots &\vdots &\vdots &\vdots &\vdots \\
\cdots& 0    & J_L    & \dfrac{J_L}{r} & 0      & 0      & \cdots \\
\cdots &J_R    & 0      & J_L            & \dfrac{J_L}{r} & 0      & \cdots \\
\cdots &J_R r  & J_R    & 0              & J_L    & \dfrac{J_L}{r} & \cdots \\
\cdots &0      & J_R r  & J_R            & 0      & J_L    & \cdots \\
\cdots &0      & 0      & J_R r          & J_R    & 0      & \cdots \\
 & \vdots & \vdots         & \vdots & \vdots & \vdots &\ddots
\end{pmatrix}.
\end{align}
Consider the similarity matrix
\begin{align}
S=\begin{pmatrix}
 r    & 0    & 0 & 0      & 0      & \cdots \\
0    & r^2      & 0            & 0 & 0      & \cdots \\
0 & 0    & r^3              & 0   & 0 & \cdots \\
0      & 0  &0            & r^4      & 0    & \cdots \\
0      & 0      & 0          & 0   & r^5      & \cdots \\
  \vdots& \vdots & \vdots         & \vdots & \vdots & \ddots
\end{pmatrix},
\end{align}
the  Hamiltonian under the similarity transformation becomes
\begin{align}
\bar{H}=S^{-1}HS=\begin{pmatrix}
\ddots & \vdots &\vdots &\vdots &\vdots &\vdots \\
\cdots& 0    & \sqrt{J_LJ_R}    & \sqrt{J_LJ_R} & 0      & 0      & \cdots \\
\cdots &\sqrt{J_LJ_R}    & 0      & \sqrt{J_LJ_R}           & \sqrt{J_LJ_R} & 0      & \cdots \\
\cdots &\sqrt{J_LJ_R}  & \sqrt{J_LJ_R}    & 0              & \sqrt{J_LJ_R}    & \sqrt{J_LJ_R} & \cdots \\
\cdots &0      & \sqrt{J_LJ_R}  & \sqrt{J_LJ_R}          & 0      & \sqrt{J_LJ_R}  & \cdots \\
\cdots &0      & 0      & \sqrt{J_LJ_R}         & \sqrt{J_LJ_R}    & 0      & \cdots \\
 & \vdots & \vdots         & \vdots & \vdots & \vdots &\ddots
\end{pmatrix},
\end{align}
which is Hermitian. The evolution equation is then 
\begin{align}
\psi_n(t)= e^{-i H t}\psi_n(0)
= S S^{-1} e^{-i H t} S S^{-1} \psi_n(0)
= S e^{-i \bar{H} t} \varphi_n(0),
\end{align}
where $\varphi_n(0)=S^{-1} \psi_n(0)$.}

\textcolor{black}{The band structure is shown in Fig.~\ref{Fig_OBC_long_range_band}(a), and the PBC and OBC spectra in the complex energy plane are shown in Fig.~\ref{Fig_OBC_long_range_band}(b). We see that the OBC spectrum is entirely real, which is different from the PBC spectrum with nontrivial point gaps. } 

\begin{figure}[h!]
\centering
\includegraphics[width=12 cm]{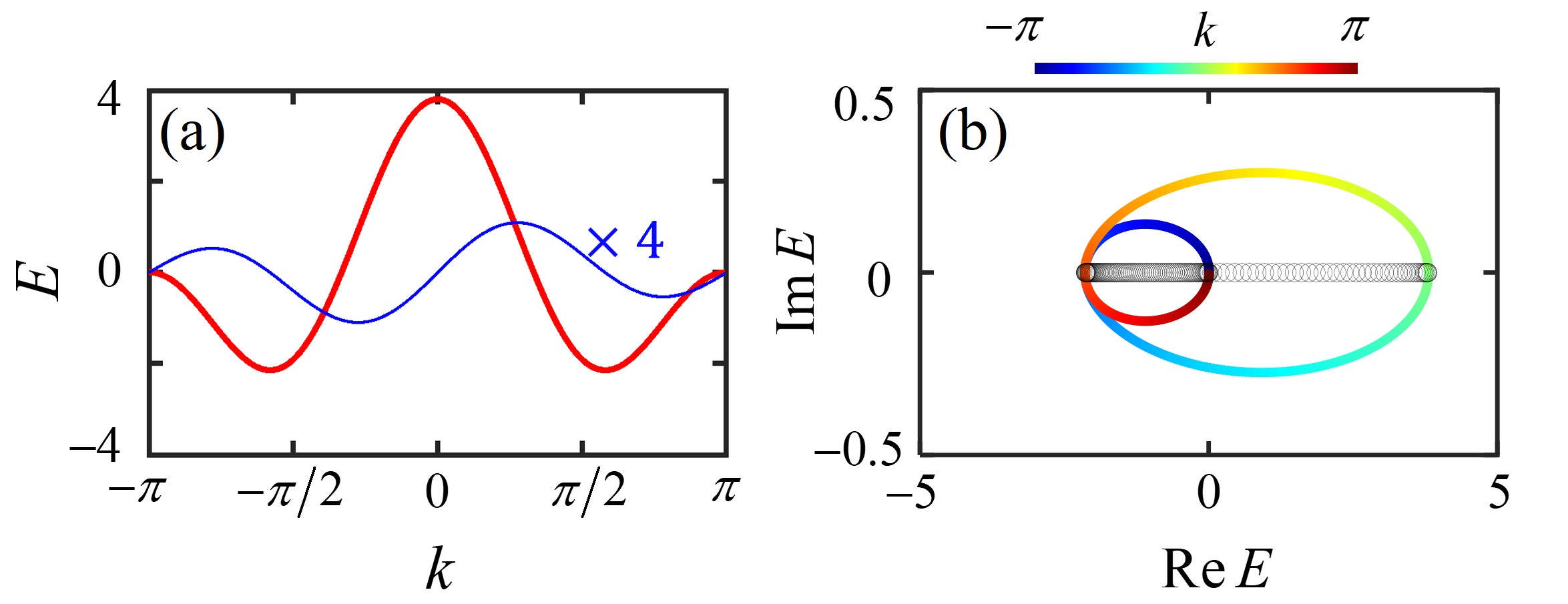}
\caption{(a) Band structure, $\mathrm{Re}(E)$ (red line) and $\mathrm{Im}(E)$ (blue line). (b) The PBC (colored) and OBC (black dots) spectra in the complex energy plane. 
The parameters are $J_1^L=J_L=1,~J_1^R=J_R=0.9,~J_2^L=J_L/r$, and $J_2^R=J_Rr$.
\label{Fig_OBC_long_range_band}}
\end{figure}

\begin{figure}[h!]
\centering
\includegraphics[width=15 cm]{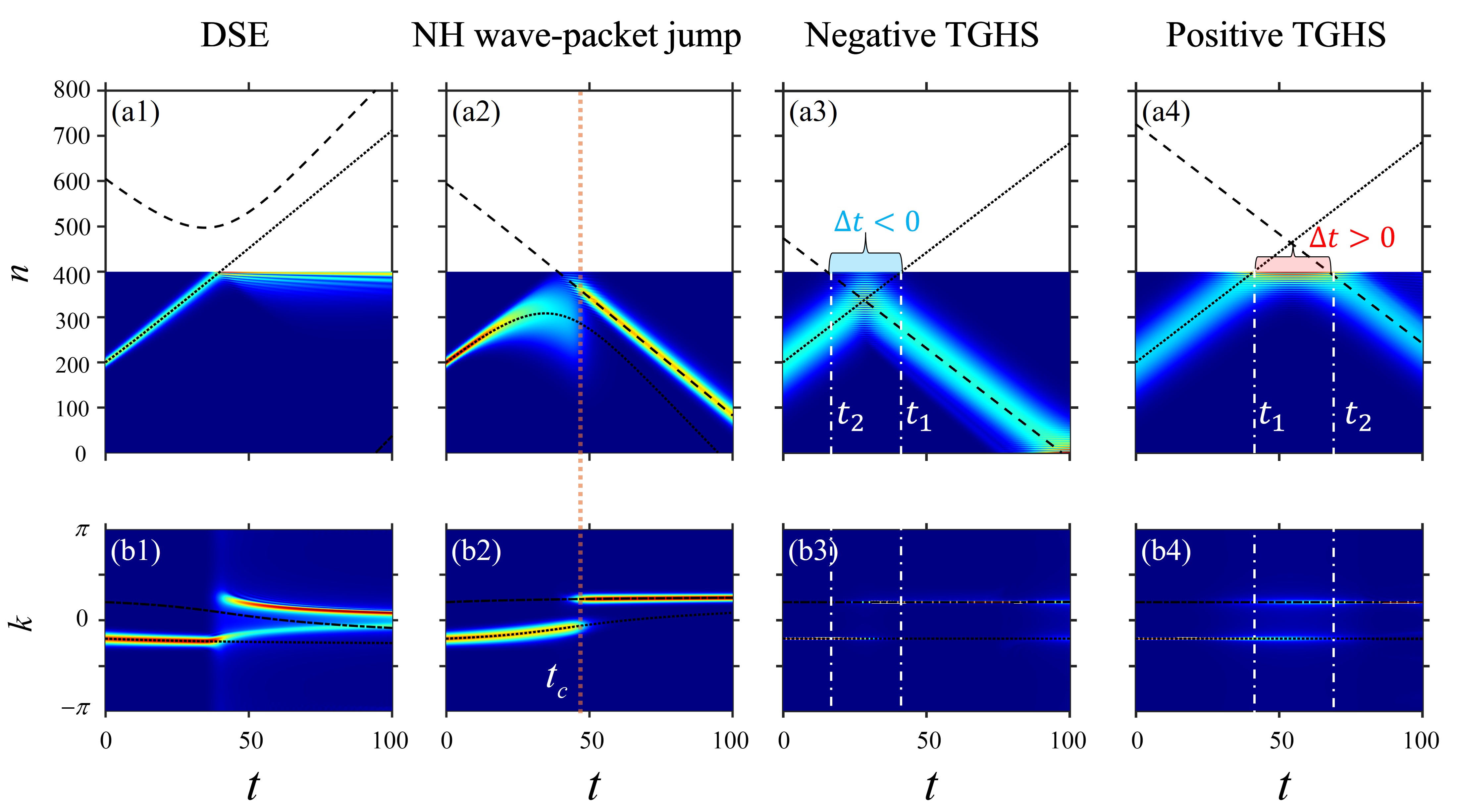}
\caption{\textcolor{black}{
(a1-a4) Time evolution of the wave packet in a finite lattice with size $N = 400$ under the OBC. (b1-b3) The evolution of the wave packet in momentum space. The parameters are $J_L=0.9,~J_R=1,~J_2^L=J_L/r,~J_2^R=J_Rr$, $k_0=-0.2\pi$, and $\sigma=5$ (a1,~b1). $J_L=1,~J_R=0.9,~J_2^L=J_L/r,~J_2^R=J_Rr$, $k_0=-0.2\pi$, and $\sigma=5$ (a2,~b2). $J_L=1,~J_R=0.95,~J_2^L=J_L/r,~J_2^R=J_Rr$, $k_0=-0.2\pi$, and $\sigma=35$ (a3,~b3). $J_L=0.95,~J_R=1,~J_2^L=J_L/r,~J_2^R=J_Rr$, $k_0=-0.2\pi$, and $\sigma=35$ (a4,~b4). }
\label{Fig_OBC_long_range_GH}}
\end{figure}

\textcolor{black}{The wave dynamics in real and momentum spaces is shown in Figs.~\ref{Fig_OBC_long_range_GH}(a2,~b2). We see that a NH wave-packet jump occurs near $t_c\approx50$ for real-space dynamics, at which the amplitude of the auxiliary wave packet becomes equal to that of the initial wave packet and then dominates the evolution. The dashed and dotted lines are the predicted center of mass of the auxiliary and initial wave packets, which agree well with the numerical results. The NH wave-packet jump in real space also indicates the jump in momentum space, as shown in Fig.~\ref{Fig_OBC_long_range_GH}(b2).}

\textcolor{black}{When the non-Hermitian direction is changed to be toward the edge, i.e., $J_L<J_R$, the NH wave-packet jump changes to the dynamic skin effect (DSE), as shown in Fig.~\ref{Fig_OBC_long_range_GH}(a1,~b1). In this case, the initial wave packet propagates through the edge, while the auxiliary wave packet does not reach the edge, which hence indicates the DSE phase.} 

\textcolor{black}{We now study the TGHSs in this lattice. We consider a larger width of the initial wave packet $\sigma=35$. For the NH direction that is opposite to the edge, i.e., $J_L>J_R$, a negative TGHS is observed with $\Delta t<0$ [see Figs.~\ref{Fig_OBC_long_range_GH}(a2)]. This negative THGS can be changed to positive one when the NH direction is changed toward the edge with $J_L<J_R$, as shown in  Fig.~\ref{Fig_OBC_long_range_GH} (a3). \textcolor{black}{We can obtain an analytical expression for the TGHS, which is the same as the one in Eq.~(6)} of the main text, 
\begin{align}
\Delta t=\frac{n_2-n_0'}{v_g(k_0)}=\frac{2\sigma^2 \mathrm{ln}({J_R/J_L})}{J_L+J_R},    
\end{align}
which shows $\Delta t>0~(<0)$ for $J_R>(<)~J_L$, and $\Delta t=0$ for the Hermitian case with $J_R=L_L$.}

\textcolor{black}{Thus, DSEs, NH wave-packet jumps, and TGHSs can arise in broad class of non-Hermitian lattices with pseudo-Hermitian Hamiltonians, and can be physically explained by the auxiliary wave packet theory.} 

\textcolor{black}{In the numerical simulations of this work, we have used the arbitrary precision library Advanpi in Matlab to improve the numerical precision. }



\end{document}